\documentclass{amsart} 
\usepackage{fullpage,psfig,amssymb,amsfonts,latexsym}
\newtheorem{rhp}{Riemann-Hilbert Problem}
\newtheorem{proposition}{Proposition}
\newtheorem{lemma}{Lemma}
\newtheorem{theorem}{Theorem}
\title[Asymptotics of Semiclassical Soliton Ensembles:  Rigorous Justification
of WKB]{Asymptotics of Semiclassical Soliton Ensembles:  Rigorous Justification
of the WKB Approximation}
\author{P. D. Miller}
\address{Department of Mathematics, The University of Michigan, East Hall, 525 E. University Avenue, Ann Arbor, MI 48109.  Email:  {\tt millerpd@umich.edu}}
\date{August 29, 2001}
\begin{document} 

\begin{abstract}
Rigorous pointwise asymptotics are established for semiclassical
soliton ensembles (SSEs) of the focusing nonlinear Schr\"odinger
equation using techniques of asymptotic analysis of matrix
Riemann-Hilbert problems.  The accumulation of poles in the
eigenfunction is handled using a new method in which the residues are
simultaneously interpolated at the poles by two distinct interpolants.
The results justify the WKB approximation for the nonselfadjoint
Zakharov-Shabat operator with real-analytic, bell-shaped, even
potentials.  The new technique introduced in this paper is applicable
to other problems as well: (i) it can be used to provide a unified
treatment by Riemann-Hilbert methods of the zero-dispersion limit of
the Korteweg-de Vries equation with negative (soliton generating)
initial data as studied by Lax, Levermore, and Venakides, and (ii) it
allows one to compute rigorous strong asymptotics for systems of
discrete orthogonal polynomials.
\end{abstract}

\maketitle

\section{Introduction}
Many important problems in the theory of integrable systems and
approximation theory can be recast as Riemann-Hilbert problems for a
matrix-valued unknown.  Via the connection with approximation theory,
and specifically the theory of orthogonal polynomials, one can also
study problems from the theory of random matrix ensembles and
combinatorics.  Roughly speaking, solving a Riemann-Hilbert problem
amounts to reconstructing a sectionally meromorphic matrix from given
homogeneous multiplicative ``jump conditions'' at the boundary
contours of the domains of meromorphy, from ``principal part data''
given at the prescribed singularities, and from a normalization
condition.  So, many asymptotic questions in integrable systems ({\em
e.g.} long time behavior and singular perturbation theory) and
approximation theory ({\em e.g.} behavior of orthogonal polynomials in
the limit of large degree) amount to determining asymptotic properties
of the solution matrix of a Riemann-Hilbert problem from given
asymptotics of the jump conditions and principal part data.

In recent years a collection of techniques has emerged for studying
certain asymptotic problems of this sort.  These techniques are
analogous to familiar asymptotic methods for expanding oscillatory
integrals, and we often refer to them as ``steepest-descent'' methods.
The basic method first appeared in the work of Deift and Zhou
\cite{DeiftZ93}.  The first applications were to Riemann-Hilbert
problems without poles, in which the solution matrix is sectionally
holomorphic.  Later, some problems were studied in which there were a
number of poles --- a number held fixed in the limit of interest ---
in the solution matrix (see, for example, the paper \cite{DeiftKKZ96}
on the long-time behavior of the Toda lattice with rarefaction initial
data).  The previous methods were extended to these more complicated
problems through the device of making a local change of variable near
each pole in some small domain containing the pole.  The change of
variable is chosen so that it has the effect of removing the pole at
the cost of introducing an explicit jump on the boundary of the domain
around the pole in which the transformation is made.  The result is a
Riemann-Hilbert problem for a sectionally holomorphic matrix, which
can be solved asymptotically by pre-existing ``steepest-descent''
methods.  Recovery of an approximation for the original sectionally
meromorphic matrix unknown involves putting back the poles by
reversing the explicit change of variables that was designed to get
rid of them to begin with.

Yet another category of Riemann-Hilbert problems consists of those
problems where the number of poles is not fixed, but becomes large in
the limit of interest, with the poles accumulating on some closed set
$F$ in the finite complex plane.  A problem of this sort has been
addressed
\cite{manifesto} by making an explicit transformation of the type
described above in a single fixed domain $G$ that contains the locus
of accumulation $F$ of all the poles.  The transformation is chosen to
get rid of all the poles at once.  In order to specify it, discrete
data related to the residues of the poles must be interpolated at the
corresponding poles by a function that is analytic and nonvanishing in
all of $G$.  Once the poles have been removed in this way, the
Riemann-Hilbert problem becomes one for a sectionally holomorphic
matrix, with a jump at the boundary of $G$ given in terms of the
explicit change of variables.  In this way, the poles are ``swept
out'' from $F$ to the boundary of $G$ resulting in an analytic jump.
There is a strong analogy in this procedure with the concept of
balayage (meaning ``sweeping out'') from potential theory.

In establishing asymptotic formulae for such Riemann-Hilbert problems,
it is essential that one make judicious use of the freedom to place
the boundary of the domain in which one removes the poles from the
problem.  Placing this boundary contour in the correct position in the
complex plane allows one to convert oscillations into exponential
decay in such a way that the errors in the asymptotics can be
rigorously controlled.  If the poles accumulate with some smooth
density on $F\subset G$, the characterization of the correct location
of the boundary of $G$ can be determined by first passing to a
continuum limit of the pole distribution in the resulting jump matrix
on the boundary of $G$, and then applying analytic techniques or
variational methods.  The continuum limit is justified as long as the
boundary of $G$ remains separated from $F$.

This idea leads to an interesting question.  What happens if the
boundary of $G$ as determined from passing to the continuum limit
turns out to intersect $F$?  Far from being a hypothetical
possibility, this situation is known to occur in at least three different
problems:
\begin{enumerate}
\item {\bf The semiclassical limit of the focusing nonlinear
Schr\"odinger hierarchy with decaying initial data.}  See
\cite{manifesto}.  This is an inverse-scattering problem for the
nonselfadjoint Zakharov-Shabat operator.  On an {\em ad hoc} basis,
one replaces the true spectral data for the given initial condition
with a formal WKB approximation.  There is no jump in the
Riemann-Hilbert problem associated with inverse-scattering for the
modified spectral data, but there are poles accumulating
asymptotically with the WKB density of states on an interval $F$ of
the imaginary axis in the complex plane of the eigenvalue.  The
methods described above turn out to yield rigorous asymptotics for
this modified inverse-scattering problem as long as the independent
time variable in the equation is not zero.  For $t=0$, the argument of
passing to the continuum limit in the pole density leads one to choose
the boundary of $G$ to {\em coincide} in part with the interval $F$.
Strangely, if one sets $t=0$ in the problem from the beginning, an
alternative method due to Lax and Levermore \cite{LaxL83} and extended
to the nonselfadjoint Zakharov-Shabat operator with real potentials by
Ercolani, Jin, Levermore, and MacEvoy \cite{ErcolaniJLM93} can be used
to carry out the asymptotic analysis in this special case; this
alternative method is not based on matrix Riemann-Hilbert problems,
and therefore when taken together with the methods described in
\cite{manifesto} does not result in a uniform treatment of the
semiclassical limit for all $x$ and $t$.  At the same time, the
Lax-Levermore method that applies when $t=0$ fails in this problem
when $t\neq 0$.
\item {\bf The zero-dispersion limit of the Korteweg-de Vries equation with
potential well initial data.}  As pointed out above, the original
treatment of this problem by Lax and Levermore \cite{LaxL83} was not
based on asymptotic analysis for a matrix-valued Riemann-Hilbert
problem.  But it is possible to pose the inverse-scattering problem
with modified (WKB) spectral data as a matrix-valued Riemann-Hilbert
problem and ask whether the ``steepest descent'' techniques for such
problems could be used to reproduce and/or strengthen the original
asymptotic results of Lax and Levermore.  In particular, we might
point out that the Lax-Levermore method only gives weak limits of the
conserved densities, and that a modification due to Venakides
\cite{Venakides90} is required to extract any pointwise asymptotics
({\em i.e.} to reconstruct the microstruture of the modulated and
rapidly oscillatory wavetrains giving rise to the leading-order weak
asymptotics).  On the other hand, ``steepest descent'' techniques for
matrix-valued Riemann-Hilbert problems typically give pointwise
asymptotics automatically.  It would therefore be most useful if these
techniques could be applied to provide a new and unified approach
to this problem.

If one tries to enclose the locus of accumulation of poles (WKB
eigenvalues for the Schr\"odinger operator with a potential well) with
a contour and determine the optimal location of this contour for
zero-dispersion asymptotics, it turns out that the contour must
contain the support of a certain weighted logarithmic equilibrium
measure.  It is a well-known consequence of the Lax-Levermore theory
that the support of this measure is a subset of the interval of
accumulation of WKB eigenvalues.  Consequently, the enclosing contour
``wants'' to lie right on top of the poles in this problem, and the
approach fails.  In a sense this failure of the ``steepest descent''
method is more serious than in the analogous problem for the focusing
nonlinear Schr\"odinger equation because the contour is in the wrong
place for all values of $x$ and $t$ (the independent variables of the
problem), whereas in the focusing nonlinear Schr\"odinger problem the
method fails generically only for $t=0$.
\item {\bf The large degree limit of certain systems of discrete 
orthogonal polynomials.}  Fokas, Its, and Kitaev \cite{FokasIK92} have
shown that the problem of reconstructing the orthogonal polynomials
associated with a given continuous weight function can be expressed as
a matrix-valued Riemann-Hilbert problem.  It is not difficult to
modify their construction to the case when the weight function is a
sum of Dirac masses.  The corresponding matrix-valued Riemann-Hilbert
problem has no jump, but has poles at the support nodes of the weight.
The solution of this Riemann-Hilbert problem gives in this case the
associated family of discrete orthogonal polynomials.  If one takes
the nodes of support of the discrete weight to be distributed
asymptotically in some systematic way, then it is natural to ask
whether ``steepest descent'' methods applied to the corresponding
Riemann-Hilbert problem with poles could yield accurate asymptotic
formulae for the discrete orthogonal polynomials in the limit of large
degree.  Indeed, similar asymptotics were obtained in the continuous
weight case \cite{DeiftKMVZ99} using precisely these methods.

Unfortunately, when the poles are encircled and the optimal contour is
sought, it turns out again to be necessary that the contour contain
the support of a certain weighted logarithmic equilibrium measure (see
\cite{KuijlaarsR98} for a description of this measure) which is
supported on a subset of the interval of accumulation of the nodes of
orthogonalization ({\em i.e.}, the poles).  For this reason, the
method based on matrix-valued Riemann-Hilbert problems would appear to
fail.
\end{enumerate}

In this paper, we present a new technique in the theory of ``steepest
descent'' asymptotic analysis for matrix Riemann-Hilbert problems that
solves all three problems mentioned above in a general framework.  We
illustrate the method in detail for the first case described above:
the inverse-scattering problem for the nonselfadjoint Zakharov-Shabat
operator with modified (WKB) spectral data, which amounts to a
treatment of the semiclassical limit for the focusing nonlinear
Schr\"odinger equation at the initial instant $t=0$.  This work thus
fills in a gap in the arguments in \cite{manifesto} connecting the
rigorous asymptotic analysis carried out there with the initial-value
problem for the focusing nonlinear Schr\"odinger equation.
Application of the same techniques to the zero dispersion limit of the
Korteweg-de Vries equation will be the topic of a future paper, and a
study of asymptotics for discrete orthogonal polynomials using these
methods is already in preparation \cite{BaikKMM01}.

The initial-value problem for the focusing nonlinear Schr\"odinger
equation is
\begin{equation}
i\hbar\frac{\partial\psi}{\partial t} + \frac{\hbar^2}{2}\frac{\partial^2\psi}
{\partial x^2} + |\psi|^2\psi = 0\,,
\label{eq:nls}
\end{equation}
subject to the initial condition $\psi(x,0)=\psi_0(x)$.  In
\cite{manifesto}, this problem is considered for cases when the
initial data $\psi_0(x)= A(x)$ where $A(x)$ is some positive real
function ${\mathbb R}\rightarrow (0,A]$.  The function $A(x)$ is taken
to decay rapidly at infinity and to be even in $x$ with a single
genuine maximum at $x=0$.  Thus $A(0)=A$, $A'(0)=0$, and $A''(0)<0$.
Also, the function $A(x)$ is taken to be real-analytic.  With this
given initial data, one has a unique solution of (\ref{eq:nls}) for
each $\hbar>0$.  To study the semiclassical limit then means
determining asymptotic properties of the family of solutions
$\psi(x,t)$ as $\hbar\downarrow 0$.

This problem is associated with the scattering and inverse-scattering theory
for the nonselfadjoint Zakharov-Shabat eigenvalue problem \cite{ZakharovS72}:
\begin{equation}
\begin{array}{rcl}
\displaystyle \hbar\frac{du}{dx} &=&-i\lambda u + A(x)v\\\\
\displaystyle \hbar\frac{dv}{dx} &=&-A(x)u + i\lambda v
\end{array}
\label{eq:ZS}
\end{equation}
for auxiliary functions $u(x;\lambda)$ and $v(x;\lambda)$.  The
complex number $\lambda$ is a spectral parameter.  Under the
conditions on $A(x)$ described above, it is known only that for each
$\hbar>0$ the discrete spectrum of this problem is invariant under
complex conjugation and reflection through the origin.  However, a
formal WKB method applied to (\ref{eq:ZS}) suggests for small $\hbar$
a distribution of eigenvalues that are confined to the imaginary axis.  
The same method suggests that the reflection coefficient for scattering
states obtained for real $\lambda$ is small beyond all orders.  

It is therefore natural to propose a modification of the problem.
Rather than studying the inverse-scattering problem for the true
spectral data (which is not known), simply replace the true spectral
data by its formal WKB approximation in which the eigenvalues are
given by a quantization rule of Bohr-Sommerfeld type, and in which the
reflection coefficient is neglected entirely.  For each $\hbar>0$,
this modified spectral data is the {\em true} spectral data for some
other ($\hbar$-dependent) initial condition $\psi^\hbar_0(x)$.  Since
there is no reflection coefficient in the modified problem, it turns
out that for each $\hbar$ the solution of (\ref{eq:nls}) corresponding
to the modified initial data $\psi^\hbar_0(x)$ is an exact $N$-soliton
solution, with $N\sim\hbar^{-1}$.  We call such a family of
$N$-soliton solutions, all obtained from the same function $A(x)$ by a
WKB procedure, a {\em semiclassical soliton ensemble}, or SSE for
short.  We will be more precise about this idea in \S~\ref{sec:SSEs}.
In \cite{manifesto}, the asymptotic behavior of SSEs was studied for
$t\neq 0$.  Although the results were rigorous, it was not possible to
deduce anything about the {\em true} initial-value problem for
(\ref{eq:nls}) with $\psi_0(x)\equiv A(x)$ because the asymptotic
method failed for $t=0$.  In this paper, we will explain the following
new result.
\begin{theorem}
Let $A(x)$ be real-analytic, even, and decaying with a single genuine
maximum at $x=0$.  Let $\psi_0^\hbar(x)$ be for each $\hbar>0$ the
exact initial value of the SSE corresponding to $A(x)$ (see \S~\ref{sec:SSEs}).
Then, there exists a sequence of values of
$\hbar$, $\hbar=\hbar_N$ for $N=1,2,3,\dots$, such that
\begin{equation}
\lim_{N\rightarrow\infty}\hbar_N = 0
\end{equation}
and such that for all $x\neq 0$, there exists a constant $K_x>0$ such that
\begin{equation}
|\psi_0^{\hbar_N}(x)-A(x)|\le K_x\hbar_N^{1/7-\nu}\,,\hspace{0.2 in}
\mbox{for} \hspace{0.2 in}N=1,2,3,\dots
\end{equation}
for all $\nu>0$.
\label{theorem:mainresult}
\end{theorem}
As $\psi_0^\hbar(x)$ is obtained by an inverse-scattering procedure
applied to WKB spectral data, this theorem establishes in a sense the
validity of the WKB approximation for the Zakharov-Shabat eigenvalue
problem (\ref{eq:ZS}).  It says that the true spectral data and the
formally approximate spectral data generate, via inverse-scattering,
potentials in the Zakharov-Shabat problem that are pointwise close.
The omission of $x=0$ is merely technical; a procedure slightly
different from that we will explain in this paper is needed to handle
this special case.  We will indicate as we proceed the modifications
that are necessary to extend the result to the whole real line.  The
pointwise nature of the asymptotics is important; the variational
methods used in \cite{ErcolaniJLM93} suggest convergence only in the
$L^2$ sense.  Rigorous statements about the nature of the WKB
approximation for the Zakharov-Shabat problem are especially
significant because the operator in (\ref{eq:ZS}) is nonselfadjoint
and the spectrum is not confined to any axis; furthermore
Sturm-Liouville oscillation theory does not apply.

\section{Characterization of SSEs}
\label{sec:SSEs}
Each $N$-soliton solution of the focusing nonlinear Schr\"odinger
equation (\ref{eq:nls}) can be found as the solution of a meromorphic
Riemann-Hilbert problem with no jumps; that is, a problem whose
solution matrix is a rational function of $\lambda\in {\mathbb C}$.
The $N$-soliton solution depends on a set of discrete data.  Given $N$
complex numbers $\lambda_0,\dots,\lambda_{N-1}$ in the upper
half-plane (these turn out to be discrete eigenvalues of the spectral
problem (\ref{eq:ZS})), and $N$ nonzero constants
$\gamma_0,\dots,\gamma_{N-1}$ (which turn out to be related to
auxiliary discrete spectrum for (\ref{eq:ZS})), and an index $J=\pm
1$, one considers the matrix ${\bf m}(\lambda)$ solving the following
problem:
\begin{rhp}[Meromorphic problem]
Find a matrix ${\bf m}(\lambda)$ with the following two properties:
\begin{enumerate}
\item
{\bf Rationality:} ${\bf m}(\lambda)$ is a rational function of
$\lambda$, with simple poles confined to the values $\{\lambda_k\}$
and the complex conjugates.  At the singularities:
\begin{equation}
\begin{array}{rcl}
\displaystyle
\mathop{\rm Res}_{\lambda=\lambda_k}
{\bf m}(\lambda)&=&\displaystyle\lim_{\lambda\rightarrow\lambda_k}{\bf m}(\lambda)
\sigma_1^{\frac{1-J}{2}}\left[\begin{array}{cc}0 & 0\\c_k(x,t) & 0\end{array}
\right]\sigma_1^{\frac{1-J}{2}}\,,\\\\
\displaystyle
\mathop{\rm Res}_{\lambda=\lambda_k^*}
{\bf m}(\lambda)&=&\displaystyle\lim_{\lambda\rightarrow\lambda_k^*}{\bf m}(\lambda)
\sigma_1^{\frac{1-J}{2}}\left[\begin{array}{cc}0 & -c_k(x,t)^*\\0 & 0\end{array}
\right]\sigma_1^{\frac{1-J}{2}}\,,
\end{array}
\label{eq:residuerelation}
\end{equation}
for $k=0,\dots,N-1$, with
\begin{equation}
c_k(x,t):=\left(\frac{1}{\gamma_k}\right)^J\frac{\displaystyle\prod_{n=0}^{N-1}
(\lambda_k-\lambda_n^*)}{\displaystyle
\mathop{\prod_{n=0}^{N-1}}_{n\neq k}(\lambda_k-\lambda_n)}\exp(2iJ(\lambda_kx+\lambda_k^2 t)/\hbar)\,.
\end{equation}
\item
{\bf Normalization:}
\begin{equation}
{\bf m}(\lambda)\rightarrow {\mathbb I}\,,\hspace{0.2 in}\mbox{as}\hspace{0.2 in} \lambda\rightarrow \infty\,.
\end{equation}
\end{enumerate}
\label{rhp:meromorphic}
\end{rhp}

\vspace{0.1 in}

Here, $\sigma_1$ denotes one of the Pauli matrices:
\begin{equation}
\sigma_1:=\left[\begin{array}{cc} 0 & 1\\1 & 0\end{array}\right]\,,
\hspace{0.2 in}
\sigma_2:=\left[\begin{array}{cc} 0 & -i\\i & 0\end{array}\right]\,,
\hspace{0.2 in}
\sigma_3:=\left[\begin{array}{cc} 1 & 0\\0 & -1\end{array}\right]\,.
\end{equation}
The function $\psi(x,t)$ defined from ${\bf m}(\lambda)$ by the limit
\begin{equation}
\psi(x,t)=2i\lim_{\lambda\rightarrow\infty}\lambda m_{12}(\lambda)
\label{eq:potentialrecovery}
\end{equation}
is the $N$-soliton solution of the focusing nonlinear Schr\"odinger
equation (\ref{eq:nls}) corresponding to the data $\{\lambda_k\}$ and
$\{\gamma_k\}$.

The index $J$ will be present throughout this work, so it is worth
explaining its role from the start.  It turns out that if $J=+1$, then
the solution ${\bf m}(\lambda)$ of Riemann-Hilbert
Problem~\ref{rhp:meromorphic} has the property that for all fixed $\lambda$
distinct from the poles, ${\bf m}(\lambda)\rightarrow{\mathbb I}$ as
$x\rightarrow +\infty$.  Likewise, if $J=-1$, then ${\bf m}(\lambda)
\rightarrow {\mathbb I}$ as $x\rightarrow -\infty$.  So as far as scattering
theory is concerned, the index $J$ indicates an arbitrary choice of
whether we are performing scattering ``from the right'' or ``from the
left''.  Both versions of scattering theory yield the same function
$\psi(x,t)$ via the relation (\ref{eq:potentialrecovery}), and are in
this sense equivalent.  However, the inverse-scattering problem
involves the independent variables $x$ and $t$ for (\ref{eq:nls}) as
parameters, and it may be the case that for different choices of $x$
and $t$, different choices of the parameter $J$ may be more convenient
for asymptotic analysis of the matrix ${\bf m}(\lambda)$ solving
Riemann-Hilbert Problem~\ref{rhp:meromorphic}.  That this is indeed
the case was observed and documented in \cite{manifesto}.  So we need
the freedom to choose the index $J$, and therefore we need to carry it
along in our calculations.

A {\em semiclassical soliton ensemble} (SSE) is a family of particular
$N$-soliton solutions of (\ref{eq:nls}) indexed by $N=1,2,3,4,\dots$
that are formally associated with given initial data $\psi_0(x)=A(x)$
via an {\em ad hoc} WKB approximation of the spectrum of
(\ref{eq:ZS}).  Note that the initial data $\psi_0(x)=A(x)$ may not
exactly correspond to a pure $N$-soliton solution of (\ref{eq:nls})
for any $\hbar$, and similarly that typically none of the $N$-soliton
solutions making up the SSE associated with $\psi_0(x)=A(x)$ will
agree with this given initial data at $t=0$.

We will now describe the discrete data $\{\lambda_k\}$ and
$\{\gamma_k\}$ that generate, via the solution of Riemann-Hilbert
Problem~\ref{rhp:meromorphic} and the subsequent use of the formula
(\ref{eq:potentialrecovery}), the SSE associated with a function
$\psi_0(x)=A(x)$.  We suppose that $A(x)$ is an even function of $x$
that has a single maximum at $x=0$, and is therefore ``bell-shaped''.
We will need $A(x)$ to be rapidly decreasing for large $x$, and we
will suppose that the maximum $A:=A(0)$ is genuine in that $A''(0)<0$.
Most importantly in what follows, we will assume that $A(x)$ is a
real-analytic function of $x$.

The starting point is the definition of the WKB eigenvalue density
function $\rho^0(\eta)$:
\begin{equation}
\rho^0(\eta):=\frac{\eta}{\pi}\int_{x_-(\eta)}^{x_+(\eta)}
\frac{dx}{\sqrt{A(x)^2+\eta^2}}\,,
\label{eq:rho0}
\end{equation}
defined for positive imaginary numbers $\eta$ in the interval
$(0,iA)$, where $x_-(\eta)$ and $x_+(\eta)$ are the (unique by our
assumptions) negative and positive values of $x$ for which
$iA(x)=\eta$.  The WKB eigenvalues asymptotically fill out the
interval $(0,iA)$, and $\rho^0(\eta)$ is their asymptotic density.
This function inherits analyticity properties in $\eta$ from those of
$A(x)$ via the functions $x_\pm(\eta)$.  Our assumption that $A(x)$ is
real-analytic makes $\rho^0(\eta)$ an analytic function of $\eta$ in
its imaginary interval of definition.  Also, our assumption that
$A(x)$ should be rapidly decreasing makes $\rho^0(\eta)$ analytic at
$\eta=0$, and our assumption that $A(x)$ have nonvanishing curvature
at its maximum makes $\rho^0(\eta)$ analytic at $\eta=iA$.  From this
function it is convenient to define a measure of the number of WKB
eigenvalues between a point $\lambda\in (0,iA)$ on the imaginary axis
and $iA$:
\begin{equation}
\theta^0(\lambda):=-\pi\int_\lambda^{iA}\rho^0(\eta)\,d\eta\,.
\label{eq:deftheta0}
\end{equation}

Now, each $N$-soliton solution in the SSE for $A(x)$ will be
associated with a particular value $\hbar=\hbar_N$, namely
\begin{equation}
\hbar=\hbar_N:=-\frac{1}{N}\int_0^{iA}\rho^0(\eta)\,d\eta = \frac{1}{N\pi}
\int_{-\infty}^\infty A(x)\,dx
\label{eq:quantization}
\end{equation}
where $N\in{\mathbb Z}_+$.  In this sense we are taking the values of
$\hbar$ themselves to be ``quantized''.  Clearly for any given $A(x)$,
$\hbar_N=O(1/N)$ which goes to zero as $N$ becomes large.  For each
$N\in{\mathbb Z}_+$, we then define the WKB eigenvalues formally
associated with $A(x)$ according to the Bohr-Sommerfeld rule
\begin{equation}
\theta^0(\lambda_k) = \pi\hbar_N(k+1/2)\,,\hspace{0.2 in}
\mbox{for}\hspace{0.2 in}k=0,1,2,\dots,N-1
\label{eq:BS}
\end{equation}
and the auxiliary scattering data by
\begin{equation}
\gamma_k:=-i(-1)^K\exp\,\left(-\frac{i(2K+1)\theta^0(\lambda_k)}{\hbar_N}
\right)\,.
\label{eq:gammainterpolate}
\end{equation}
Here, $K$ is an arbitrary integer.  Clearly the Bohr-Sommerfeld rule
(\ref{eq:BS}) implies that choosing different integer values of $K$ in
(\ref{eq:gammainterpolate}) will yield the same set of numbers
$\{\gamma_k\}$.  However, we take the point of view that the
right-hand side of (\ref{eq:gammainterpolate}) furnishes an analytic
function that interpolates the $\{\gamma_k\}$ at the $\{\lambda_k\}$;
for different $K\in{\mathbb Z}$ these are different interpolating
functions which is a freedom that we will exploit to our advantage.
In fact, we will only need to consider $K=0$ or $K=-1$.

For $A(x)$ given as above, the SSE is a sequence of exact solutions of
(\ref{eq:nls}) such that the $N$th element $\psi^{\hbar_N}(x,t)$ of
the SSE (i) solves (\ref{eq:nls}) with $\hbar=\hbar_N$ as given by
(\ref{eq:quantization}) and (ii) is defined as the $N$-soliton
solution corresponding to the eigenvalues $\{\lambda_k\}$ given by
(\ref{eq:BS}) and the auxiliary spectrum $\{\gamma_k\}$ given by
(\ref{eq:gammainterpolate}) via the solution of Riemann-Hilbert
Problem~\ref{rhp:meromorphic} with $\hbar=\hbar_N$.  For each $N$, we
restrict the SSE to $t=0$ to obtain functions
\begin{equation}
\psi_0^{\hbar_N}(x):=\psi^{\hbar_N}(x,0)\,.
\label{eq:psi0define}
\end{equation}
It is this sequence of functions that is the subject of
Theorem~\ref{theorem:mainresult}.  In the following sections we will
set up a new framework for the asymptotic analysis of SSEs in the
limit $N\rightarrow\infty$, a problem closely related to the
computation of asymptotics of solutions of (\ref{eq:nls}) for fixed
initial data $\psi_0(x)=A(x)$ in the semiclassical limit.  

\section{Removal of the Poles}
The asymptotic method we will now develop for studying Riemann-Hilbert
Problem~\ref{rhp:meromorphic} for SSEs is especially well-adapted to
studying the case of $t=0$, where the method described in detail in
\cite{manifesto} fails.  To illustrate the new method, we therefore set
$t=0$ in the rest of this paper.  Also, we anticipate the utility of
tying the value of the parameter $J=\pm 1$ to the remaining
independent variable $x$ by setting
\begin{equation}
J:={\rm sign}(x)\,.
\label{eq:Jdef}
\end{equation}
In all subsequent formulae in which the index $J$ appears it should
be assumed to be assigned a definite value according to (\ref{eq:Jdef}).

We now want to convert Riemann-Hilbert Problem~\ref{rhp:meromorphic}
into a new Riemann-Hilbert problem for a sectionally holomorphic
matrix so that the ``steepest-descent'' methods can be applied.  As
mentioned in the introduction, in \cite{manifesto} this transformation
can be accomplished by encircling the locus of accumulation of the poles,
here the imaginary interval $(0,iA)$, with a loop contour in the upper
half-plane and making a specific change of variables based on the
interpolation formula (\ref{eq:gammainterpolate}) for some value of
$K\in{\mathbb Z}$ in the interior of the region enclosed by the loop
and also in the complex-conjugate region.  One then tries to choose
the position of the loop contour in the complex plane that is best
adapted to asymptotic analysis of the resulting holomorphic
Riemann-Hilbert problem.  The trouble with this approach is that it
turns out that for $t=0$ the ``correct'' placement of the contour
requires that part of it should lie on a subset of the imaginary
interval $(0,iA)$, that is, right on top of the accumulating poles!
For such a choice of the loop contour, the boundary values taken by
the transformed matrix on the outside of the loop would be singular
and the ``steepest descent'' theory would not apply.

\begin{figure}[h] 
\begin{center} 
\mbox{\psfig{file=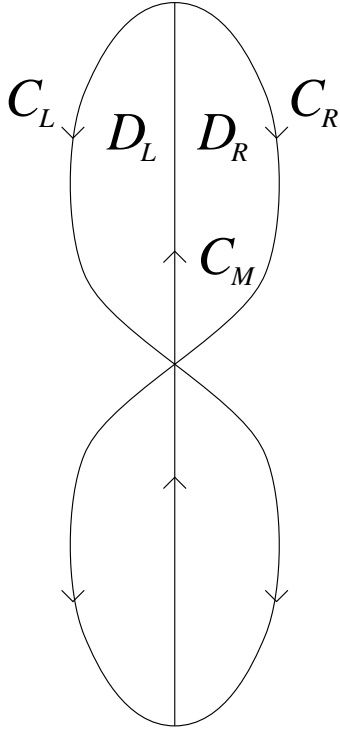,width=1.75 in}}
\end{center} 
\caption{\em The geometry of contours introduced in the
complex $\lambda$-plane.  The uppermost common point of the contours
$C_L$ $C_M$ and $C_R$ is $\lambda=iA$.  The six-fold self-intersection
point is the origin $\lambda=0$.  }
\label{fig:twohalves} 
\end{figure}

So taking the point of view that making any particular choice of
$K\in{\mathbb Z}$ in (\ref{eq:gammainterpolate}) leads to problems, we
propose to simultaneously make use of {\em two distinct values of $K$}
in passing to a Riemann-Hilbert problem for a sectionally holomorphic
matrix.  Consider the contours illustrated in
Figure~\ref{fig:twohalves},
arranged such that $\{\lambda_0,\dots,\lambda_{N-1}\}\subset D_L\cup
D_R$.  For $\lambda\in D_L$, set
\begin{equation}
{\bf M}(\lambda):={\bf m}(\lambda)\sigma_1^{\frac{1-J}{2}}
\left[\begin{array}{cc}
1 & 0 \\\\
\displaystyle i\left(\prod_{k=0}^{N-1}\frac{\lambda-\lambda_k^*}
{\lambda-\lambda_k}\right)
\exp\,\left(\frac{2i\lambda|x|-i\theta^0(\lambda)}{\hbar_N}
\right)
& 1\end{array}\right]\sigma_1^{\frac{1-J}{2}}\,.
\label{eq:lefttransform}
\end{equation}
For $\lambda\in D_R$, set
\begin{equation}
{\bf M}(\lambda):={\bf m}(\lambda)\sigma_1^{\frac{1-J}{2}}
\left[\begin{array}{cc}
1 & 0 \\\\
\displaystyle -i\left(\prod_{k=0}^{N-1}\frac{\lambda-\lambda_k^*}
{\lambda-\lambda_k}\right)
\exp\,\left(\frac{2i\lambda|x|+i\theta^0(\lambda)}{\hbar_N}
\right)
& 1\end{array}\right]\sigma_1^{\frac{1-J}{2}}\,.
\label{eq:righttransform}
\end{equation}
For $\lambda\in D_L^*\cup D_R^*$ set ${\bf M}(\lambda):=\sigma_2{\bf
M}(\lambda^*)^*\sigma_2$, and for all other complex $\lambda$ set
${\bf M}(\lambda)={\bf m}(\lambda)$.  So rather than enclosing the
poles in a loop and making a single change of variables inside, we are
splitting the region inside the loop in half, and we are using
different interpolants (\ref{eq:gammainterpolate}) of the
$\{\gamma_k\}$ at the $\{\lambda_k\}$ in each half of the loop.  Some
of the properties of the transformed matrix ${\bf M}(\lambda)$ are the
following.

\begin{proposition}
The matrix ${\bf M}(\lambda)$ is analytic in ${\mathbb C}\setminus
\Sigma$ where $\Sigma$ is the union of the contours $C_L$, $C_R$, and
$C_M$, and their complex conjugates.  Moreover, ${\bf M}(\lambda)$ takes
continuous boundary values on $\Sigma$.  
\label{prop:Manalytic}
\end{proposition}

{\em Proof:} The function $\theta^0(\lambda)$ is analytic in $D_L$ and
$D_R$ if $C_L$ and $C_R$ are chosen close enough to the imaginary axis
since $\rho^0(\eta)$ is analytic there.  By using the residue relation
(\ref{eq:residuerelation}) and the interpolation formula
(\ref{eq:gammainterpolate}) alternatively for $K=0$ and $K=-1$, one
checks directly that the poles of ${\bf m}(\lambda)$ are canceled by
the explicit Blaschke factors in (\ref{eq:lefttransform}) and
(\ref{eq:righttransform}). $\Box$

\vspace{0.1 in}

\begin{proposition}
Let ${\bf M}_\pm(\lambda)$ denote the boundary values taken on the
oriented contour $\Sigma$, where the subscript ``$+$'' (respectively
``$-$'') indicates the boundary value taken from the left
(respectively from the right).  Then for $\lambda\in C_L$,
\begin{equation}
{\bf M}_-(\lambda)^{-1}{\bf M}_+(\lambda)=
\sigma_1^{\frac{1-J}{2}}\left[\begin{array}{cc}
1 & 0\\\\
\displaystyle i
\left(\prod_{k=0}^{N-1}\frac{\lambda-\lambda_k^*}{\lambda-\lambda_k}\right)
\exp\,\left(\frac{2i\lambda |x| -i\theta^0(\lambda)}{\hbar_N}
\right) & 1
\end{array}\right]\sigma_1^{\frac{1-J}{2}}\,.
\label{eq:MjumpCL}
\end{equation}
For $\lambda\in C_R$,
\begin{equation}
{\bf M}_-(\lambda)^{-1}{\bf M}_+(\lambda)=
\sigma_1^{\frac{1-J}{2}}\left[\begin{array}{cc}
1 & 0\\\\
\displaystyle i
\left(\prod_{k=0}^{N-1}\frac{\lambda-\lambda_k^*}{\lambda-\lambda_k}\right)
\exp\,\left(\frac{2i\lambda |x| +i\theta^0(\lambda)}{\hbar_N}
\right) & 1
\end{array}\right]\sigma_1^{\frac{1-J}{2}}\,.
\label{eq:MjumpCR}
\end{equation}
For $\lambda\in C_M$, 
\begin{equation}
{\bf M}_-(\lambda)^{-1}{\bf M}_+(\lambda)=
\sigma_1^{\frac{1-J}{2}}\left[\begin{array}{cc}
1 & 0\\\\
\displaystyle i
\left(\prod_{k=0}^{N-1}\frac{\lambda-\lambda_k^*}{\lambda-\lambda_k}\right)
\exp\,\left(\frac{2i\lambda |x|}{\hbar_N}
\right)\cdot 2\cos\,
\left(\frac{\theta^0(\lambda)}{\hbar_N}\right) & 1
\end{array}\right]\sigma_1^{\frac{1-J}{2}}\,.
\label{eq:MjumpCM}
\end{equation}
On the contours in the lower half-plane the jump relations are
determined by the symmetry ${\bf M}(\lambda)=\sigma_2{\bf
M}(\lambda^*)^*\sigma_2$.  All jump matrices are
analytic functions in the vicinity of their respective contours.
\label{prop:Mjump}
\end{proposition}

{\em Proof:} This is also a direct consequence of
(\ref{eq:lefttransform}) and (\ref{eq:righttransform}). The
analyticity is clear on $C_L$ and $C_R$ since $\theta^0(\lambda)$ is
analytic there, while on $C_M$ one observes that as a consequence of
the Bohr-Sommerfeld quantization condition (\ref{eq:BS}), the cosine
factor precisely cancels the poles on $C_M$ contributed by the product
of Blaschke factors.  $\Box$

\vspace{0.1 in}

Although we have specified the contour $C_M$ to coincide with a
segment of the imaginary axis, the reader will see that the same
statements concerning the analyticity of ${\bf M}(\lambda)$ and the
continuity of the boundary values on $\Sigma$ also hold when $C_M$ is
taken to be absolutely any smooth contour in the upper half-plane
connecting $\lambda=0$ to $\lambda=iA$.  Given a choice of $C_M$, the
contours $C_L$ and $C_R$ must be such that the topology of
Figure~\ref{fig:twohalves} is preserved.  We also have specified that
$C_L$ and $C_R$ should lie sufficiently close to $C_M$ (a distance
independent of $\hbar_N$) so that $\theta^0(\lambda)$ is analytic in
$D_L$ and $D_R$.  Later we will also exploit the proximity of these
two contours to $C_M$ to deduce decay properties of certain analytic
functions on these contours from their oscillation properties on $C_M$
by the Cauchy-Riemann equations.

Taken together, Proposition~\ref{prop:Manalytic} and
Proposition~\ref{prop:Mjump} indicate that the matrix ${\bf
M}(\lambda)$ satisfies a Riemann-Hilbert problem without poles, but
instead having explicit homogeneous jump relations on $\Sigma$ given
by the matrix functions on the right-hand sides of (\ref{eq:MjumpCL}),
(\ref{eq:MjumpCR}), and (\ref{eq:MjumpCM}).  The normalization of
${\bf M}(\lambda)$ at infinity is the same as that of ${\bf
m}(\lambda)$ since no transformation has been made outside a compact
set, so if ${\bf M}(\lambda)$ can be recovered from its jump relations
and normalization condition, then the SSE itself can be obtained for
$t=0$ from (\ref{eq:potentialrecovery}) with ${\bf m}(\lambda)$
replaced by ${\bf M}(\lambda)$.

\section{The Complex Phase Function}
We now introduce a further change of dependent variable involving a
scalar function that is meant to capture the dominant asymptotics for
the problem.  Let $g(\lambda)$ be a complex-valued function that is
independent of $\hbar$, analytic for $\lambda\in {\mathbb C}\setminus
(C_M\cup C_M^*)$ taking continuous boundary values, satisfies
$g(\lambda)+g(\lambda^*)^*=0$, and $g(\infty)=0$.  Setting
\begin{equation}
{\bf N}(\lambda):={\bf M}(\lambda)\exp(-g(\lambda)\sigma_3/\hbar)
\end{equation}
we find that for $\lambda\in C_L$,
\begin{equation}
{\bf N}_-(\lambda)^{-1}{\bf N}_+(\lambda)=\sigma_1^{\frac{1-J}{2}}
\left[\begin{array}{cc} 1 & 0\\
a_L(\lambda)
& 1\end{array}\right]
\sigma_1^{\frac{1-J}{2}}\,,
\label{eq:CLjumpN}
\end{equation}
where
\begin{equation}
a_L(\lambda):=
i\left(\prod_{k=0}^{N-1}\frac{\lambda-\lambda_k^*}{\lambda-\lambda_k}\right)
\exp\,\left(\frac{2i\lambda |x| - i\theta^0(\lambda)
-2Jg(\lambda)}{\hbar_N}\right)\,.
\end{equation}
Similarly, for $\lambda\in C_R$, we find
\begin{equation}
{\bf N}_-(\lambda)^{-1}{\bf N}_+(\lambda)=\sigma_1^{\frac{1-J}{2}}
\left[\begin{array}{cc} 1 & 0\\a_R(\lambda)
& 1\end{array}\right]
\sigma_1^{\frac{1-J}{2}}\,,
\label{eq:CRjumpN}
\end{equation}
where
\begin{equation}
a_R(\lambda):=
i\left(\prod_{k=0}^{N-1}\frac{\lambda-\lambda_k^*}{\lambda-\lambda_k}\right)
\exp\,\left(\frac{2i\lambda |x| +i\theta^0(\lambda)
-2Jg(\lambda)}{\hbar_N}\right)\,. 
\end{equation}
Finally, for $\lambda\in C_M$, 
\begin{equation}
{\bf N}_-(\lambda)^{-1}{\bf N}_+(\lambda)=\sigma_1^{\frac{1-J}{2}}
\left[\begin{array}{cc} \exp(i\theta(\lambda)/\hbar_N) & 0\\
a_M(\lambda)& 
\exp(-i\theta(\lambda)/\hbar_N)\end{array}\right]
\sigma_1^{\frac{1-J}{2}}\,,
\label{eq:CMjumpN}
\end{equation}
where
\begin{equation}
\displaystyle 
a_M(\lambda):=i\left(\prod_{k=0}^{N-1}\frac{\lambda-\lambda_k^*}{\lambda-\lambda_k}\right)
\exp\,\left(\frac{2i\lambda |x| 
-Jg_+(\lambda)-Jg_-(\lambda)}{\hbar_N}\right) \cdot 2\cos\,\left(
\frac{\theta^0(\lambda)}{\hbar_N}
\right)\,,
\end{equation}
and
\begin{equation}
\theta(\lambda):=iJ(g_+(\lambda)-g_-(\lambda))\,.
\label{eq:defthetaoriginal}
\end{equation}

This means that given a function $g(\lambda)$ with the properties
described above, one finds that the matrix ${\bf N}(\lambda)$
satisfies another holomorphic Riemann-Hilbert problem with jump
conditions determined from (\ref{eq:CLjumpN}), (\ref{eq:CRjumpN}), and
(\ref{eq:CMjumpN}).  Because $g(\infty)=0$ and $g(\lambda)$ is
analytic near infinity, it follows that the correct normalization
condition for ${\bf N}(\lambda)$ is again that ${\bf
N}(\lambda)\rightarrow{\mathbb I}$ as $\lambda\rightarrow\infty$.
These same conditions on $g(\lambda)$ show that if ${\bf N}(\lambda)$
can be found from its jump conditions and normalization condition,
then the SSE can be found via (\ref{eq:potentialrecovery}) with ${\bf
m}(\lambda)$ replaced by ${\bf N}(\lambda)$.  

The function $g(\lambda)$ is called a {\em complex phase function}.
The advantage of introducing it into the problem is that by choosing
it correctly, the jump matrices (\ref{eq:CLjumpN}),
(\ref{eq:CRjumpN}), and (\ref{eq:CMjumpN}) can be cast into a form
that is especially convenient for analysis in the semiclassical limit
of $\hbar_N\rightarrow 0$.  The idea of introducing the complex phase
function to assist in finding the leading-order asymptotics and
controlling the error in this way first appeared in the paper
\cite{DeiftVZ94} as a modification of the ``steepest-descent'' method
proposed in
\cite{DeiftZ93}.

\section{Pointwise Semiclassical Asymptotics of the Jump Matrices}
For our purposes, we would like to have each element of the jump
matrix for ${\bf N}(\lambda)$ of the form $\exp(f(\lambda)/\hbar_N)$
for some appropriate function $f(\lambda)$ that is independent of
$\hbar_N$.  While this is not true strictly speaking, it becomes a
good approximation in the limit $\hbar_N\rightarrow 0$ with $\lambda$
held fixed (the approximation is not uniform near $\lambda=0$ or
$\lambda=iA$).  In this section, we describe the pointwise asymptotics
of the jump matrix for ${\bf N}(\lambda)$ with the aim of writing all
nonzero matrix elements asymptotically in the form
$\exp(f(\lambda)/\hbar_N)$ with a small relative error whose magnitude
we can estimate.

Roughly speaking, the intuition is that the product over $k$ of
Blaschke factors should be replaced with an exponential of a sum over
$k$ of logarithms.  The latter sum goes over to an integral that
scales like $\hbar_N^{-1}$ in the semiclassical limit.  On the contour
$C_M$, the cosine that cancels the poles must also be encorporated
into the asymptotics.

The branch of the logarithm that is convenient to use here is most conveniently
viewed as a function of two complex variables:
\begin{equation}
L^0_\eta(\lambda):=\log(-i(\lambda-\eta))+\frac{i\pi}{2}\,.
\end{equation}
As a function of $\lambda$ for fixed $\eta$, it is a logarithm that is
cut downwards in the negative imaginary direction from the logarithmic
pole at $\lambda=\eta$.  Equivalently, $L^0_\eta(\lambda)$ can be
viewed as the branch of the multivalued function $\log(\lambda-\eta)$
for which $\arg(\lambda-\eta)\in (-\pi/2,3\pi/2)$.  Suppose $\eta\in
C_M$.  The boundary value of $L_\eta^0(\lambda)$ taken on $C_M$
as $\lambda$ approaches from the left (respectively right) side is
denoted by $L_{\eta +}^0(\lambda)$ (respectively $L_{\eta
-}^0(\lambda)$).  
The average of these two boundary values is denoted by
$\overline{L}^0_\eta(\lambda)$.  

All the results we need will come from studying the asymptotic behavior
of two quotients:
\begin{equation}
S(\lambda):=
\left(\prod_{k=0}^{N-1}\frac{\lambda-\lambda_k^*}{\lambda-\lambda_k}\right)
\exp\,\left(-\frac{1}{\hbar_N}\left(\int_0^{iA}L^0_\eta(\lambda)
\rho^0(\eta)\,d\eta + 
\int_{-iA}^0L^0_\eta(\lambda)\rho^0(\eta^*)^*\,d\eta\right)\right)
\label{eq:S}
\end{equation}
and
\begin{equation}
T(\lambda):= \left(\prod_{k=0}^{N-1}
\frac{\lambda-\lambda_k^*}{\lambda-\lambda_k}\right)
\exp\,\left(-\frac{1}{\hbar_N}\left(
\int_0^{iA}\overline{L}^0_\eta(\lambda)\rho^0(\eta)\,d\eta +
\int_{-iA}^0\overline{L}^0_\eta(\lambda)\rho^0(\eta^*)^*\,d\eta\right)\right)
\cdot
 2\cos\,\left(\frac{\theta^0(\lambda)}{\hbar_N}
\right)\,.
\label{eq:T}
\end{equation}
The function $S(\lambda)$ is analytic and nonvanishing for $\lambda\in
{\mathbb C}_+\setminus C_M$.  Let us denote by $\Omega\subset {\mathbb
C}_+$ the domain of analyticity of $\rho^0(\lambda)$ restricted to the
upper half-plane, so that by our assumptions on $A(x)$,
$C_M\subset\Omega$.  Then, due to the zeros of the cosine on the
imaginary axis, which match the poles of the product below
$\lambda=iA$ and are not cancelled above $\lambda=iA$, $T(\lambda)$ is
analytic and nonvanishing for $\lambda\in \Omega\setminus V$, where
$V$ is the vertical ray from $\lambda=iA$ to infinity along the
positive imaginary axis.  The domain of analyticity for $T(\lambda)$
is a subset of $\Omega$ rather than of the whole upper half-plane due
to the presence of the averages of the logarithms in the integrand of
(\ref{eq:T}).  Whereas these are boundary values defined {\em a
priori} only on $C_M$, the integrals extend from $C_M$ to analytic
functions in the domain $\Omega_+\setminus V$ via the introduction of
the function $\theta^0(\lambda)$ ({\em cf.} equation (\ref{eq:logjump})).
\begin{lemma}
For all $\lambda$ in the upper half-plane with $\hbar_N\le
|\Re(\lambda)|\le B$, where $B$ is positive and sufficiently small,
but fixed as $\hbar_N\rightarrow 0$,
\begin{equation}
S(\lambda)=1+O\left(\frac{\hbar_N}{|\Re(\lambda)|}\right)\,.
\end{equation}
\label{lemma:Souter}
\end{lemma}

{\em Proof:}  
Let us define the function $m(\eta)$ by
\begin{equation}
m(\eta):=-\int_0^\eta\rho^0(\xi)\,d\xi\,.
\end{equation}
This analytic function takes the imaginary interval $[0,iA]$ to the real
interval $[0,M]$ where 
\begin{equation}
M=m(iA)=\frac{1}{\pi}\int_{-\infty}^\infty A(x)\,dx\,.
\end{equation}
Since $\rho^0(\xi)$ does not vanish on $C_M$, we have the inverse
function $\eta=e(m)$ defined for $m$ near the real interval $[0,M]$.
Using these tools, we get the following representation
for $S(\lambda)$:
\begin{equation}
S(\lambda)=\exp(-\tilde{I}(\lambda))\hspace{0.2 in}\mbox{where}
\hspace{0.2 in}\tilde{I}(\lambda)=\sum_{k=0}^{N-1}\tilde{I}_k(\lambda)\,,
\end{equation}
and
\begin{equation}
\tilde{I}_k(\lambda):=\frac{1}{\hbar_N}\int_{m_k-\hbar_N/2}^{m_k+\hbar_N/2}\left[L^0_{-e(m)}(\lambda)-L^0_{e(m)}(\lambda)\right]\,dm -
\left[L^0_{-e(m_k)}(\lambda)-L^0_{e(m_k)}(\lambda)\right]\,,
\end{equation}
with $m_k:=M-\hbar_N(k+1/2)$.  Expanding the logarithms, we find that
\begin{equation}
\tilde{I}_k(\lambda)=\frac{1}{\hbar_N}\int_{m_k-\hbar_N/2}^{m_k+\hbar_N/2}
\,dm\int_{m_k}^m\,d\zeta \int_{m_k}^\zeta\,d\xi
\left[\frac{2e''(\xi)\lambda^3-2e''(\xi)e(\xi)^2\lambda+4e'(\xi)^2e(\xi)\lambda}{(\lambda^2-e(\xi)^2)^2}\right]\,.
\label{eq:integralformula}
\end{equation}
This quantity is clearly $O(\hbar_N^2)$ for $\lambda$ fixed away from
$C_M$.  Now, when $|\Re(\lambda)|=o(1)$ as $\hbar_N\downarrow 0$, we
can estimate the denominator in the integrand to obtain two different
bounds:
\begin{equation}
\frac{2e''(\xi)\lambda^3-2e''(\xi)e(\xi)^2\lambda+4e'(\xi)^2e(\xi)\lambda}{(\lambda^2-e(\xi)^2)^2}=O\left(\frac{1}{\Re(\lambda)^2}\right)
\label{eq:lambdadenominator}
\end{equation}
and
\begin{equation}
\frac{2e''(\xi)\lambda^3-2e''(\xi)e(\xi)^2\lambda+4e'(\xi)^2e(\xi)\lambda}{(\lambda^2-e(\xi)^2)^2}=O\left(\frac{1}{|i\Im(\lambda)-e(\xi)|^2}\right)\,.
\label{eq:xidenominator}
\end{equation}

The idea is to use the estimate (\ref{eq:lambdadenominator}) when
$e(m_k)$ is close to $i\Im(\lambda)$ and to use the estimate
(\ref{eq:xidenominator}) for the remaining terms.  Suppose first
$\Im(\lambda)$ is bounded between $0$ and $A$, {\em i.e.} there are
small fixed positive numbers $\delta_1$ and $\delta_2$ so that
$\delta_1\le \Im(\lambda)\le A-\delta_2$, and let
$\epsilon=\epsilon(\hbar_N)$ be a small positive scale tied to $\hbar$
and satisfying $\hbar_N\ll\epsilon\ll 1$, and let $L_1$ be chosen from
$0,\dots,N-1$ so that $e(m_{L_1})$ is as close as possible to
$i(\Im(\lambda)+\epsilon)$, and likewise let $L_2$ be chosen from
$0,\dots,N-1$ so that $e(m_{L_2})$ is as close as possible to
$i(\Im(\lambda)-\epsilon)$.  Using (\ref{eq:lambdadenominator}) we
then find that
\begin{equation}
\sum_{k=L_1}^{L_2-1} \tilde{I}_k(\lambda) =
O\left(\frac{\hbar_N\epsilon}{\Re(\lambda)^2}\right)
\end{equation}
because the sum contains $O(\epsilon/\hbar_N)$ terms and the volume of
the region of integration for each term is $O(\hbar_N^3)$, and we must
take into account the overall factor of $1/\hbar_N$.  Now in each of
the remaining terms $\tilde{I}_k(\lambda)$, we have
\begin{equation}
\frac{1}{|i\Im(\lambda)-e(\xi)|^2} = O\left(\frac{1}{(m_k-m(i\Im(\lambda)))^2}\right)
\end{equation}
so using (\ref{eq:xidenominator}) and summing over $k$ we get both
\begin{equation}
\sum_{k=0}^{L_1-1}\tilde{I}_k(\lambda)=O\left(\frac{\hbar_N}{\epsilon}\right)
\hspace{0.2 in}
\mbox{and}\hspace{0.2 in}
\sum_{k=L_2}^{N-1}\tilde{I}_k(\lambda)=O\left(\frac{\hbar_N}{\epsilon}\right)\,.
\end{equation}
The total estimate of $\tilde{I}(\lambda)$ is then optimized by a
dominant balance among the three partial sums.  This balance requires
taking $\epsilon\sim |\Re(\lambda)|$, upon which we deduce that under
our assumptions on $\lambda$, we indeed have
\begin{equation}
\tilde{I}(\lambda)=O\left(\frac{\hbar_N}{|\Re(\lambda)|}\right)\hspace{0.2 in}
\mbox{and consequently} \hspace{0.2 in}
S(\lambda)-1 = O\left(\frac{\hbar_N}{|\Re(\lambda)|}\right)\,,
\label{eq:outerestimate}
\end{equation}
when $\Im(\lambda)$ is bounded between $0$ and $A$.  When
$\Im(\lambda)\approx 0$ or $\Im(\lambda)\approx A$, the estimate
(\ref{eq:lambdadenominator}) should be used only for those terms that
correspond to $m$ near zero or $m$ near $M$ respectively.  In both of
these exceptional cases, the same estimate is found.  When
$\Im(\lambda)$ is bounded below by $A$, there is no need to use the
estimate (\ref{eq:lambdadenominator}) at all, and the relative error
is of order $\hbar_N$ uniformly in $\Re(\lambda)$.  This completes the
proof.  $\Box$

\vspace{0.1 in}

We now use this information about $S(\lambda)$ to effectively replace
the sums of logarithms by integrals, at least on some portions of the
contour $\Sigma$.  
\begin{proposition}
Suppose that the contour $C_L$ is independent of $\hbar_N$ and that for
some sufficiently small positive number $B$, $C_L$ lies in the strip
$-B\le\Re(\lambda)\le 0$ and meets the imaginary axis only at its
endpoints and does so transversely.  Then 
\begin{equation}
\begin{array}{rcl}
a_L(\lambda)&=&\displaystyle i\exp\,\left(\frac{1}{\hbar_N}\left(2i\lambda |x|
+\int_0^{iA}L^0_\eta(\lambda)\rho^0(\eta)\,d\eta +
\int_{-iA}^{0}L^0_\eta(\lambda)\rho^0(\eta^*)^*\,d\eta
-2Jg(\lambda)\right)\right)\\\\
&&\displaystyle\,\,\,\times\,\,\,
\exp\,\left(-\frac{i\theta^0(\lambda)}{\hbar_N}
\right)
\left(1+O\left(\frac{\hbar_N}{|\lambda|}\right)+
O\left(\frac{\hbar_N}{|\lambda-iA|}\right)\right)\,,
\end{array}
\label{eq:Lasymp}
\end{equation}
as $\hbar_N$ goes to zero through positive values, for all $\lambda\in
C_L$ with $|\lambda|>\hbar_N$ and $|\lambda-iA|>\hbar_N$.
\label{proposition:CLout}
\end{proposition}
\begin{proposition}
Suppose that the contour $C_R$ is independent of $\hbar_N$ and that for
some sufficiently small positive number $B$, $C_R$ lies in the strip
$0\le\Re(\lambda)\le B$ and meets the imaginary axis only at its
endpoints and does so transversely.  Then 
\begin{equation}
\begin{array}{rcl}
a_R(\lambda)&=&\displaystyle i\exp\,\left(\frac{1}{\hbar_N}\left(2i\lambda |x|
+\int_0^{iA}L^0_\eta(\lambda)\rho^0(\eta)\,d\eta +
\int_{-iA}^0L^0_\eta(\lambda)\rho^0(\eta^*)^*\,d\eta
-2Jg(\lambda)\right)\right)\\\\
&&\displaystyle\,\,\,\times\,\,\,
\exp\,\left(\frac{i\theta^0(\lambda)}{\hbar_N}
\right)
\left(1+O\left(\frac{\hbar_N}{|\lambda|}\right)+
O\left(\frac{\hbar_N}{|\lambda-iA|}\right)\right)\,,
\end{array}
\label{eq:Rasymp}
\end{equation}
as $\hbar_N$ goes to zero through positive values, for all $\lambda\in
C_R$ with $|\lambda|>\hbar_N$ and $|\lambda-iA|>\hbar_N$.
\label{proposition:CRout}
\end{proposition}

{\em Proof of Propositions~\ref{proposition:CLout}
and~\ref{proposition:CRout}:} 
These propositions follow directly from Lemma~\ref{lemma:Souter} upon
using the transversality of the intersections with the imaginary axis to
replace $O(1/|\Re(\lambda)|)$ by $O(1/|\lambda|)+O(1/|\lambda-iA|)$.
$\Box$

\vspace{0.1 in}

We notice that the first factor on the second line in
(\ref{eq:Lasymp}) and the first factor on the second line in
(\ref{eq:Rasymp}) are both exponentially small as $\hbar_N$ goes to
zero through positive values, as a consequence of the fact that
$\rho^0(\eta)\,d\eta$ is an analytic negative real measure on $C_M$.
This follows from the Cauchy-Riemann equations and the geometry of
Figure~\ref{fig:twohalves}.  It will be a very useful fact for us
shortly.  

Now we turn our attention to the function $T(\lambda)$.  The result
analogous to Lemma~\ref{lemma:Souter} is the following.
\begin{lemma}
For all $\lambda$ in the upper half-plane with 
$\hbar_N\le
|\Re(\lambda)|\le B$, where $B$ is positive and sufficiently small,
but fixed as $\hbar_N\rightarrow 0$,
\begin{equation}
T(\lambda)=1+O\left(\frac{\hbar_N}{|\Re(\lambda)|}\right)\,.
\end{equation}
\label{lemma:Touter}
\end{lemma}

{\em Proof:}  We begin with the jump condition 
\begin{equation}
\begin{array}{c}
\displaystyle \int_0^{iA}L_{\eta +}^0(\lambda)\rho^0(\eta)\,d\eta +
\int_{-iA}^0L_{\eta +}^0(\lambda)\rho^0(\eta^*)^*\,d\eta \,\,\,=
\hspace{2 in}\\\\ 
\displaystyle
\hspace{1 in}
\int_0^{iA}L_{\eta -}^0(\lambda)\rho^0(\eta)\,d\eta +
\int_{-iA}^0L_{\eta -}^0(\lambda)\rho^0(\eta^*)^*\,d\eta  
-
2i\theta^0(\lambda)\,.
\end{array}
\label{eq:logjump}
\end{equation}
relating the boundary values of the logarithm $L^0_\eta(\lambda)$ on
the imaginary axis.  Using this jump relation and the definition of
$\overline{L}^0_\eta(\lambda)$ as the average of the boundary values of
$L^0_{\eta+}(\lambda)$ and $L^0_{\eta-}(\lambda)$, we see that for
$\Re(\lambda)<0$, we have
\begin{equation}
T(\lambda)=S(\lambda)\left(1+\exp\,\left(-\frac{2i\theta^0(\lambda)}
{\hbar_N}\right)\right)
\label{eq:Tleft}
\end{equation}
while for $\Re(\lambda)>0$, we have
\begin{equation}
T(\lambda)=S(\lambda)\left(1+\exp\,\left(\frac{2 i\theta^0(\lambda)}
{\hbar_N}\right)\right)\,.
\label{eq:Tright}
\end{equation}
Now, using the fact that $\rho^0(\eta)$ is an analytic function
satisfying $\rho^0(\eta)\in i{\mathbb R}_+$ for $\eta\in C_M$, we see
by the Cauchy-Riemann equations that in both cases, the exponential
relative error term is of the order $e^{-K|\Re(\lambda)|/\hbar_N}$ for
some $K>0$.  Since this is negligible compared with the relative error
associated with the asymptotic approximation of $S(\lambda)$ given in
Lemma~\ref{lemma:Souter}, the proof is complete.  $\Box$

\vspace{0.1 in}

Unfortunately, we need asymptotic information about $T(\lambda)$ right
on the imaginary axis, which contains the contour $C_M$, so we need to
improve upon Lemma~\ref{lemma:Touter}.  We begin to extract this
additional information by noting that under some circumstances, it is
easy to show that $T(\lambda)$ remains bounded in the vicinity of the
imaginary axis.
\begin{lemma}
If either (i) $\lambda$ is real or (ii) $|\lambda|=A$ and
$\Im(\lambda)>0$, and if for some $B>0$ sufficiently small
$|\Re(\lambda)|<B$, then $T(\lambda)$ is uniformly bounded as
$\hbar_N\rightarrow 0$.  
\label{lemma:Tbounded}
\end{lemma}

{\em Proof:} It suffices to show that $S(\lambda)$ is bounded under
the same assumptions, because from (\ref{eq:Tleft}) and
(\ref{eq:Tright}) and the Cauchy-Riemann equations, we see easily that
$|T(\lambda)|\le 2|S(\lambda)|$.

Using the function $m(\cdot)$ and its inverse
$e(\cdot)$, we have the following
\begin{equation}
\hbar_N\log|S(\lambda)| = \sum_{k=0}^{N-1}H(m_k)
\hbar_N - \int_0^M H(m)\,dm\,,
\label{eq:difference}
\end{equation}
where
\begin{equation}
H(m):=\log\left|\frac{\lambda+e(m)}
{\lambda-e(m)}\right|\,.
\end{equation}
When $\lambda\in{\mathbb R}$, we see immediately that $H(m)\equiv 0$,
and therefore $|S(\lambda)|\equiv 1$ and hence $|T(\lambda)|\le 2$.

Now consider $\lambda=iAe^{i\theta}$ with $\theta$ sufficiently small
independent of $\hbar_N$.  The idea is that of the terms on the
right-hand side of (\ref{eq:difference}), the discrete sum is a
Riemann sum approximation to the integral.  The Riemann sum is
constructed using the midpoints of $N$ equal subintervals as sample
points.  If $H''(m)$ is bounded uniformly, then this sort of Riemann
sum provides an approximation to the integral that is of order
$N^{-2}$ or equivalently $\hbar_N^2$.  In this case, we deduce that
$S(\lambda)=1+O(\hbar_N)$ and in particular this is bounded as
$\hbar_N$ tends to zero.  But as $\lambda$ approaches the imaginary
axis, the accuracy of the approximation is lost.

\begin{figure}[h]
\begin{center}
\mbox{\psfig{file=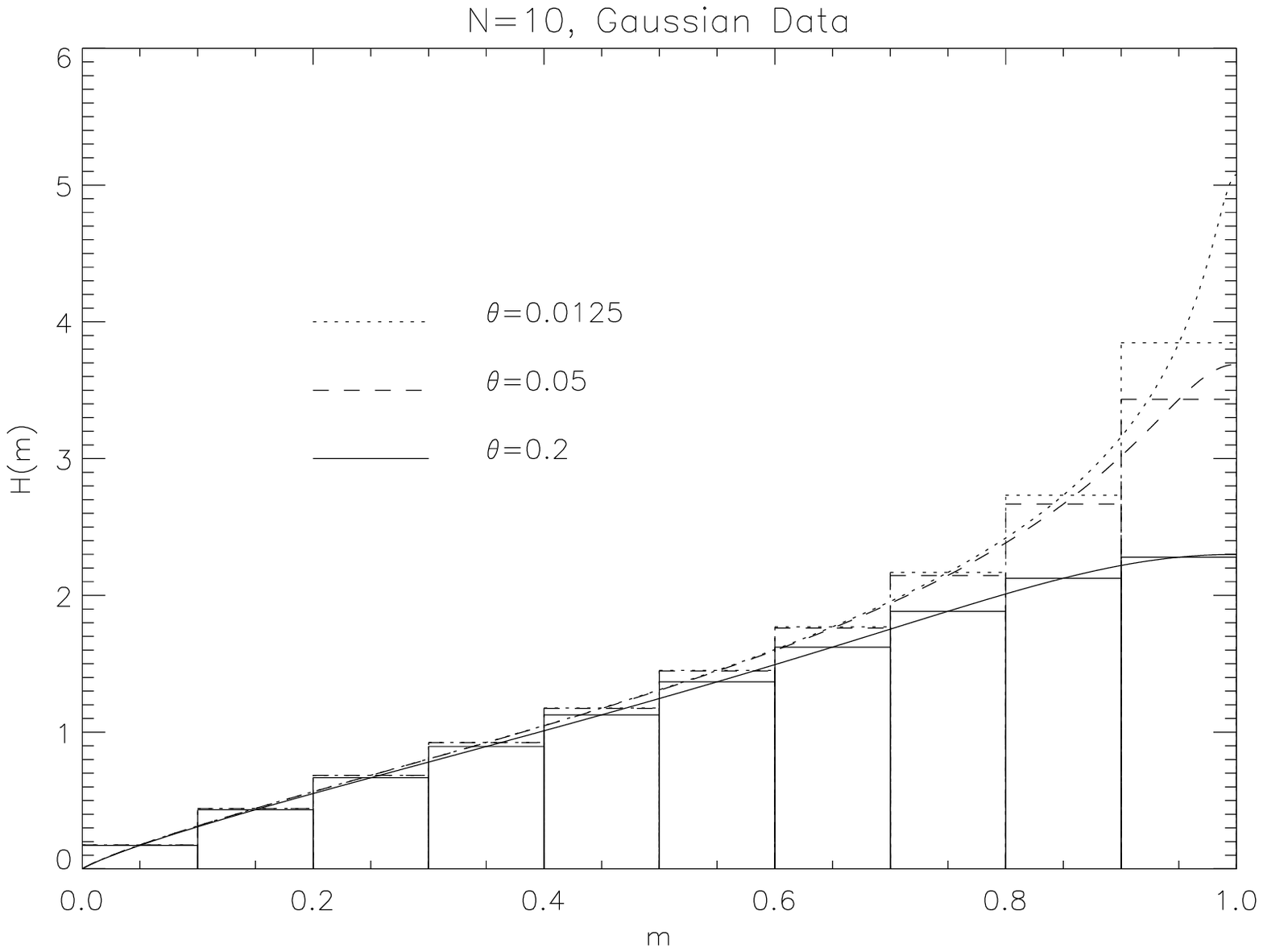,width=3 in}\psfig{file=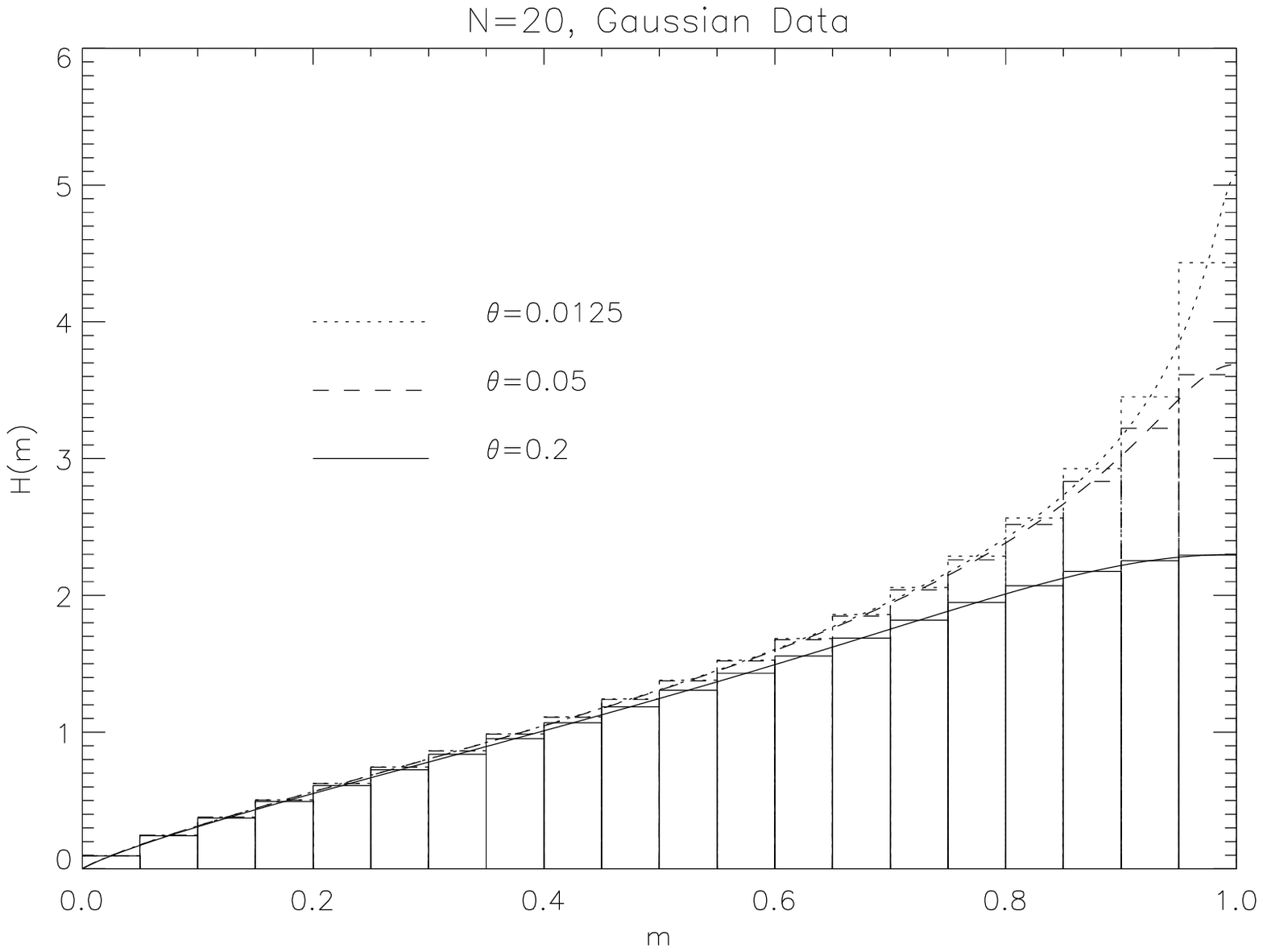,width=3 in}}
\end{center}
\caption{\em The midpoint rule Riemann sums approximating the
integral, pictured here for the Gaussian initial data
$A(x)=\sqrt{\pi}e^{-x^2}$.  When the peak of $H(m)$ becomes
underresolved for small $\theta$, the Riemann sums underestimate the
value of the integral by an amount that is of the order $\hbar_N$.}
\label{fig:riemannsums}
\end{figure}

For $\lambda=iAe^{i\theta}$, the function $H(m)$ satisfies $H(0)=H'(M)=0$
and takes its
maximum when $m=M$, with a maximum value
\begin{equation}
H(M)=\log|\cot(\theta/2)|\,.
\end{equation}
Therefore, as $\theta$ tends to zero, $H(m)$ becomes unbounded,
growing logarithmically in $\theta$.  As a consequence of this blowup
the approximation of the integral by the Riemann sum based on
midpoints for $|\lambda|=A$ fails to be second-order accurate
uniformly in $\theta$.  However, because the maximum of $H(m)$ always
occurs at the right endpoint, it is easy to see that when the error
becomes larger than $O(\hbar_N^2)$ in magnitude its sign is such that
the Riemann sum is always an {\em underestimate} of the value of the
integral, and consequently the right-hand side of
(\ref{eq:difference}) is negative.  This is concretely illustrated in
Figure~\ref{fig:riemannsums}
where we have taken the example of the Gaussian function
$A(x)=\sqrt{\pi}e^{-x^2}$ in order to supply the function
$\rho^0(\eta)$ and therefore the function $e(m)$ needed to build
$H(m)$.  In this case, $A=\sqrt{\pi}$ and $M=1$.  The error of the
Riemann sum is worst when $\theta=0$.  In this case it is easy to see
that the discrepancy contributed by only the subinterval adjacent to
the logarithmic singularity of $H(m)$ is $(1-\log 2)\hbar_N +
O(\hbar_N^2)$, which clearly dominates the $O(\hbar_N^2)$ error
contributed by the majority of the subintervals bounded away from
$m=M$.  Consequently, for those $\lambda$ on the circle $|\lambda|=A$
for which $\log|S(\lambda)|$ is not asymptotically small in $\hbar_N$,
it is negative, and therefore $S(\lambda)$ is uniformly bounded for
$|\lambda|=A$, as is $T(\lambda)$.  $\Box$

Using this information, we can finally extract enough information
about $T(\lambda)$ on the imaginary axis to approximate $a_M(\lambda)$
for $\lambda\in C_M$.
\begin{proposition}
Let $C_M$ be a fixed contour from $\lambda=0$ to $\lambda=iA$ lying between
$C_L$ and $C_R$, possibly coinciding with the imaginary axis.  Then, for
$\mu>0$ arbitrarily small,
\begin{equation}
\begin{array}{rcl}
a_M(\lambda)&=&\displaystyle i\exp\,\left(\frac{1}{\hbar_N}\left(2i\lambda |x|
+\int_0^{iA}\overline{L}^0_\eta(\lambda)\rho^0(\eta)\,d\eta +
\int_{-iA}^{0}\overline{L}^0_\eta(\lambda)\rho^0(\eta^*)^*\,d\eta
-Jg_+(\lambda)-Jg_-(\lambda)\right)\right)\\\\
&&\displaystyle\,\,\,\times\,\,\,
\left(1+O\left(\frac{\hbar_N^{1-\mu}}{|\lambda|}\right)
+O\left(\frac{\hbar_N^{1-\mu}}{|\lambda-iA|}\right)\right)\,,
\end{array}
\label{eq:Masymp}
\end{equation}
as $\hbar_N$ goes to zero through positive values, for all $\lambda\in C_M$
with $|\lambda|>\hbar_N$ and $|\lambda-iA|>\hbar_N$.
\label{prop:a_M}
\end{proposition}

\begin{figure}[h]
\begin{center}
\mbox{\psfig{file=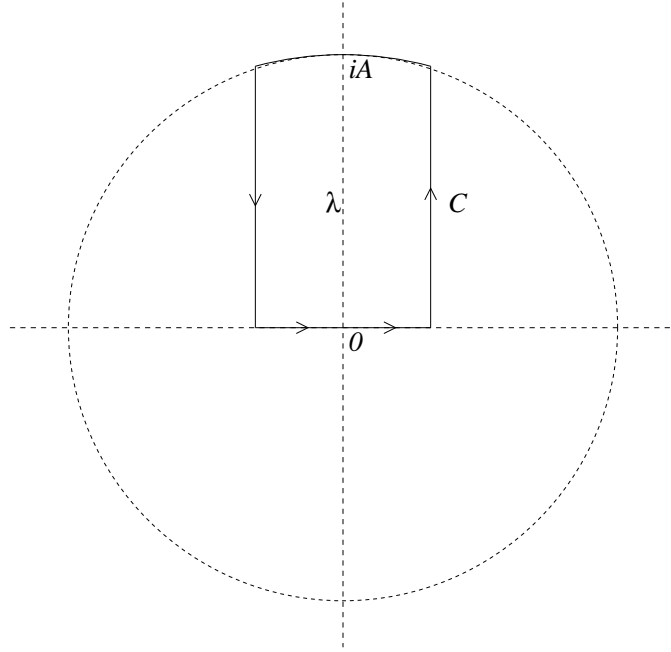,width=3.5 in}}
\end{center}
\caption{\em The contour $C$ of the Cauchy integral argument.}
\label{fig:C}
\end{figure}

{\em Proof:}  Let $C$ be the closed contour illustrated in Figure~\ref{fig:C}.
This counter-clockwise oriented contour consists of two vertical
segments, one horizontal segment that lies on the real axis, and an
arc of the circle of radius $A$ centered at the origin.  The function
$T(\lambda)$ is analytic on the interior of $C$ and is continuous on
$C$ itself.  In fact it is analytic on most of the boundary, failing
to be analytic only at $\lambda=0$ and $\lambda=iA$.  Therefore for
any $\lambda$ in the interior, we may write
\begin{equation}
T(\lambda)=1+\frac{1}{2\pi i}\oint_C \frac{T(s)-1}{s-\lambda}\,ds\,.
\end{equation}
If we let $C_{\rm in}$ denote the part of $C$ with $|\Re(s)|<\hbar_N$,
and let $C_{\rm out}$ denote the remaining portion of $C$, then we get
\begin{equation}
|T(\lambda)-1|\le \frac{1}{2\pi}\int_{C_{\rm in}}\frac{|T(s)-1|}{|s-\lambda|}
\,|ds| + \frac{1}{2\pi}\int_{C_{\rm out}}\frac{|T(s)-1|}{|s-\lambda|}
\,|ds|\,.
\end{equation}
Using the estimate guaranteed by Lemma~\ref{lemma:Touter} in the
integral over $C_{\rm out}$, and the uniform boundedness of $T(s)$
(and therefore of $T(s)-1$) guaranteed by Lemma~\ref{lemma:Tbounded}
in the integral over $C_{\rm in}$, we find
\begin{equation}
|T(\lambda)-1|\le K_{\rm in}\hbar_N\sup_{s\in C_{\rm in}}\frac{1}{|s-\lambda|}
-K_{\rm out}\hbar_N\log\hbar_N\sup_{s\in C_{\rm out}}\frac{1}{|s-\lambda|}
\end{equation}
for some positive constants $K_{\rm in}$ and $K_{\rm out}$.  Replacing
the logarithm by a slightly cruder estimate of $\hbar_N^{-\mu}$ for
arbitrarily small positive $\mu$ completes the proof.  $\Box$

\vspace{0.1 in}

We have therefore succeeded in showing that, at least away from the
self-intersection points of the contour $\Sigma$, the jump matrices for
${\bf N}(\lambda)$ as defined by (\ref{eq:CLjumpN}) for $\lambda\in
C_L$, (\ref{eq:CRjumpN}) for $\lambda\in C_R$, and (\ref{eq:CMjumpN})
for $\lambda\in C_M$ are well-approximated in the semiclassical limit
$\hbar_N\rightarrow 0$ by matrices in which all nonzero matrix elements
are of the form $\exp(f(\lambda)/\hbar_N)$ with $f(\lambda)$ being
independent of $\hbar_N$.  The fact that this approximation is valid
even when the ``active'' contour $C_M$ is taken to be right on top of
the poles of the meromorphic Riemann-Hilbert problem for ${\bf
m}(\lambda)$ is an advantage over the approach taken in
\cite{manifesto}.

Using these approximations, we can introduce an {\em ad hoc}
approximation of the matrix ${\bf N}(\lambda)$.  First, define
\begin{equation}
\tilde{\phi}(\lambda):=2i\lambda |x| + \int_0^{iA}\overline{L}_\eta^0(\lambda)
\rho^0(\eta)\,d\eta + \int_{-iA}^0\overline{L}^0_\eta(\lambda)
\rho^0(\eta^*)^*\,d\eta -Jg_+(\lambda) -Jg_-(\lambda)\,,\hspace{0.2 in}
\mbox{for}\hspace{0.2 in}\lambda\in C_M\,,
\end{equation}
and for $\lambda\in C_L$ or $C_R$, define
\begin{equation}
\tau(\lambda):= 2i\lambda |x| + \int_0^{iA}L_\eta^0(\lambda)
\rho^0(\eta)\,d\eta + \int_{-iA}^0L^0_\eta(\lambda)
\rho^0(\eta^*)^*\,d\eta -2Jg(\lambda)\,.
\end{equation}
Then we pose the following problem.
\begin{rhp}[Formal Continuum Limit]
Given a complex phase function $g(\lambda)$ find a matrix $\tilde{\bf
N}(\lambda)$ satisfying:
\begin{enumerate}
\item {\bf Analyticity:}  $\tilde{\bf N}(\lambda)$ is analytic for
$\lambda\in{\mathbb C}\setminus \Sigma$.
\item {\bf Boundary behavior:}  $\tilde{\bf N}(\lambda)$ assumes continuous
boundary values on $\Sigma$.
\item {\bf Jump conditions:}  The boundary values taken on $\Sigma$
satisfy
\begin{equation}
\tilde{\bf N}_+(\lambda)=\tilde{\bf N}_-(\lambda)\sigma_1^{\frac{1-J}{2}}
\left[\begin{array}{cc} 1 & 0\\\\\displaystyle
i\exp\,\left(\frac{\tau(\lambda) - i\theta^0(\lambda)}{\hbar_N}
\right) & 1\end{array}
\right]\sigma_1^{\frac{1-J}{2}}
\end{equation}
for $\lambda\in C_L$,
\begin{equation}
\tilde{\bf N}_+(\lambda)=\tilde{\bf N}_-(\lambda)\sigma_1^{\frac{1-J}{2}}
\left[\begin{array}{cc} 1 & 0\\\\\displaystyle
i\exp\,\left(\frac{\tau(\lambda) +i\theta^0(\lambda)}{\hbar_N}
\right) & 1\end{array}
\right]\sigma_1^{\frac{1-J}{2}}
\end{equation}
for $\lambda\in C_R$, and
\begin{equation}
\tilde{\bf N}_+(\lambda)=\tilde{\bf N}_-(\lambda)\sigma_1^{\frac{1-J}{2}}
\left[\begin{array}{cc} \displaystyle \exp\,\left(\frac{i\theta(\lambda)}{\hbar_N}\right) & 0\\\\
i\displaystyle\exp\,\left(\frac{\tilde{\phi}(\lambda)}{\hbar_N}\right) &
\displaystyle\exp\,\left(-\frac{i\theta(\lambda)}{\hbar_N}\right)
\end{array}\right]\sigma_1^{\frac{1-J}{2}}
\label{eq:ntildemiddle}
\end{equation}
for $\lambda\in C_M$.  For all other $\lambda\in \Sigma$ (that is, in
the lower half-plane), the jump is determined by the symmetry
$\tilde{\bf N}(\lambda)=\sigma_2\tilde{\bf N}(\lambda^*)^*\sigma_2$.
\item {\bf Normalization:}  $\tilde{\bf N}(\lambda)$ is normalized at
infinity:
\begin{equation}
\tilde{\bf N}(\lambda)\rightarrow{\mathbb I}\hspace{0.2 in}\mbox{as}
\hspace{0.2 in}\lambda\rightarrow\infty\,.
\end{equation}
\end{enumerate}
\end{rhp}

\section{Choosing $g(\lambda)$ to Arrive at an Outer Model}
Let $R(\lambda)$ be defined by the equation $R(\lambda)^2 =
\lambda^2+A(x)^2$, the fact that $R(\lambda)$ is an analytic function
for $\lambda$ away from the imaginary interval 
$I:=[-iA(x),iA(x)]$, and the normalization that for large $\lambda$,
$R(\lambda)\sim -\lambda$.  For $\eta\in I\cap C_M$, let
\begin{equation}
\rho(\eta):=\rho^0(\eta)+\frac{R_+(\eta)}{\pi i}\int_{-iA}^{-iA(x)}\frac{\rho^0(s^*)^*\,ds}{(\eta-s)R(s)} + 
\frac{R_+(\eta)}{\pi i}\int_{iA(x)}^{iA}\frac{\rho^0(s)\,ds}{(\eta-s)R(s)}\,.
\label{eq:rhodefine}
\end{equation}
It is easy to check directly that for all $\eta\in I\cap C_M$, we have
$\rho(\eta)\in i{\mathbb R}_+$.  Also, using the fact that $\rho^0(s)$ is
purely imaginary on the imaginary axis, and that $R(s)$ is purely imaginary
in the domain of integration, where it satisfies $R(-s)=-R(s)$, we see
that 
\begin{equation}
\rho(0)=\rho^0(0)\,.
\label{eq:rhosequal}
\end{equation}
Furthermore, it follows easily from (\ref{eq:rhodefine}) that for all
$\eta\in I\cap C_M$, we have 
\begin{equation}
0\le -i\rho(\eta)\le -i\rho^0(\eta)\,,
\label{eq:upperlowerconstraints}
\end{equation}
with the lower constraint being achieved only at the
endpoint\footnote{It is often convenient to think of the function
$\rho(\eta)$ being extended to all of $C_M$ by setting
$\rho(\eta)\equiv 0$ for $\lambda$ above the endpoint $iA(x)$.  In
this case one views the lower constraint as being active on the whole
imaginary interval $[iA(x),iA]$.} of $I$, $\lambda=iA(x)$, and the
upper constraint being achieved only at the origin in accordance with
(\ref{eq:rhosequal}).

Now, set
\begin{equation}
g(\lambda):=\frac{J}{2}\int_{-iA(x)}^{0}L_\eta^0(\lambda)\rho(\eta^*)^*\,d\eta
+\frac{J}{2}\int_{0}^{iA(x)}L_\eta^0(\lambda)\rho(\eta)\,d\eta\,.
\end{equation}
This function satisfies all of the basic criteria set out earlier: it
is analytic in ${\mathbb C}\setminus (C_M\cup C_M^*)$ and takes
continuous boundary values, it satisfies $g(\lambda)+g(\lambda^*)^*=0$, and
it satisfies $g(\infty)=0$ because
\begin{equation}
\int_{-iA(x)}^{0}\rho(\eta^*)^*\,d\eta +\int_0^{iA(x)}\rho(\eta)\,d\eta = 0\,.
\end{equation}
Note that $g(\lambda)$ is analytic across $C_M$ for $\lambda$ above
$iA(x)$.  Consequently $\theta(\lambda)=0$ for all such $\lambda$.
For $\lambda\in C_M$ below $iA(x)$, $\theta(\lambda)$ becomes ({\em
cf.} equation (\ref{eq:defthetaoriginal}))
\begin{equation}
\theta(\lambda)=-\pi\int_\lambda^{iA(x)}\rho(\eta)\,d\eta\,.
\label{eq:deftheta}
\end{equation}

We now describe a number of important consequences of our choice of
$g(\lambda)$.
\begin{proposition}
For all $\lambda\in I\cap C_M=[0,iA(x)]$, $\tilde{\phi}(\lambda)=0$.
\label{prop:phitildezero}
\end{proposition}
To prove the proposition, we first point out that 
\begin{equation}
\mathop{\lim_{\lambda\rightarrow 0}}_{\lambda\in C_M}\tilde{\phi}(\lambda)=0\,,
\end{equation}
simply as a consequence of the fact that both $\rho^0(\eta)$ and $\rho(\eta)$
are purely imaginary on $C_M$.  Next we point out that
\begin{equation}
\tilde{\phi}'(\lambda)=0
\end{equation}
whenever $\lambda\in[0,iA(x)]$.  This follows from a direct calculation in
which all integrals are evaluated by residues and the formula (\ref{eq:rho0})
is used.

Next we consider $\tilde{\phi}(\lambda)$ for $\lambda\in
C_M\setminus[0,iA(x)]$, that is, above the endpoint of the support.
Clearly, $\tilde{\phi}(\lambda)+i\theta(\lambda)$ is the boundary value
on $C_M$ of an analytic function defined near $C_M$ in $D_L$.  Since the
boundary value taken below the endpoint is $i\theta(\lambda)$ because
$\tilde{\phi}(\lambda)\equiv 0$ there, and the boundary value taken above
the endpoint is $\tilde{\phi}(\lambda)$ because $\theta(\lambda)\equiv 0$
there, we obtain the formula
\begin{equation}
\tilde{\phi}(\lambda)=i\theta_+(\lambda)=
-i\pi\int_\lambda^{iA(x)}\rho_+(\eta)\,d\eta
\label{eq:phifromtheta}
\end{equation}
valid for $\lambda\in C_M$ above $iA(x)$, where by $\rho_+(\eta)$ for
$\eta$ in the imaginary interval $(iA(x),iA)$ we mean the function
$\rho(\eta)$ defined by (\ref{eq:rhodefine}) for $\eta$ in the imaginary
interval $(0,iA(x))$, analytically continued from $(0,iA(x))$ in the
clockwise direction about the endpoint $\lambda=iA(x)$.  In particular
for such $\lambda$ we have
\begin{equation}
\tilde{\phi}'(\lambda)=i\pi\rho_+(\lambda)\,.
\label{eq:phiprime}
\end{equation}
Carrying out the analytic continuation, we find from (\ref{eq:rhodefine})
that for $\eta\in (iA(x),iA)$,
\begin{equation}
\rho_+(\lambda)=\frac{R(\lambda)}{\pi i}\int_{-iA}^{-iA(x)}
\frac{\rho^0(s^*)^*\,ds}{(\lambda-s)R(s)} +
\frac{R(\lambda)}{\pi i}\mbox{P.V.}\int_{iA(x)}^{iA}
\frac{\rho^0(s)\,ds}{(\lambda-s)R(s)}\,.
\end{equation}
From this formula we see easily that for all $\lambda$ strictly above
the endpoint $iA(x)$, $\rho_+(\lambda)$ is positive real.
Consequently, from (\ref{eq:phiprime}) and since
$\tilde{\phi}(\lambda)=0$ for $\lambda=iA(x)$, we get the following
result.
\begin{proposition}
The function $\tilde{\phi}(\lambda)$ is negative real and decreasing in
the positive imaginary direction for $\lambda\in C_M\setminus [0,iA(x)]$.
\label{prop:tildephineg}
\end{proposition}

Now we consider the behavior of the function $\tau(\lambda)$ on $C_L$
and $C_R$.  From the definitions of the functions $\tau(\lambda)$ and
$\tilde{\phi}(\lambda)$, we see that for $\lambda\in C_L$,
\begin{equation}
\tau(\lambda)=\tilde{\phi}(\lambda)+i\theta(\lambda)-i\theta^0(\lambda)\,,
\label{eq:tauleft}
\end{equation}
and for $\lambda\in C_R$,
\begin{equation}
\tau(\lambda)=\tilde{\phi}(\lambda)-i\theta(\lambda)+i\theta^0(\lambda)\,.
\label{eq:tauright}
\end{equation}
That is, the analytic function $\tau(\lambda)$ takes boundary values
from the left on $C_M$ equal to
$\tilde{\phi}(\lambda)+i\theta(\lambda)-i\theta^0(\lambda)$ and from
the right on $C_M$ equal to
$\tilde{\phi}(\lambda)-i\theta(\lambda)+i\theta^0(\lambda)$.  First
consider the situation to the left or right of the imaginary interval
$[0,iA(x)]$.  Since $\tilde{\phi}(\lambda)\equiv 0$ in $[0,iA(x)]$,
the function $\tau(\lambda)$ on $C_L$ will be the analytic
continuation of $i\theta(\lambda)-i\theta^0(\lambda)$ from $C_M$ and
the function $\tau(\lambda)$ on $C_R$ will be the analytic
continuation of $-i\theta(\lambda)+i\theta^0(\lambda)$ from $C_M$.
From (\ref{eq:upperlowerconstraints}) we see that for $\eta\in
[0,iA(x)]$ one has $\rho^0(\eta)-\rho(\eta)\in i{\mathbb R}_+$.
Therefore, it follows from the Cauchy-Riemann equations that for
$\lambda$ in portions of $C_L$ and $C_R$ close enough (independently
of $\hbar_N$) to the interval $[0,iA(x)]$ one has
\begin{equation}
\Re(\tau(\lambda))<0
\end{equation}
for $\lambda$ on {\em both} $C_L$ and $C_R$.  Furthermore, it
follows from the fact that $\rho^0(\eta)\in i{\mathbb R}_+$
that $\Re(-i\theta^0(\lambda))<0$ for $\lambda\in C_L$
and $\Re(i\theta^0(\lambda))<0$ for $\lambda\in C_R$.  Therefore
\begin{equation}
\Re(\tau(\lambda)-i\theta^0(\lambda))<0
\label{eq:decayleft}
\end{equation}
for $\lambda\in C_L$ near the portion of $C_M$ below $iA(x)$, and
\begin{equation}
\Re(\tau(\lambda)+i\theta^0(\lambda))<0
\label{eq:decayright}
\end{equation}
for $\lambda$ in the analogous portion of $C_R$.  Next consider the
situation to the left or right of the portion of $C_M$ lying above the
endpoint $\lambda=iA(x)$.  Since $\theta(\lambda)\equiv 0$ and
$\Re(\tilde{\phi}(\lambda))<0$ for $\lambda\in [iA(x),iA]$ we see that
for $C_L$ and $C_R$ close enough (again independently of $\hbar_N$) to
this part of $C_M$ we again find that we have (\ref{eq:decayleft}) on
$C_L$ and (\ref{eq:decayright}) on $C_R$.  This shows that the jump
matrix on both contours $C_L$ and $C_R$ is an exponentially small
perturbation of the identity for small positive $\hbar_N$, pointwise
in $\lambda$ bounded away from the origin and $iA$.

\begin{figure}[h]
\begin{center}
\mbox{\psfig{file=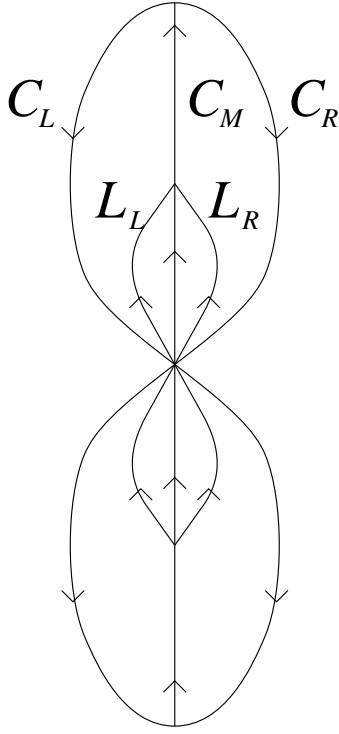,width=1.75 in}}
\end{center}
\caption{\em Introduction of the lens boundaries $L_L$ and $L_R$.}
\label{fig:twohalveslens}
\end{figure}

For $\lambda\in [0,iA(x)]$, the jump matrix for $\tilde{\bf N}(\lambda)$
factors (recall $\tilde{\phi}(\lambda)\equiv 0$ here):
\begin{equation}
\left[\begin{array}{cc}
\displaystyle\exp\,\left(\frac{i\theta(\lambda)}{\hbar_N}\right) & 0\\\\
\displaystyle i &
\displaystyle\exp\,\left(-\frac{i\theta(\lambda)}{\hbar_N}\right)
\end{array}\right] = \left[\begin{array}{cc}
1 &\displaystyle -i\exp\,\left(\frac{i\theta(\lambda)}{\hbar_N}\right)\\\\
0 & 1\end{array}\right]
\left[\begin{array}{cc}
0 & i\\\\i&0\end{array}\right]
\left[\begin{array}{cc}
1 &\displaystyle -i\exp\,\left(-\frac{i\theta(\lambda)}{\hbar_N}\right)\\\\
0 & 1\end{array}\right]\,.
\label{eq:factorization}
\end{equation}
Let $L_L$ and $L_R$ be two boundaries of a lens surrounding $[0,iA]$.  See
Figure~\ref{fig:twohalveslens}.
Using the factorization (\ref{eq:factorization}), we new define a new
matrix function ${\bf O}(\lambda)$.  In the region between $L_L$ and
$C_M$ set
\begin{equation}
{\bf O}(\lambda):=\tilde{\bf N}(\lambda)\sigma_1^{\frac{1-J}{2}}
\left[\begin{array}{cc}1 & 
\displaystyle i\exp\,\left(-\frac{i\theta(\lambda)}{\hbar_N}\right)\\\\
0 & 1\end{array}\right]\sigma_1^{\frac{1-J}{2}}\,.
\label{eq:ONleft}
\end{equation}
In the region between $C_M$ and $L_R$, set
\begin{equation}
{\bf O}(\lambda):=\tilde{\bf N}(\lambda)
\sigma_1^{\frac{1-J}{2}}\left[\begin{array}{cc}1 & 
\displaystyle -i\exp\,\left(\frac{i\theta(\lambda)}{\hbar_N}\right)\\\\
0 & 1\end{array}\right]\sigma_1^{\frac{1-J}{2}}\,.
\label{eq:ONright}
\end{equation}
Elsewhere in the upper half-plane set ${\bf O}(\lambda):=\tilde{\bf
N}(\lambda)$.  And in the lower half-plane define ${\bf O}(\lambda)$ by
symmetry:  ${\bf O}(\lambda)=\sigma_2{\bf O}(\lambda^*)^*\sigma_2$.  

These transformations imply jump conditions satisfied by ${\bf
O}(\lambda)$ on the contours in Figure~\ref{fig:twohalveslens} since
the jump conditions for $\tilde{\bf N}(\lambda)$ are given.  For
$\lambda\in L_L$ we have
\begin{equation}
{\bf O}_+(\lambda)={\bf O}_-(\lambda)
\sigma_1^{\frac{1-J}{2}}\left[\begin{array}{cc}1 & 
\displaystyle -i\exp\,\left(-\frac{i\theta(\lambda)}{\hbar_N}\right)\\\\
0 & 1\end{array}\right]\sigma_1^{\frac{1-J}{2}}
\end{equation}
which is an exponentially small perturbation of the identity except near
the endpoints.  And for $\lambda\in L_R$ we have
\begin{equation}
{\bf O}_+(\lambda)={\bf O}_-(\lambda)
\sigma_1^{\frac{1-J}{2}}\left[\begin{array}{cc}1 & 
\displaystyle -i\exp\,\left(\frac{i\theta(\lambda)}{\hbar_N}\right)\\\\
0 & 1\end{array}\right]\sigma_1^{\frac{1-J}{2}}
\end{equation}
which is also a jump that is exponentially close to the identity.  
For $\lambda\in [0,iA(x)]$ we get
\begin{equation}
{\bf O}_+(\lambda)={\bf O}_-(\lambda)\left[\begin{array}{cc} 0&i\\\\i&0\end{array}\right]
\end{equation}
as a consequence of the factorization (\ref{eq:factorization}).  Since
${\bf O}(\lambda):=\tilde{\bf N}(\lambda)$ for all $\lambda$ in the
upper half-plane outside the lens bounded by $L_L$ and $L_R$, we see
that ${\bf O}(\lambda)$ satisfies the following jump condition on $C_L$:
\begin{equation}
{\bf O}_+(\lambda)={\bf O}_-(\lambda)\sigma_1^{\frac{1-J}{2}}
\left[\begin{array}{cc} 1 & 0\\\\\displaystyle
i\exp\,\left(\frac{\tau(\lambda) - i\theta^0(\lambda)}{\hbar_N}
\right) & 1\end{array}
\right]\sigma_1^{\frac{1-J}{2}}
\end{equation}
the following jump relation on $C_R$:
\begin{equation}
{\bf O}_+(\lambda)={\bf O}_-(\lambda)\sigma_1^{\frac{1-J}{2}}
\left[\begin{array}{cc} 1 & 0\\\\\displaystyle
i\exp\,\left(\frac{\tau(\lambda) +i\theta^0(\lambda)}{\hbar_N}
\right) & 1\end{array}
\right]\sigma_1^{\frac{1-J}{2}}\,,
\end{equation}
and the following jump relation on the imaginary interval
$[iA(x),iA]\subset C_M$:
\begin{equation}
{\bf O}_+(\lambda)={\bf O}_-(\lambda)\sigma_1^{\frac{1-J}{2}}
\left[\begin{array}{cc} 1 & 0\\\\\displaystyle i\exp\,\left(\frac{\tilde{\phi}(\lambda)}{\hbar_N}\right) & 1\end{array}\right]
\sigma_1^{\frac{1-J}{2}}\,.
\end{equation}
All three of these matrices are exponentially close to the identity
matrix pointwise in $\lambda$ for interior points of their respective
contours.

The matrix ${\bf O}(\lambda)$ is related to $\tilde{\bf N}(\lambda)$
by explicit transformations.  However, taking the pointwise limit of
the jump matrix for ${\bf O}(\lambda)$ leads us to the following {\em
ad hoc} model problem.
\begin{rhp}[Outer model problem]
Find a matrix $\tilde{\bf O}(\lambda)$ satisfying:
\begin{enumerate}
\item
{\bf Analyticity:} $\tilde{\bf O}(\lambda)$ is analytic for
$\lambda\in{\mathbb C}\setminus I$, where $I$ is the imaginary
interval $[-iA(x),iA(x)]$.
\item
{\bf Boundary behavior:} $\tilde{\bf O}(\lambda)$ assumes boundary values
that are continuous except at $\lambda=\pm iA(x)$, where at worst inverse
fourth-root singularities are admitted.
\item
{\bf Jump condition:}  For $\lambda\in I$,
\begin{equation}
\tilde{\bf O}_+(\lambda)=\tilde{\bf O}_-(\lambda)\left[\begin{array}{cc}
0&i\\\\i&0\end{array}\right]\,.
\end{equation}
\item
{\bf Normalization:}  $\tilde{\bf O}(\lambda)$ is normalized at infinity:
\begin{equation}
\tilde{\bf O}(\lambda)\rightarrow{\mathbb I}\hspace{0.2 in}\mbox{as}
\hspace{0.2 in}\lambda\rightarrow\infty\,.
\end{equation}
\end{enumerate}
\end{rhp}

This model problem is easily solved explicitly.
\begin{proposition}
The unique solution of this Riemann-Hilbert problem is
\begin{equation}
\tilde{\bf O}(\lambda):=\frac{1}{2R(\lambda)\beta(\lambda)}\left[\begin{array}{cc} R(\lambda)-\lambda-iA(x)& R(\lambda)+\lambda+iA(x)\\
R(\lambda)+\lambda+iA(x) & R(\lambda)-\lambda-iA(x)\end{array}\right]\,,
\label{eq:otildesolution}
\end{equation}
where $R(\lambda)^2=\lambda^2+A(x)^2$ and 
\begin{equation}
\beta(\lambda)^4 = \frac{\lambda+iA(x)}{\lambda-iA(x)}\,,
\end{equation}
with both functions $R(\lambda)$ and $\beta(\lambda)$ being analytic
in ${\mathbb C}\setminus I$, normalized according to
$R(\lambda)\sim-\lambda$ and $\beta(\lambda)\sim 1$ as
$\lambda\rightarrow\infty$.  
\end{proposition}

Using this matrix, we define an ``outer'' model for the matrix ${\bf
N}(\lambda)$ as follows.  The idea is to recall the relationship
between the matrix $\tilde{\bf N}(\lambda)$ and ${\bf O}(\lambda)$,
and simply substitute $\tilde{\bf O}(\lambda)$ for ${\bf O}(\lambda)$
in these formulae.  For $\lambda$ in between $L_L$ and $C_M$, we use
(\ref{eq:ONleft}) to set
\begin{equation}
\hat{\bf N}_{\rm out}(\lambda):=\tilde{\bf O}(\lambda)
\sigma_1^{\frac{1-J}{2}}\left[\begin{array}{cc}
1 & \displaystyle -i\exp\,\left(-\frac{i\theta(\lambda)}{\hbar_N}\right)\\\\
0 & 1\end{array}\right]\sigma_1^{\frac{1-J}{2}}\,.
\end{equation}
For $\lambda$ in between $C_M$ and $L_R$, we use (\ref{eq:ONright}) to
set
\begin{equation}
\hat{\bf N}_{\rm out}(\lambda):=\tilde{\bf O}(\lambda)
\sigma_1^{\frac{1-J}{2}}\left[\begin{array}{cc}
1 & \displaystyle i\exp\,\left(\frac{i\theta(\lambda)}{\hbar_N}\right)\\\\
0 & 1\end{array}\right]\sigma_1^{\frac{1-J}{2}}\,.
\end{equation}
For all other $\lambda$ in the upper half-plane, set $\hat{\bf N}_{\rm
out}(\lambda):=\tilde{\bf O}(\lambda)$, and in the lower half-plane
set $\hat{\bf N}_{\rm out}(\lambda):=\sigma_2\hat{\bf N}_{\rm
out}(\lambda^*)^* \sigma_2$.  The important properties of this matrix
are the following.
\begin{proposition}
The matrix $\hat{\bf N}_{\rm out}(\lambda)$ is analytic for all
complex $\lambda$ except at the contours $L_L$, $L_R$, the imaginary
interval $[0,iA(x)]$, and their complex-conjugates.  It satisfies the
following jump conditions:
\begin{equation}
\hat{\bf N}_{{\rm out},+}(\lambda)=\hat{\bf N}_{{\rm out},-}(\lambda)
\sigma_1^{\frac{1-J}{2}}\left[\begin{array}{cc}
1 & \displaystyle i\exp\,\left(-\frac{i\theta(\lambda)}{\hbar_N}
\right)\\\\
0 & 1\end{array}\right]\sigma_1^{\frac{1-J}{2}}\,,
\hspace{0.2 in}\mbox{for}\hspace{0.2 in}
\lambda\in L_L\,,
\end{equation}
\begin{equation}
\hat{\bf N}_{{\rm out},+}(\lambda)=\hat{\bf N}_{{\rm out},-}(\lambda)
\sigma_1^{\frac{1-J}{2}}\left[\begin{array}{cc}
1 & \displaystyle i\exp\,\left(\frac{i\theta(\lambda)}{\hbar_N}
\right)\\\\
0 & 1\end{array}\right]\sigma_1^{\frac{1-J}{2}}\,,
\hspace{0.2 in}\mbox{for}\hspace{0.2 in}
\lambda\in L_R\,,
\end{equation}
\begin{equation}
\hat{\bf N}_{{\rm out},+}(\lambda)=\hat{\bf N}_{{\rm out},-}(\lambda)
\sigma_1^{\frac{1-J}{2}}
\left[\begin{array}{cc}
\displaystyle\exp\,\left(\frac{i\theta(\lambda)}{\hbar_N}\right)&0\\\\
i &\displaystyle\exp\,\left(-\frac{i\theta(\lambda)}{\hbar_N}\right)
\end{array}\right]
\sigma_1^{\frac{1-J}{2}}\,,\hspace{0.2 in}
\mbox{for}\hspace{0.2 in}\lambda\in [0,iA(x)]\,,
\end{equation}
with the jump matrices on the conjugate contours in the lower half-plane
being obtained from these by the symmetry $\hat{\bf N}_{\rm out}(\lambda^*)=
\sigma_2\hat{\bf N}_{\rm out}(\lambda)^*\sigma_2$.  
In particular, note that for $\lambda\in [0,iA(x)]$, we have 
$\hat{\bf N}_{{\rm out},-}(\lambda)^{-1}\hat{\bf N}_{{\rm out},+}(\lambda)=
\tilde{\bf N}_-(\lambda)^{-1}\tilde{\bf N}_+(\lambda)$.
Also, if $D$ is any given open set containing the endpoint
$\lambda=iA(x)$, then $\hat{\bf N}_{\rm out}(\lambda)$ is uniformly
bounded for $\lambda\in {\mathbb C}\setminus (D\cup D^*)$ with a bound
that depends only on $D$ and not on $\hbar_N$.
\label{prop:Noutproperties}
\end{proposition}

\section{Local Analysis}
In justifying formally the local model $\hat{\bf N}_{\rm
out}(\lambda)$, we ignored the fact that the pointwise asymptotics for
the jump matrices for ${\bf O}(\lambda)$ that we used to obtain the
matrix $\tilde{\bf O}(\lambda)$ were not uniform near the origin or
near the moving endpoint $\lambda=iA(x)$.  We also neglected the
breakdown of the asymptotics for $a_L(\lambda)$, $a_R(\lambda)$, and
$a_M(\lambda)$ near the points $\lambda=0$ and $\lambda=iA$.
Consequently, we do not expect the outer model $\hat{\bf N}_{\rm
out}(\lambda)$ to be a good approximation to ${\bf N}(\lambda)$ near
$\lambda=0$, $\lambda=iA(x)$, or $\lambda=iA$.  In this section, we
examine the neighborhoods of these three points in more detail, and we
will obtain accurate local models for ${\bf N}(\lambda)$ in the
corresponding neighborhoods.

\subsection{Local analysis near $\lambda=0$.}
\label{sec:origin}
\subsubsection{Local behavior of the matrix elements $a_L(\lambda)$, 
$a_R(\lambda)$, and $a_M(\lambda)$}.  Let $\epsilon$ and $\delta$ be
small scales tied to $\hbar_N$ such that $\hbar_N\ll
\delta\ll\epsilon\ll 1$ as $\hbar_N\downarrow 0$.  Let $L$ be defined
as the unique integer for which exactly $N-L$ of the numbers
$\lambda_0,\dots,\lambda_{N-1}$ lie strictly below $i\epsilon$ on the
positive imaginary axis.  We want to compute uniform asymptotics for
$S(\lambda)$ defined by (\ref{eq:S}) for $\lambda\in C_L\cup C_R$, and
for $T(\lambda)$ defined by (\ref{eq:T}) for $\lambda\in C_M$ when
$|\lambda|\le\delta$.
\begin{lemma}
\label{lemma:originpeelfirst}
When $\Im(\lambda)\ge 0$ and $|\lambda|\le\delta$ and with $L$
defined as above,
\begin{equation}
\exp\,\left(-\sum_{k=0}^{L-1}\tilde{I}_k(\lambda)\right)=1+O\left(\frac{\hbar_N
}{\epsilon}\right)\,.
\end{equation}
\end{lemma}

{\em Proof:} We
recall the integral formula ({\em cf.} equation (\ref{eq:integralformula}))
\begin{equation}
\tilde{I}_k(\lambda)=\frac{1}{\hbar_N}\int_{m_k-\hbar_N/2}^{m_k+\hbar_N/2}
\,dm\int_{m_k}^m\,d\zeta \int_{m_k}^\zeta\,d\xi\,g(\lambda,\xi)
\,,
\label{eq:integralformulag}
\end{equation}
in which we expand the integrand in partial fractions:
\begin{equation}
g(\lambda,\xi)=\frac{e''(\xi)}{\lambda+e(\xi)}+\frac{e''(\xi)}{\lambda-e(\xi)}
-\frac{e'(\xi)^2}{(\lambda+e(\xi))^2}+\frac{e'(\xi)^2}{(\lambda-e(\xi))^2}\,.
\label{eq:glambdaxi}
\end{equation}
Since $\Im(\lambda)\ge 0$, for $m_k-\hbar_N/2\le\xi\le m_k+\hbar_N/2$
and $k=0,\dots,L-1$, we get
\begin{equation}
\frac{1}{|\lambda+e(\xi)|}\le
\frac{1}{|\lambda-e(\xi)|}\le 
\frac{1}{|i\delta-e(\xi)|} \le 
\frac{1}{|i\delta-e(m_k-\hbar_N/2)|}= 
O\left(\frac{1}{|m(\delta)-m_k+\hbar_N/2|}\right)\,.
\end{equation}
For such $\xi$ we therefore have
\begin{equation}
g(\lambda,\xi) = O\left(\frac{1}{|m(\delta)-m_k+\hbar_N/2|^2}\right)\,,
\end{equation}
so summing over $k$ gives
\begin{equation}
\sum_{k=0}^{L-1}\tilde{I}_k(\lambda)=O\left(\hbar_N^2\sum_{k=0}^{L-1} \frac{1}
{|m(\delta)-m_k+\hbar_N/2|^2}\right) = 
O\left(\hbar_N\int_{m(\epsilon)}^M\frac{dm}{(m-m(\delta))^2}\right)=O\left(\frac{\hbar_N}{\epsilon}\right)\,,
\end{equation}
because $\delta\ll\epsilon$, which proves the lemma.  $\Box$

\vspace{0.1 in}

So only the fraction of terms $\tilde{I}_k(\lambda)$ with $k\ge L$
contribute significantly to the sum for $\tilde{I}(\lambda)$.  It is
easy to check directly that $\exp(-\tilde{I}_k(\lambda))$ is an
analytic function for $|\lambda|\le\delta$ whenever $0\le k\le L-1$,
so it makes no difference in these terms whether it is
$L^0_\eta(\lambda)$ or $\overline{L}^0_\eta(\lambda)$ that appears in
the definition of $\tilde{I}_k$.  Therefore, the terms in $S(\lambda)$
and $T(\lambda)$ that can be significant for $\lambda$ near the origin
are thus
\begin{equation}
S_1^{(0)}(\lambda):=
\left(\prod_{k=L}^{N-1}\frac{\lambda-\lambda_k^*}
{\lambda-\lambda_k}\right)\exp\,\left(\frac{1}{\hbar_N}
\int_0^{m_L+\hbar_N/2}\left(L^0_{e(m)}(\lambda)
-L^0_{-e(m)}(\lambda)\right)\,dm
\right)
\end{equation}
and
\begin{equation}
T_1^{(0)}(\lambda):=
\left(\prod_{k=L}^{N-1}\frac{\lambda-\lambda_k^*}
{\lambda-\lambda_k}\right)\exp\,\left(\frac{1}{\hbar_N}
\int_0^{m_L+\hbar_N/2}\left(\overline{L}^0_{e(m)}(\lambda)
-\overline{L}^0_{-e(m)}(\lambda)\right)\,dm
\right)\cdot
 2\cos\,
\left(\frac{\theta^0(\lambda)}{\hbar_N}
\right)\,.
\end{equation}
Here we have written the integrals in the exponent using the change of
variables $m=m(\eta)$.  So Lemma~\ref{lemma:originpeelfirst} simply
says that $S(\lambda)=S_1^{(0)}(\lambda)(1+O(\hbar_N/\epsilon))$ and
$T(\lambda)=T_1^{(0)}(\lambda)(1+O(\hbar_N/\epsilon))$ uniformly for
$|\lambda|<\delta$.  When $\lambda$ is close to the origin along with
the points $\lambda_k$ contributing to $T(\lambda)$, the ladder of
discrete nodes appears to become equally spaced.  The next lemma shows
that this is indeed the case.
\begin{lemma}
\label{lemma:originpeelsecond}
Let $\tilde{\lambda}_{N-k}$ for $k=1,2,3,\dots$ be the sequence of
numbers defined by the relation:
\begin{equation}
\tilde{\lambda}_{N-k}:=-\frac{\hbar_N}{\rho^0(0)}(k-1/2)\,,
\end{equation}
which results from expanding the Bohr-Sommerfeld relation (\ref{eq:BS})
for $\lambda_{N-k}$ small, and keeping only the dominant terms.  Define
\begin{equation}
S_2^{(0)}(\lambda):=\left(\prod_{k=L}^{N-1}\frac{\lambda-\tilde{\lambda}_k^*}
{\lambda-\tilde{\lambda}_k}\right)
\exp\,\left(\frac{1}{\hbar_N}
\int_0^{m_L+\hbar_N/2}\left(L^0_{e'(0)m}(\lambda)
-L^0_{-e'(0)m}(\lambda)\right)\,dm
\right)\,,
\end{equation}
and
\begin{equation}
\begin{array}{rcl}
\displaystyle T_2^{(0)}(\lambda)&:=&\displaystyle
\left(\prod_{k=L}^{N-1}\frac{\lambda-\tilde{\lambda}_k^*}
{\lambda-\tilde{\lambda}_k}\right)
\exp\,\left(\frac{1}{\hbar_N}
\int_0^{m_L+\hbar_N/2}\left(\overline{L}^0_{e'(0)m}(\lambda)
-\overline{L}^0_{-e'(0)m}(\lambda)\right)\,dm
\right)\\\\
&&\,\,\,\displaystyle\times\,\,\,
 2\cos\,
\left(\frac{\pi\rho^0(0)}{\hbar_N}(iA-\lambda)\right)\,.
\end{array}
\end{equation}
Then, for $\Im(\lambda)\ge 0$ and $|\lambda|\le\delta$, 
\begin{equation}
T_1^{(0)}(\lambda)=T_2^{(0)}(\lambda)\left(1+O\left(\frac{\epsilon^2}{\hbar_N}\log\left(\frac{\epsilon}{\hbar_N}\right)\right)\right)\,,
\end{equation}
where we suppose that the scale $\epsilon$ is further constrained so that
the relative error is asymptotically small.  If $\lambda$ is additionally
bounded outside of some sector containing the positive imaginary axis, then
\begin{equation}
S_1^{(0)}(\lambda)=S_2^{(0)}(\lambda)\left(1+O\left(\frac{\epsilon^2}{\hbar_N}\right)\right)\,.
\end{equation}
\end{lemma}

{\em Proof:} We begin by observing that for $k=L,\dots,N-1$, the
distance between $\lambda_k$ and $\tilde{\lambda}_k$ is much smaller
than the distance between $\lambda_k$ and $\lambda_{k+1}$, as long as
$\epsilon\ll\hbar_N^{1/2}$.  More precisely, we have
\begin{equation}
|\tilde{\lambda}_k-\lambda_k|=O(\hbar_N^2(N-k)^2)\,.
\label{eq:twinning}
\end{equation}

Decompose the quotients as follows:
\begin{equation}
\frac{T_1^{(0)}(\lambda)}{T_2^{(0)}(\lambda)}=D(\lambda)C(\lambda)\overline{L}(\lambda)
\hspace{0.2 in}\mbox{and}\hspace{0.2 in}
\frac{S_1^{(0)}(\lambda)}{S_2^{(0)}(\lambda)}=D(\lambda)L(\lambda)\,,
\end{equation}
where
\begin{equation}
D(\lambda):=\prod_{k=L}^{N-1}\frac{\lambda-\lambda_k^*}{\lambda-\lambda_k}
\frac{\lambda-\tilde{\lambda}_k}{\lambda-\tilde{\lambda}_k^*}\,,
\end{equation}
\begin{equation}
C(\lambda):=\cos\,\left(\frac{\pi}{\hbar_N}
\int_\lambda^{iA}\rho^0(\eta)\,d\eta\right)\,\sec\,
\left(-\pi N-\frac{\pi}{\hbar_N}
\rho^0(0)\lambda\right)\,,
\end{equation}
\begin{equation}
\overline{L}(\lambda):=\exp\,\left(\frac{1}{\hbar_N}
\int_0^{m_L+\hbar_N/2}\left([\overline{L}^0_{e(m)}(\lambda)-
\overline{L}^0_{e'(0)m}(\lambda)]
-[\overline{L}^0_{-e(m)}(\lambda)-\overline{L}^0_{-e'(0)m}(\lambda)]\right)\,dm
\right)\,,
\label{eq:ellbar}
\end{equation}
and
\begin{equation}
L(\lambda):=\exp\,\left(\frac{1}{\hbar_N}
\int_0^{m_L+\hbar_N/2}\left([L^0_{e(m)}(\lambda)-L^0_{e'(0)m}(\lambda)]
-[L^0_{-e(m)}(\lambda)-L^0_{-e'(0)m}(\lambda)]\right)\,dm
\right)\,.
\label{eq:ell}
\end{equation}

First we deal with $L(\lambda)$ and $\overline{L}(\lambda)$.  Since
$e(m)$ is differentiable and $m$ is small we have $e(m)-e'(0)m=O(\epsilon)$.
Also, the interval of integration is $O(\epsilon)$ in length.
Although the integrands in (\ref{eq:ellbar}) and (\ref{eq:ell})
are not pointwise small, upon integration it follows that
\begin{equation}
L(\lambda)=1+O\left(\frac{\epsilon^2}{\hbar_N}\right)\hspace{0.2 in}
\mbox{and}\hspace{0.2 in}
\overline{L}(\lambda)=1+O\left(\frac{\epsilon^2}{\hbar_N}\right)\,,
\label{eq:elelbar}
\end{equation}
uniformly for all $\lambda$ in the upper half-plane satisfying
$|\lambda|\le\delta$. Here we are assuming that
$\epsilon\ll\hbar_N^{1/2}$.

For the moment, let's drop the conditions $\Im(\lambda)\ge 0$ and
$|\lambda|\le\delta$ and instead consider $\lambda$ to lie on the
sides of the square centered at the origin, one of whose sides is
parallel to the real axis and intersects the positive imaginary axis
halfway between the points $\lambda=\tilde{\lambda}_L$ and
$\lambda=\tilde{\lambda}_{L-1}$.  Note that the estimate
(\ref{eq:twinning}) implies that the sides of the square intersect the
real and imaginary axes a distance from the origin that is
approximately $\epsilon$.  Therefore the square asymptotically
contains the closed disk $|\lambda|\le\delta$ because
$\delta\ll\epsilon$.  We will show that for $\lambda$ on the four
sides of the square, both $D(\lambda)$ and $C(\lambda)$ are very
close to one.  We write $D(\lambda)$ in the form
\begin{equation}
D(\lambda)=
\prod_{k=L}^{N-1}\left(1+\frac{\tilde{\lambda}_k^*-\lambda_k^*}
{\lambda-\tilde{\lambda}_k^*}\right)
\left(1+\frac{\tilde{\lambda}_k-\lambda_k}
{\lambda-\tilde{\lambda}_k}\right)^{-1}\,.
\end{equation}
First consider the top of the square: for
$\Im(\lambda)=-i(\tilde{\lambda}_L+\tilde{\lambda}_{L-1})/2$, we
easily see that
\begin{equation}
|\lambda-\tilde{\lambda}_k|\ge\frac{i\hbar_N}{\rho^0(0)}(k-L+1/2)
\hspace{0.2 in}\mbox{and}\hspace{0.2 in} \frac{1}{|\lambda-\tilde{\lambda}_k^*|}=O\left(\frac{1}{\epsilon}\right)\,,
\end{equation}
for $k=L,\dots,N-1$.  Combining this with (\ref{eq:twinning}), we get
\begin{equation}
\frac{\tilde{\lambda}^*_k-\lambda_k^*}{\lambda-\tilde{\lambda}_k^*}=
O\left(\frac{\hbar_N^2(N-k)^2}{\epsilon}\right)\hspace{0.2 in}
\mbox{and}\hspace{0.2 in}
\frac{\tilde{\lambda}_k-\lambda_k}{\lambda-\tilde{\lambda}_k}=
O\left(\frac{\hbar_N^2(N-k)^2}{\hbar_N(k-L+1/2)}\right)\,.
\end{equation}
Summing these estimates over $k$ (it is convenient to approximate sums
by integrals in doing so), we find that
\begin{equation}
\prod_{k=L}^{N-1}\left(1+\frac{\tilde{\lambda}_k^*-\lambda_k^*}{\lambda-\tilde{\lambda}_k^*}\right) = 1+O\left(\frac{\epsilon^2}{\hbar_N}\right)
\hspace{0.2 in}\mbox{and}\hspace{0.2 in}
\prod_{k=L}^{N-1}\left(1+\frac{\tilde{\lambda}_k-\lambda_k}{\lambda-\tilde{\lambda}_k}\right)^{-1}=1+
O\left(\frac{\epsilon^2}{\hbar_N}\log\left(\frac{\epsilon}{\hbar_N}\right)\right)\,.
\end{equation}
Consequently, for $\lambda$ on the top of the square, 
\begin{equation}
D(\lambda)=1+
O\left(\frac{\epsilon^2}{\hbar_N}\log\left(\frac{\epsilon}{\hbar_N}\right)
\right)\,.
\end{equation}
An estimate of the same form holds when $\lambda$ is on the bottom of the
square, where $\Im(\lambda)=
i(\tilde{\lambda}_L+\tilde{\lambda}_{L-1})/2$.  When $\lambda$ is on the
left or right side of the square, so that $|\Re(\lambda)|=-
i(\tilde{\lambda}_L+\tilde{\lambda}_{L-1})/2$, both 
$|\lambda-\tilde{\lambda}^*_k|^{-1}$ and $|\lambda-\tilde{\lambda}_k|^{-1}$
are $O(\epsilon^{-1})$.  By the same arguments as above, we then have for
such $\lambda$ that
\begin{equation}
D(\lambda)=1+O\left(\frac{\epsilon^2}{\hbar_N}\right)\,.
\end{equation}
Now we look at $C(\lambda)$ on the same square.  Generally, for such $\lambda$
which are of order $\epsilon$ in magnitude, we have
\begin{equation}
C(\lambda)=1+O\left(\frac{\epsilon^2}{\hbar_N}\right)\sec\,\left(-\pi N -
\frac{\pi}{\hbar_N}\rho^0(0)\lambda\right)\,.
\end{equation}
When $\lambda$ is on the top or bottom of the square, we have
\begin{equation}
\left|\sec\,\left(-\pi N -
\frac{\pi}{\hbar_N}\rho^0(0)\lambda\right)\right|\le 1\,,
\end{equation}
and when $\lambda$ is on the left or right sides of the square, the
same quantity is exponentially small.  It follows easily that for
$\lambda$ on any of the sides of the square,
\begin{equation}
C(\lambda)=1+O\left(\frac{\epsilon^2}{\hbar_N}\right)\,.
\end{equation}

So uniformly on the four sides of the square, we have
\begin{equation}
D(\lambda)C(\lambda)=1+
O\left(\frac{\epsilon^2}{\hbar_N}\log\left(\frac{\epsilon}{\hbar_N}\right)
\right)\,.
\end{equation}
But the product $D(\lambda)C(\lambda)$ is analytic within the square,
so by the maximum principle it follows that the same estimate holds
for all $\lambda$ on the interior of the square, and in particular for
all $\lambda$ in the upper half-plane with $|\lambda|\le\delta$.
This shows that 
\begin{equation}
T_1^{(0)}(\lambda)=T_2^{(0)}(\lambda)\left(1+O\left(\frac{\epsilon^2}{\hbar_N}\log\left(
\frac{\epsilon}{\hbar_N}\right)\right)\right)
\end{equation}
holds for all such $\lambda$.  

Now to control the relationship between
$S_1^{(0)}(\lambda)$ and $S_2^{(0)}(\lambda)$ we consider $\lambda$ to lie outside
of some symmetrical sector about the positive imaginary axis, of
arbitrarily small nonzero opening angle $2\alpha$ independent of
$\hbar_N$.  Since $\Im(\lambda)\ge 0$, we get
\begin{equation}
|\lambda-\tilde{\lambda}_k^*|\ge |\lambda-\tilde{\lambda}_k|\ge
\frac{|\tilde{\lambda}_k|}{\sin(\alpha)} = \frac{i\hbar_N (N-k-1/2)}{\rho^0(0)
|\sin(\alpha)|}\,.
\end{equation}
Combining this result with (\ref{eq:twinning}), we find
\begin{equation}
\frac{\tilde{\lambda}_k^*-\lambda_k^*}{\lambda-\tilde{\lambda}_k^*}=
O(\hbar_N(N-k))\hspace{0.2 in}\mbox{and}\hspace{0.2 in}
\frac{\tilde{\lambda}_k-\lambda_k}{\lambda-\tilde{\lambda}_k}=
O(\hbar_N(N-k))\,.
\end{equation}
Summing these estimates over $k$ one finds that
\begin{equation}
D(\lambda)=1+O\left(\frac{\epsilon^2}{\hbar_N}\right)\,.
\end{equation}
Combining this with the estimate (\ref{eq:elelbar}) of $L(\lambda)-1$, 
we find that
\begin{equation}
S_1^{(0)}(\lambda)=S_2^{(0)}(\lambda)\left(1+O\left(\frac{\epsilon^2}{\hbar_N}\right)\right)\,,
\end{equation}
for all $\lambda$ in the upper half-plane with $|\lambda|<\delta$ and
bounded outside of the sector of opening angle $2\alpha$ about the
positive imaginary axis.  This completes the proof.  $\Box$

Without any approximation, $S_2^{(0)}(\lambda)$ can be rewritten in the form:
\begin{equation}
S_2^{(0)}(\lambda)=(-i\zeta)^{-i\zeta}(i\zeta)^{-i\zeta}\frac{\Gamma(1/2+i\zeta)
(\overline{N}+i\zeta)^{\overline{N}+i\zeta}\Gamma(\overline{N}+1/2-i\zeta)}
{\Gamma(1/2-i\zeta)(\overline{N}-i\zeta)^{\overline{N}-i\zeta}
\Gamma(\overline{N}+1/2+i\zeta)}
\end{equation}
and $T_2^{(0)}(\lambda)$ can be rewritten in the
form:
\begin{equation}
T_2^{(0)}(\lambda)=\frac{2\pi}{\Gamma(1/2-i\zeta)^2}(-i\zeta)^{-2i\zeta}\frac{(\overline{N}+i\zeta)^{\overline{N}+i\zeta}\Gamma(\overline{N}+1/2-i\zeta)}
{(\overline{N}-i\zeta)^{\overline{N}-i\zeta}\Gamma(\overline{N}+1/2+i\zeta)}\,,
\end{equation}
where $\overline{N}:=N-L$ and we are introducing a transformation $\varphi_0$
to a local variable $\zeta$ given by
\begin{equation}
\zeta=\varphi_0(\lambda):=-i\rho^0(0)\lambda/\hbar_N\,.
\label{eq:varphi0}
\end{equation}
These formulae come from evaluating the logarithmic integrals exactly,
which is possible because $e(m)$ has been replaced by the
linear function $e'(0)m$, taking advantage of the equal spacing of the
$\tilde{\lambda}_k$ to write the product explicitly in terms of gamma
functions, and then using the reflection identity for the gamma
function to eliminate the cosine from $T_2^{(0)}(\lambda)$.  Now, the integer
$\overline{N}$ is large, approximately of size $\epsilon/\hbar_N$.
But for $|\lambda|\le\delta$, $\overline{N}$ is asymptotically large
compared to $\zeta$ because $\delta\ll\epsilon$.  These observations
allow us to apply Stirling-type asymptotics to $S_2^{(0)}(\lambda)$ and
$T_2^{(0)}(\lambda)$.
\begin{lemma}
\label{lemma:originpeelthird}
In addition to all prior hypotheses, suppose that $\delta^2\ll\epsilon\hbar_N$.
Then, 
\begin{equation}
S_2^{(0)}(\lambda)=e^{2i\zeta}(-i\zeta)^{-i\zeta}(i\zeta)^{-i\zeta}\frac{\Gamma(1/2-i\zeta)}{\Gamma(1/2+i\zeta)}\left(1+O\left(\frac{\delta^2}{\epsilon\hbar_N}\right)\right)\,,
\end{equation}
and
\begin{equation}
T_2^{(0)}(\lambda)=\frac{2\pi e^{2i\zeta}(-i\zeta)^{-2i\zeta}}
{\Gamma(1/2-i\zeta)^2}\left(1+O\left(
\frac{\delta^2}{\epsilon\hbar_N}\right)\right)\,.
\end{equation}
\end{lemma}

{\em Proof:} Asymptotically expanding the gamma functions for large
$\overline{N}$, we find that
\begin{equation}
S_2^{(0)}(\lambda)=e^{2i\zeta}(-i\zeta)^{-i\zeta}(i\zeta)^{-i\zeta}\frac{\Gamma(1/2+i\zeta)}{\Gamma(1/2-i\zeta)}\cdot\Delta(\zeta,\overline{N})\cdot\left(1+
O\left(\frac{1}{\overline{N}}\right)\right)\,,
\end{equation}
and
\begin{equation}
T_2^{(0)}(\lambda)=\frac{2\pi e^{2i\zeta}(-i\zeta)^{-2i\zeta}}{\Gamma(1/2-i\zeta)^2}\cdot\Delta(\zeta,
\overline{N})\cdot
\left(1+O\left(\frac{1}{\overline{N}}\right)\right)\,.
\end{equation}
where
\begin{equation}
\Delta(\zeta,\overline{N}):=
\frac{(\overline{N}+i\zeta)^{\overline{N}+i\zeta}}
{(\overline{N}+i\zeta + 1/2)^{\overline{N}+i\zeta}}
\frac{(\overline{N}-i\zeta+1/2)^{\overline{N}-i\zeta}}
{(\overline{N}-i\zeta)^{\overline{N}-i\zeta}}\,.
\end{equation}
Next, expanding $\Delta(\zeta,\overline{N})$, one gets worse error
terms:
\begin{equation}
\Delta(\zeta,\overline{N})=
1+O\left(\left(\frac{\delta}{\hbar_N}\right)^2\frac{1}{\overline{N}}\right)\,.
\end{equation}
Combining these estimates and noting that $1/\overline{N}=O(\hbar_N/\epsilon)$
completes the proof of the lemma.  $\Box$

With these results in hand, we can easily establish the following.
\begin{proposition}
Let $\lambda$ be in the upper half-plane, with
$|\lambda|\le\hbar_N^\alpha$, where $3/4<\alpha<1$, and let $\lambda$
be bounded outside of some fixed symmetrical sector containing the
positive imaginary axis.  Then
\begin{equation}
S(\lambda)=e^{2i\zeta}(-i\zeta)^{-i\zeta}(i\zeta)^{-i\zeta}
\frac{\Gamma(1/2+i\zeta)}{\Gamma(1/2-i\zeta)}
\left(1+O(\hbar_N^{4\alpha/3-1})\right)\,,
\end{equation}
where $\zeta=\varphi_0(\lambda):=-i\rho^0(0)\lambda/\hbar_N$.
\label{prop:Sinner}
\end{proposition}

{\em Proof:} According to Lemmas~\ref{lemma:originpeelfirst},
\ref{lemma:originpeelsecond}, and \ref{lemma:originpeelthird}, the
total relative error is a sum of three terms:
\begin{equation}
O\left(\frac{\hbar_N}{\epsilon}\right)\hspace{0.2 in}\mbox{and}\hspace{0.2 in}
O\left(\frac{\epsilon^2}{\hbar_N}\right)
\hspace{0.2 in}\mbox{and}\hspace{0.2 in}
O\left(\frac{\delta^2}{\epsilon\hbar_N}\right)\,.
\end{equation}
Note that since $\hbar_N\ll\delta$, the order $\hbar_N/\epsilon$ term
is always dominated asymptotically by the order
$\delta^2/\epsilon\hbar_N$ term.  The error is optimized by picking $\epsilon$
so that the two possibly dominant terms are in balance.  This forces us
to choose $\epsilon\sim\delta^{2/3}$.  The proposition follows upon taking
$\delta=\hbar_N^{\alpha}$.  $\Box$

\vspace{0.1 in}

\begin{proposition}
Let $\lambda$ be in the upper half-plane, with
$|\lambda|\le\hbar_N^\alpha$, where $3/4<\alpha<1$.  Then for all $\nu>0$,
however small,
\begin{equation}
T(\lambda)=\frac{2\pi e^{2i\zeta}(-i\zeta)^{-2i\zeta}}{\Gamma(1/2-i\zeta)^2}
\left(1+O(\hbar_N^{4\alpha/3-1-\nu})\right)\,,
\end{equation}
where $\zeta=\varphi_0(\lambda):=-i\rho^0(0)\lambda/\hbar_N$.
\label{prop:Tinner}
\end{proposition}

{\em Proof:} In this case, according to Lemmas~\ref{lemma:originpeelfirst},
\ref{lemma:originpeelsecond}, and \ref{lemma:originpeelthird}, 
the total relative error is a sum of three
different terms:
\begin{equation}
O\left(\frac{\hbar_N}{\epsilon}\right)\hspace{0.2 in}\mbox{and}\hspace{0.2 in}
O\left(\frac{\epsilon^2}{\hbar_N}\log\left(\frac{\epsilon}{\hbar_N}\right)\right)\hspace{0.2 in}\mbox{and}\hspace{0.2 in}
O\left(\frac{\delta^2}{\epsilon\hbar_N}\right)\,.
\end{equation}
Again, since $\hbar_N\ll\delta$, the order $\hbar_N/\epsilon$ term is
always dominated asymptotically by the order
$\delta^2/\epsilon\hbar_N$ term.  For any $\sigma>0$, we have
\begin{equation}
\frac{\epsilon^2}{\hbar_N}\log\left(\frac{\epsilon}{\hbar_N}\right)=O\left(
\frac{\epsilon^2}{\hbar_N}\left(\frac{\epsilon}{\hbar_N}\right)^\sigma\right)\,.
\label{eq:cruder}
\end{equation}
so we can eliminate the logarithm at the expense of a slightly larger
error.  Taking $\delta=\hbar_N^\alpha$ as in the statement of the
proposition, and using the cruder estimate (\ref{eq:cruder}), the
nearly optimal value of $\epsilon$ to minimize the total relative
error is achieved by a dominant balance between the right-hand side of
(\ref{eq:cruder}) and the term of order $\delta^2/\epsilon\hbar_N$.
The balance gives $\epsilon=\hbar_N^\beta$, with
\begin{equation}
\beta=\frac{2\alpha+\sigma}{3+\sigma}\,.
\end{equation}
With this choice of $\epsilon$, the total relative error is of the
order $\hbar_N^\gamma$, with
\begin{equation}
\gamma=2\alpha-1-\beta=\frac{4\alpha+2(\alpha-1)\sigma-3}{3+\sigma}<\frac{4}{3}\alpha-1
\end{equation}
with the inequality following because $\sigma>0$ and $\alpha<1$.  The
inequality fails in the limit $\sigma\rightarrow 0$.  Therefore, for
each arbitrarily small $\nu>0$, we can find a $\sigma>0$ sufficiently
small that $\gamma>4\alpha/3-1-\nu$.  This gives us a slightly less
optimal estimate of the relative error: simply
$O(\hbar_N^{4\alpha/3-1-\nu})$, which completes the proof.  $\Box$

\subsubsection{The model Riemann-Hilbert problem.}
To repair the flaw in our model $\hat{\bf N}_{\rm out}(\lambda)$ for
the matrix ${\bf N}(\lambda)$ related to the nonuniformity of the
approximation of the jump matrices near the origin, we need to provide
a different approximation of ${\bf N}(\lambda)$ that will be valid
when $|\lambda|\le\hbar_N^\alpha$ for some $\alpha\in (3/4,1)$.  The
local failure of the approximation is gauged by the deviation of the
matrix quotient ${\bf N}(\lambda)\hat{\bf N}_{\rm out}(\lambda)^{-1}$
from the identity matrix near the origin.  It turns out to be more
convenient to study a conjugated form of this matrix (which also
deviates from the identity for $\lambda$ near the origin).  Namely,
for $|\lambda|\le\hbar_N^\alpha$, set
\begin{equation}
{\bf F}(\lambda):=e^{-i\theta(0)\sigma_3/(2\hbar_N)}
\sigma_1^{\frac{1-J}{2}}(i\sigma_1)
\tilde{\bf O}(\lambda)^{-1}
{\bf N}(\lambda)\hat{\bf N}_{\rm out}(\lambda)^{-1}\tilde{\bf O}(\lambda)
(-i\sigma_1)\sigma_1^{\frac{1-J}{2}}e^{i\theta(0)\sigma_3/(2\hbar_N)}
\label{eq:Fdefleft}
\end{equation}
if $\Re(\lambda)<0$ and
\begin{equation}
{\bf F}(\lambda):=e^{-i\theta(0)\sigma_3/(2\hbar_N)}
\sigma_1^{\frac{1-J}{2}}\tilde{\bf O}(\lambda)^{-1}
{\bf N}(\lambda)\hat{\bf N}_{\rm out}(\lambda)^{-1}\tilde{\bf O}(\lambda)
\sigma_1^{\frac{1-J}{2}}e^{i\theta(0)\sigma_3/(2\hbar_N)}
\label{eq:Fdefright}
\end{equation}
if $\Re(\lambda)>0$.  It is easy to check that as a consequence of the boundary
conditions satisfied by the matrix $\tilde{\bf O}(\lambda)$ on the imaginary
axis near the origin, the conjugating factors are analytic throughout the
disk $|\lambda|\le\hbar_N^\alpha$.

For later convenience, let us assume without loss of generality that
the auxiliary contours $C_L$, $C_R$, $L_L$ and $L_R$ are straight rays in
some $\hbar_N$-independent neighborhood of the origin.  It is easy to write
down the jump conditions satisfied by ${\bf F}(\lambda)$ on these four
rays and also on the positive imaginary axis.  We find:
\begin{equation}
{\bf F}_+(\lambda)={\bf F}_-(\lambda)\left[\begin{array}{cc}
1 & a_L(\lambda)e^{-i\theta(0)/\hbar_N}\\\\
0 & 1\end{array}\right]\,,\hspace{0.2 in}
\mbox{for}\hspace{0.2 in}\lambda\in C_L\,,
\end{equation}
\begin{equation}
{\bf F}_+(\lambda)={\bf F}_-(\lambda)\left[\begin{array}{cc}
1 & 0\\\\
a_R(\lambda)e^{i\theta(0)/\hbar_N} & 1\end{array}\right]\,,\hspace{0.2 in}
\mbox{for}\hspace{0.2 in}\lambda\in C_R\,,
\end{equation}
\begin{equation}
{\bf F}_+(\lambda)={\bf F}_-(\lambda)\left[\begin{array}{cc}
1 & 0\\\\
-ie^{-i(\theta(\lambda)-\theta(0))/\hbar_N} & 1\end{array}\right]\,,
\hspace{0.2 in}
\mbox{for}\hspace{0.2 in}\lambda\in L_L\,,
\end{equation}
\begin{equation}
{\bf F}_+(\lambda)={\bf F}_-(\lambda)\left[\begin{array}{cc}
1 & -ie^{i(\theta(\lambda)-\theta(0))/\hbar_N}\\\\
0 & 1\end{array}\right]\,,
\hspace{0.2 in}
\mbox{for}\hspace{0.2 in}\lambda\in L_R\,,
\end{equation}
and for $\lambda\in C_M$,
\begin{equation}
{\bf F}_+(\lambda)={\bf F}_-(\lambda)\left[\begin{array}{cc}
1 +[ia_M(\lambda)+e^{\tilde{\phi}(\lambda)/\hbar_N}]& 
-ie^{i(\theta(\lambda)-\theta(0))/\hbar_N}
[ia_M(\lambda)+e^{\tilde{\phi}(\lambda)/\hbar_N}]\\\\
-ie^{-i(\theta(\lambda)-\theta(0))/\hbar_N}
[ia_M(\lambda)+e^{\tilde{\phi}(\lambda)/\hbar_N}] & 
1-[ia_M(\lambda)+e^{\tilde{\phi}(\lambda)/\hbar_N}]\end{array}\right]\,.
\label{eq:FjumpCM}
\end{equation}
The jump relations satisfied by ${\bf F}(\lambda)$ on the complex
conjugate contours in the lower half-plane follow from these by the
symmetry ${\bf F}(\lambda)=\sigma_2{\bf F}(\lambda^*)^*\sigma_2$.

Now, for $\lambda\in C_L$ with $|\lambda|\le\hbar_N^\alpha$,
\begin{equation}
\begin{array}{rcl}
\displaystyle a_L(\lambda)e^{-i\theta(0)/\hbar_N}&=&
\displaystyle
i\exp\,\left(\frac{\tau(\lambda)-i\theta^0(\lambda)-i\theta(0)}
{\hbar_N}\right)\cdot S(\lambda)\\\\ &=&\displaystyle
i\exp\,\left(\frac{\tilde{\phi}(\lambda)+i(\theta(\lambda)-\theta(0))-2i\theta^0(\lambda)}{\hbar_N}\right)\cdot
S(\lambda)\\\\ &=&\displaystyle
i\exp\,\left(\frac{i(\theta(\lambda)-\theta(0))-2i(\theta^0(\lambda)-\theta^0(0))}{\hbar_N}\right)\cdot
S(\lambda)\\\\ &=&\displaystyle
ie^{\pi\zeta}S(\lambda)\left(1+O\left(\hbar_N^{2\alpha-1}\right)\right)\\\\
&=&\displaystyle
ie^{(2i+\pi)\zeta}(-i\zeta)^{-i\zeta}(i\zeta)^{-i\zeta}\frac{\Gamma(1/2+i\zeta)}
{\Gamma(1/2-i\zeta)}\left(1+
O\left(\hbar^{4\alpha/3-1}_N\right)\right)\,,
\end{array}
\label{eq:aLmodel}
\end{equation}
where in the last line $\zeta=\varphi_0(\lambda)$ with the change of
coordinate being given by (\ref{eq:varphi0}).  In these steps, we used
the relation (\ref{eq:tauleft}), the fact that from
Proposition~\ref{prop:phitildezero} we get
$\tilde{\phi}(\lambda)\equiv 0$, and, according to
(\ref{eq:deftheta0}) and the quantization condition
(\ref{eq:quantization}) on $\hbar_N$, $2\theta^0(0)/\hbar_N=2\pi N\in
2\pi {\mathbb Z}$.  We have also used the fact that
\begin{equation}
\frac{1}{\rho^0(0)}\left(2\frac{d\theta^0}{d\lambda}(0)-\frac{d\theta}
{d\lambda}(0)\right) = \pi\,.
\end{equation}
which follows directly from the definition (\ref{eq:deftheta0}) of
$\theta^0(\lambda)$, the definition (\ref{eq:deftheta}) of
$\theta(\lambda)$, and the relation (\ref{eq:rhosequal}).  
In a similar way, for $\lambda\in C_R$ with $|\lambda|\le\hbar_N^\alpha$,
we get
\begin{equation}
a_R(\lambda)e^{i\theta(0)/\hbar_N}=ie^{(2i-\pi)\zeta}(-i\zeta)^{-i\zeta}
(i\zeta)^{-i\zeta}\frac{\Gamma(1/2+i\zeta)}{\Gamma(1/2-i\zeta)}
\left(1+O\left(\hbar_N^{4\alpha/3-1}\right)\right)\,,
\label{eq:aRmodel}
\end{equation}
and for $\lambda\in L_L\cup C_M \cup L_R$ with $|\lambda|\le\hbar_N^\alpha$,
\begin{equation}
e^{\pm i(\theta(\lambda)-\theta(0))/\hbar_N}=
e^{\mp\pi\zeta}\left(1+
O\left(\hbar_N^{2\alpha-1}\right)\right)\,,
\end{equation}
with $\zeta=\varphi_0(\lambda)$.  Finally, when $\lambda\in C_M$ and
$|\lambda|\le\hbar_N^\alpha$ we have for arbitrarily small $\nu>0$,
\begin{equation}
\begin{array}{rcl}
\displaystyle ia_M(\lambda)+e^{\tilde{\phi}(\lambda)/\hbar_N}&=&
\displaystyle e^{\tilde{\phi}(\lambda)/\hbar_N}\left[1-T(\lambda)
\right]\\\\
&=&\displaystyle 1-T(\lambda)\\\\
&=&\displaystyle 1-\frac{2\pi e^{2i\zeta}(-i\zeta)^{-2i\zeta}}
{\Gamma(1/2-i\zeta)^2}
\left(1+O\left(\hbar_N^{4\alpha/3-1-\nu}\right)\right) \\\\
&=&\displaystyle 
1-\frac{2\pi e^{2i\zeta}(-i\zeta)^{-2i\zeta}}{\Gamma(1/2-i\zeta)^2} + 
O\left(\hbar_N^{4\alpha/3-1-\nu}\right)\,,
\end{array}
\label{eq:aMmodel}
\end{equation}
again with $\zeta=\varphi_0(\lambda)$.  The last step follows because
$2\pi e^{2i\zeta}(-i\zeta)^{-2i\zeta}/\Gamma(1/2-i\zeta)^2$ is
uniformly bounded on $C_M$.

Let $\vec{C_L}$, $\vec{C_R}$, $\vec{L_L}$, $\vec{L_R}$, and
$\vec{C_M}$ denote the straight rays that agree with the corresponding
contours in a fixed neighborhood of the origin in the $\lambda$-plane,
but lying in the $\zeta$-plane (according to (\ref{eq:varphi0}),
$\zeta$ is a simple rescaling of $\lambda$ by a positive number).
These rays are oriented contours, with the same orientation as the
original contours: $\vec{C_M}$, $\vec{L_L}$, and $\vec{L_R}$ are
oriented outwards from the origin toward infinity, and $\vec{C_L}$ and
$\vec{C_R}$ are oriented inwards from infinity toward the origin.  Let
the union of these contours with their complex conjugates be denoted
$\Sigma_0$.  Consider the following Riemann-Hilbert problem.
\begin{rhp}[Local Model for the Origin]
Find a matrix $\hat{\bf F}(\zeta)$ with the following properties:
\begin{enumerate}
\item{\bf Analyticity:}  $\hat{\bf F}(\zeta)$ is analytic for
$\zeta\in {\mathbb C}\setminus \Sigma_0$.
\item
{\bf Boundary behavior:}  $\hat{\bf F}(\zeta)$ assumes continuous
boundary values on $\Sigma_0$.
\item
{\bf Jump conditions:}  The boundary values taken on $\Sigma_0$
satisfy
\begin{equation}
\hat{\bf F}_+(\zeta)=\hat{\bf F}_-(\zeta)\left[\begin{array}{cc}
1 & \displaystyle ie^{(2i+\pi)\zeta}(-i\zeta)^{-i\zeta}(i\zeta)^{-i\zeta}
\frac{\Gamma(1/2+i\zeta)}{\Gamma(1/2-i\zeta)}\\\\
0 & 1\end{array}\right]\,,\hspace{0.2 in}\zeta\in \vec{C_L}\,,
\end{equation}
\begin{equation}
\hat{\bf F}_+(\zeta)=\hat{\bf F}_-(\zeta)\left[\begin{array}{cc}
1 & 0\\\\
\displaystyle ie^{(2i-\pi)\zeta}(-i\zeta)^{-i\zeta}(i\zeta)^{-i\zeta}
\frac{\Gamma(1/2+i\zeta)}{\Gamma(1/2-i\zeta)} & 1\end{array}\right]\,,\hspace{0.2 in}\zeta\in \vec{C_R}\,,
\end{equation}
\begin{equation}
\hat{\bf F}_+(\zeta)=\hat{\bf F}_-(\zeta)\left[\begin{array}{cc}
1 & 0\\\\ -ie^{\pi\zeta}&1
\end{array}\right]\,,\hspace{0.2 in}\zeta\in \vec{L_L}\,,
\end{equation}
\begin{equation}
\hat{\bf F}_+(\zeta)=\hat{\bf F}_-(\zeta)\left[\begin{array}{cc}
1 & -ie^{-\pi\zeta}\\\\ 0&1
\end{array}\right]\,,\hspace{0.2 in}\zeta\in \vec{L_R}\,,
\end{equation}
and
\begin{equation}
\hat{\bf F}_+(\zeta)=\hat{\bf F}_-(\zeta)\left[\begin{array}{cc}
\displaystyle 2-\frac{2\pi e^{2i\zeta}(-i\zeta)^{-2i\zeta}}
{\Gamma(1/2-i\zeta)^2} &
\displaystyle ie^{-\pi\zeta}
\left[\frac{2\pi e^{2i\zeta}(-i\zeta)^{-2i\zeta}}
{\Gamma(1/2-i\zeta)^2}-1\right]
\\\\
\displaystyle ie^{\pi\zeta} 
\left[\frac{2\pi e^{2i\zeta}(-i\zeta)^{-2i\zeta}}
{\Gamma(1/2-i\zeta)^2}-1\right]
&\displaystyle
\frac{2\pi e^{2i\zeta}(-i\zeta)^{-2i\zeta}}{\Gamma(1/2-i\zeta)^2} 
\end{array}\right]\,,\hspace{0.2 in}\zeta\in \vec{C_M}\,.
\label{eq:FhatjumpCM}
\end{equation}
On the contours in the lower half-plane, the jump conditions are
implied by the symmetry $\hat{\bf F}(\zeta^*)=\sigma_2\hat{\bf
F}(\zeta)^*\sigma_2$.
\item{\bf Normalization:}  $\hat{\bf F}(\zeta)$ is normalized
at infinity:
\begin{equation}
\hat{\bf F}(\zeta)\rightarrow {\mathbb I}\hspace{0.2 in}\mbox{as}
\hspace{0.2 in}\zeta\rightarrow\infty\,.
\label{eq:Fhatnorm}
\end{equation}
\end{enumerate}
\label{rhp:localorigin}
\end{rhp}

\vspace{0.1 in}

Unfortunately, we cannot solve Riemann-Hilbert
Problem~\ref{rhp:localorigin} explicitly.  Luckily, we will not
require an explicit solution.  However, existence of a solution is an
issue that must be resolved, and we need to obtain a decay estimate
that quantifies the normalization condition (\ref{eq:Fhatnorm}).
These questions are addressed via the abstract theory of Riemann-Hilbert
problems.
\begin{proposition}
The local model Riemann-Hilbert Problem~\ref{rhp:localorigin}
has a unique solution satisfying $\hat{\bf F}(\zeta)={\mathbb I}+
O(1/\zeta)$ as $\zeta\rightarrow\infty$, uniformly with respect
to direction.  Also, $\det(\hat{\bf F}(\zeta))\equiv 1$.
\end{proposition}

{\em Proof:} Each Riemann-Hilbert problem is equivalent to an
inhomogeneous system of linear singular integral equations.  It must
be shown that the matrix singular integral operator associated with
these equations is of Fredholm type, with index zero.  Then it must be
shown that there are no homogeneous solutions, at which point one has
existence and uniqueness of a solution to the inhomogeneous system.
Finally, one maps the solution of the integral equations to the unique
solution of the Riemann-Hilbert problem and it remains to verify the
rate of decay to the identity matrix as $\zeta\rightarrow\infty$.

The theory we will use is the theory of matrix Riemann-Hilbert
problems on self-intersecting contours, with boundary values taken in
spaces of H\"older continuous functions.  This theory is summarized in
a self-contained way in the appendix of \cite{manifesto}.

The first step is to establish that the operator of the associated
system of singular integral equations is Fredholm index zero on an
appropriate space of functions.  As described in \cite{manifesto},
this follows from two facts.  First, on each ray of the contour
$\Sigma_0$ the jump matrix ${\bf v}_{\hat{\bf F}}(\zeta):=\hat{\bf
F}_-(\zeta)^{-1}\hat{\bf F}_+(\zeta)$ is uniformly Lipschitz with
respect to $\zeta$, and differs from the identity by a quantity that
is $O(1/\zeta)$ for large $\zeta$.  Second, the limiting values of the
jump matrix, taken as $\zeta\rightarrow 0$ along each ray of
$\Sigma_0$, are consistent with a bounded solution $\hat{\bf
F}(\zeta)$ near $\zeta=0$.  This means the following.  Suppose that
$\hat{\bf F}(\zeta)$ has a limiting value, say a matrix $\hat{\bf
F}_0$, as $\zeta\rightarrow 0$ inside one of the sectors of ${\mathbb
C}\setminus\Sigma_0$.  Using the limiting value of the jump matrix
${\bf v}_{\hat{\bf F}}(\zeta)$ at the origin along one of the rays of
$\Sigma_0$ bounding that sector, one can compute the limiting value of
$\hat{\bf F}(\zeta)$ at the origin in the neighboring sector.  This
procedure can be continued, moving from sector to sector of ${\mathbb
C}\setminus\Sigma_0$ in the same direction, until one arrives once
again in the original sector, with a matrix $\hat{\bf F}_1$.  The
consistency condition is simply that $\hat{\bf F}_1=\hat{\bf F}_0$,
which upon elimination of $\hat{\bf F}_0$ can be viewed as a cyclic
relation among the limiting values of the jump matrix ${\bf
v}_{\hat{\bf F}}(\zeta)$ taken along each ray of $\Sigma_0$ as
$\zeta\rightarrow 0$.  It is easily checked that this cyclic relation
indeed holds for Riemann-Hilbert Problem~\ref{rhp:localorigin}.

The second step is to establish existence and uniqueness of the
solution $\hat{\bf F}(\zeta)$.  The fact that the associated singular
integral equations are Fredholm index zero means that, in a certain
precise sense, the Fredholm alternative applies to our Riemann-Hilbert
problem.  The inhomogeneity is the normalization to the identity
matrix at $\zeta=\infty$.  The corresponding homogeneous
Riemann-Hilbert problem has exactly the same form except that the
normalization condition is replaced by the condition $\hat{\bf
F}(\zeta)\rightarrow {\bf 0}$ as $\zeta\rightarrow\infty$.  We will
have a unique solution of Riemann-Hilbert
Problem~\ref{rhp:localorigin} if it can be shown that no such
homogeneous solutions exist.  For this purpose, it is sufficient that
the jump matrix ${\bf v}_{\hat{\bf F}}(\zeta)$ should have a certain
symmetry with respect to Schwartz reflection through the real axis in
the $\zeta$-plane.  For the orientation of $\Sigma_0$ described above,
the required relation is:
\begin{equation}
{\bf v}_{\hat{\bf F}}(\zeta^*)^{-1} = 
{\bf v}_{\hat{\bf F}}(\zeta)^\dagger
\end{equation}
for $\zeta\in\Sigma_0\cap{\mathbb C}_+$.  It is easily checked, using
the symmetry ${\bf v}_{\hat{\bf F}}(\zeta^*)=\sigma_2{\bf v}_{\hat{\bf
F}}(\zeta)^*\sigma_2$ and structural details of ${\bf v}_{\hat{\bf
F}}(\zeta)$ in the upper half-plane, that this relation holds, and
this means that Riemann-Hilbert Problem~\ref{rhp:localorigin} has a
unique solution $\hat{\bf F}(\zeta)$.

The third step is to establish that the unique solution $\hat{\bf
F}(\zeta)$ decays to the identity for large $\zeta$ like $1/\zeta$.
The H\"older theory that we have been using generally provides a
solution $\hat{\bf F}(\zeta)$ under these circumstances that takes
boundary values on $\Sigma_0$ that are H\"older continuous with
exponent $\mu$ and that differs from the identity matrix by
$O(1/\zeta^\mu)$ as $\zeta\rightarrow \infty$, for all $\mu$ strictly
less than 1.  This fact can be traced to the compact embedding of each
H\"older space into all H\"older spaces with strictly smaller
exponents.  The compactness is needed to establish the Fredholm
property of the Riemann-Hilbert problem.  So to obtain the required
decay, we need an additional argument.  The condition that is required
to obtain the $O(1/\zeta)$ decay is that a signed sum of the mean
values of $\zeta\cdot({\bf v}_{\hat{\bf F}}(\zeta)-{\mathbb I})$ taken
as $\zeta\rightarrow\infty$ along each ray of $\Sigma_0$ (the signs
are related to the orientation of the individual rays) is zero
\cite{manifesto}.  Now along each ray of $\Sigma_0$ except for
$\vec{C_M}$ and its conjugate ({\em i.e.} the imaginary axis in the
$\zeta$-plane), ${\bf v}_{\hat{\bf F}}(\zeta)$ decays to the identity
exponentially fast as $\zeta\rightarrow\infty$.  So these rays to do
not contribute to the sum and it is only necessary to check the
imaginary axis.  When $\zeta\in \vec{C_M}$,
\begin{equation}
\zeta\cdot({\bf v}_{\hat{\bf F}}(\zeta)-{\mathbb I})=
\frac{i}{12}\left[\begin{array}{cc} -1 & ie^{-\pi\zeta}\\
ie^{\pi\zeta} & 1\end{array}\right]+O\left(\frac{1}{\zeta}\right)\,,
\end{equation}
as $\zeta\rightarrow\infty$, and for $\zeta$ on the negative imaginary
axis oriented upwards,
\begin{equation}
\zeta\cdot({\bf v}_{\hat{\bf F}}(\zeta)-{\mathbb I})=
\frac{i}{12}\left[\begin{array}{cc} -1 & -ie^{\pi\zeta}\\
-ie^{-\pi\zeta} & 1\end{array}\right]+O\left(\frac{1}{\zeta}\right)\,,
\end{equation}
as $\zeta\rightarrow\infty$.  The limits of these quantities do not
exist as $\zeta\rightarrow\infty$ due to the oscillations on the
off-diagonal.  But the mean values exist and are equal, and it turns
out that they enter the sum with opposite signs due to the orientation
of the contour rays.  Thus, the required sum of signed mean values
indeed vanishes.  This, along with the analyticity of the jump matrix
${\bf v}_{\hat{\bf F}}(\zeta)$ along each ray of $\Sigma_0$
establishes that $\hat{\bf F}(\zeta)-{\mathbb I}=O(1/\zeta)$ as
$\zeta\rightarrow\infty$.  

Finally, we check that $\det(\hat{\bf F}(\zeta))=1$.  Taking
determinants in the jump relations we see that on all rays of the
contour, $\det(\hat{\bf F}_+(\zeta))=\det(\hat{\bf F}_-(\zeta))$.
Since the boundary values taken by $\hat{\bf F}(\zeta)$ on $\Sigma_0$
are continuous, we discover that $\det(\hat{\bf F}(\zeta))$ is an
entire function.  Since this function tends to one at infinity, it
follows from Liouville's theorem that $\det(\hat{\bf F}(\zeta))\equiv
1$.  This completes the proof of the proposition.  $\Box$

\subsubsection{The local model for ${\bf N}(\lambda)$ near $\lambda=0$.}
From (\ref{eq:Fdefleft}) and (\ref{eq:Fdefright}) we can express ${\bf
N}(\lambda)$ in terms of ${\bf F}(\lambda)$ for $|\lambda|\le\hbar_N^\alpha$.
For $\Re(\lambda)<0$ we have
\begin{equation}
{\bf N}(\lambda)=\tilde{\bf O}(\lambda)(i\sigma_1)\sigma_1^{\frac{1-J}{2}}
e^{i\theta(0)\sigma_3/(2\hbar_N)}{\bf F}(\lambda)
e^{-i\theta(0)\sigma_3/(2\hbar_N)}\sigma_1^{\frac{1-J}{2}}(-i\sigma_1)
\tilde{\bf O}(\lambda)^{-1}\hat{\bf N}_{\rm out}(\lambda)
\end{equation}
and for $\Re(\lambda)>0$ we have
\begin{equation}
{\bf N}(\lambda)=\tilde{\bf O}(\lambda)\sigma_1^{\frac{1-J}{2}}
e^{i\theta(0)\sigma_3/(2\hbar_N)}{\bf F}(\lambda)
e^{-i\theta(0)\sigma_3/(2\hbar_N)}\sigma_1^{\frac{1-J}{2}}
\tilde{\bf O}(\lambda)^{-1}\hat{\bf N}_{\rm out}(\lambda)\,.
\end{equation}
To obtain a local model for ${\bf N}(\lambda)$ near the origin, we
simply replace ${\bf F}(\lambda)$ in these formulae by the
approximation $\hat{\bf F}(\varphi_0(\lambda))$.  With
$|\lambda|\le\hbar_N^\alpha$, we set for $\Re(\lambda)<0$,
\begin{equation}
\hat{\bf N}_{\rm origin}(\lambda):=
\tilde{\bf O}(\lambda)(i\sigma_1)\sigma_1^{\frac{1-J}{2}}
e^{i\theta(0)\sigma_3/(2\hbar_N)}\hat{\bf F}(\varphi_0(\lambda))
e^{-i\theta(0)\sigma_3/(2\hbar_N)}\sigma_1^{\frac{1-J}{2}}(-i\sigma_1)
\tilde{\bf O}(\lambda)^{-1}\hat{\bf N}_{\rm out}(\lambda)
\end{equation}
and for $\Re(\lambda)>0$ we set
\begin{equation}
\hat{\bf N}_{\rm origin}(\lambda):=
\tilde{\bf O}(\lambda)\sigma_1^{\frac{1-J}{2}}
e^{i\theta(0)\sigma_3/(2\hbar_N)}\hat{\bf F}(\varphi_0(\lambda))
e^{-i\theta(0)\sigma_3/(2\hbar_N)}\sigma_1^{\frac{1-J}{2}}
\tilde{\bf O}(\lambda)^{-1}\hat{\bf N}_{\rm out}(\lambda)\,.
\end{equation}
The most important properties of this matrix function are easily seen
to be the following.
\begin{proposition}
The matrix $\hat{\bf N}_{\rm origin}(\lambda)$ is a piecewise analytic
function of $\lambda$ in the disk $|\lambda|<\hbar_N^\alpha$, with
jumps only on the locally straight-line contours $C_L$, $C_R$, $L_L$, 
$L_R$, and $C_M$, and their conjugates in the lower half-disk.  The jump
relations satisfied by $\hat{\bf N}_{\rm origin}(\lambda)$ on these
contours are the following:
\begin{equation}
\hat{\bf N}_{{\rm origin},-}(\lambda)^{-1}\hat{\bf N}_{{\rm origin},+}(\lambda)
= 
\sigma_1^{\frac{1-J}{2}}
\left[\begin{array}{cc}
1 & 0 \\\\
\displaystyle ie^{i\theta(0)/\hbar_N}e^{(2i+\pi)\zeta}(-i\zeta)^{-i\zeta}
(i\zeta)^{-i\zeta}\frac{\Gamma(1/2+i\zeta)}{\Gamma(1/2-i\zeta)} & 1
\end{array}\right]
\sigma_1^{\frac{1-J}{2}}\,,\hspace{0.2 in}\lambda\in C_L\,,
\end{equation}
\begin{equation}
\hat{\bf N}_{{\rm origin},-}(\lambda)^{-1}\hat{\bf N}_{{\rm origin},+}(\lambda)
= 
\sigma_1^{\frac{1-J}{2}}
\left[\begin{array}{cc}
1 & 0 \\\\
\displaystyle ie^{-i\theta(0)/\hbar_N}e^{(2i-\pi)\zeta}(-i\zeta)^{-i\zeta}
(i\zeta)^{-i\zeta}\frac{\Gamma(1/2+i\zeta)}{\Gamma(1/2-i\zeta)} & 1
\end{array}\right]
\sigma_1^{\frac{1-J}{2}}\,,\hspace{0.2 in}\lambda\in C_R\,,
\end{equation}
\begin{equation}
\hat{\bf N}_{{\rm origin},-}(\lambda)^{-1}\hat{\bf N}_{{\rm origin},+}(\lambda)
= 
\sigma_1^{\frac{1-J}{2}}
\left[\begin{array}{cc}
1 & i(e^{-i\theta(\lambda)/\hbar_N}-e^{\pi\zeta}e^{-i\theta(0)/\hbar_N})\\\\
0 & 1
\end{array}\right]\sigma_1^{\frac{1-J}{2}}\,,
\hspace{0.2 in}
\lambda\in L_L\,,
\end{equation}
\begin{equation}
\hat{\bf N}_{{\rm origin},-}(\lambda)^{-1}\hat{\bf N}_{{\rm origin},+}(\lambda)
= 
\sigma_1^{\frac{1-J}{2}}
\left[\begin{array}{cc}
1 & i(e^{i\theta(\lambda)/\hbar_N}-e^{-\pi\zeta}e^{i\theta(0)/\hbar_N})\\\\
0 & 1
\end{array}\right]\sigma_1^{\frac{1-J}{2}}\,,
\hspace{0.2 in}
\lambda\in L_R\,,
\end{equation}
\begin{equation}
\hat{\bf N}_{{\rm origin},-}(\lambda)^{-1}\hat{\bf N}_{{\rm origin},+}(\lambda)
= 
\sigma_1^{\frac{1-J}{2}}
{\bf v}(\lambda)
\sigma_1^{\frac{1-J}{2}}\,,\hspace{0.2 in}
\lambda\in C_M\,,
\end{equation}
where
\begin{equation}
\begin{array}{rcl}
v_{11}&:=&e^{i\theta(\lambda)/\hbar_N}
(1+(1-e^{-\pi\zeta-i(\theta(\lambda)-\theta(0))/\hbar_N})Z)\,,  \\\\
v_{12}&:=&iZ (e^{\pi\zeta+i(\theta(\lambda)-\theta(0))/\hbar_N}+
e^{-\pi\zeta-i(\theta(\lambda)-\theta(0))/\hbar_N}-2)\,,\\\\
v_{21}&:=&i+iZ\,,  \\\\
v_{22}&:=&e^{-i\theta(\lambda)/\hbar_N}(1+(1-e^{\pi\zeta+i(\theta(\lambda)-\theta(0))/\hbar_N})Z)\,,
\end{array}
\end{equation}
with
\begin{equation}
Z:=\frac{2\pi e^{2i\zeta}(-i\zeta)^{-2i\zeta}}{\Gamma(1/2-i\zeta)^2}-1\,,
\label{eq:Delta}
\end{equation}
and where $\zeta=\varphi_0(\lambda)$.  The jumps on the corresponding
contours in the lower half-plane are obtained from the symmetry
$\hat{\bf N}_{\rm origin}(\lambda^*)=\sigma_2\hat{\bf N}_{\rm
origin}(\lambda)^*\sigma_2$.  The matrix $\hat{\bf N}_{\rm origin}(\lambda)$
is uniformly bounded for $|\lambda|<\hbar_N^\alpha$, with a bound that is
independent of $\hbar_N$.  Also, $\det(\hat{\bf N}_{\rm origin}(\lambda))
\equiv 1$ and when $|\lambda|=\hbar_N^\alpha$,
\begin{equation}
\hat{\bf N}_{\rm origin}(\lambda)\hat{\bf N}_{\rm out}(\lambda)^{-1}=
{\mathbb I}+O(\hbar_N^{1-\alpha})\,.
\end{equation}
\label{prop:originmodelproperties}
\end{proposition}

\subsection{Local analysis near $\lambda=iA$.}
\label{sec:local_iA}
\subsubsection{Local behavior of $a_L(\lambda)$, $a_R(\lambda)$, and
$a_M(\lambda)$.}  As before, we suppose that $\epsilon$ and $\delta$
are small scales satisfying $\hbar_N\ll\delta\ll\epsilon\ll 1$ as
$\hbar_N$ tends to zero.  We redefine the integer $L$ so that exactly
the first $L$ of the numbers $\lambda_0,\dots,\lambda_{N-1}$ lie on
the positive imaginary axis above $i(A-\epsilon)$.  We will suppose
that $\Im(\lambda)\le A$, and $|\lambda-iA|\le\delta$ and we will
deduce asymptotic formulae for $T(\lambda)$ given by (\ref{eq:T})
valid for such $\lambda$, and for $S(\lambda)$ given by (\ref{eq:S})
when $\lambda$ is also bounded outside of some downward-opening sector
with vertex at $iA$, in the semiclassical limit $\hbar_N\rightarrow
0$. First, we establish a result that is the analogue of
Lemma~\ref{lemma:originpeelfirst}.
\begin{lemma}
\label{lemma:iApeelfirst}
When $\Im(\lambda)\le A$ and $|\lambda-iA|\le\delta$ and with $L$ defined
as above,
\begin{equation}
\exp\,\left(-\sum_{k=L}^{N-1}\tilde{I}_k(\lambda)\right)=1+O\left(\frac{\hbar_N}{\epsilon}\right)\,.
\end{equation}
\end{lemma}

{\em Proof:}  We again estimate $\tilde{I}_k(\lambda)$ using the integral 
formula (\ref{eq:integralformulag}) with integrand $g(\lambda,\xi)$ given
by (\ref{eq:glambdaxi}).  Given our conditions on $\lambda$, for
$m_k-\hbar_N/2\le\xi\le m_k+\hbar_N/2$ and $k\ge L$ we have
\begin{equation}
\frac{1}{|\lambda+e(\xi)|}\le\frac{1}{|\lambda-e(\xi)|}
\le\frac{1}{|i(A-\delta)-e(\xi)|}\le
\frac{1}{|i(A-\delta)-e(m_k+\hbar_N/2)|}=
O\left(\frac{1}{|m(iA-i\delta)-m_k-\hbar_N/2|}\right)\,.
\end{equation}
For all such $\xi$ we therefore have the estimate
\begin{equation}
g(\lambda,\xi)=O\left(\frac{1}{|m(iA-i\delta)-m_k-\hbar_N/2|}\right)\,.
\end{equation}
Summing over $k$ gives
\begin{equation}
\sum_{k=L}^{N-1}\tilde{I}_k(\lambda)=O\left(\hbar_N^2\sum_{k=L}^{N-1}
\frac{1}{|m(iA-i\delta)-m_k-\hbar_N/2|^2}\right) =
O\left(\hbar_N\int_0^{m(iA-\epsilon)}\frac{dm}{(m(iA-i\delta)-m)^2}
\right)
\end{equation}
which is $O(\hbar_N/\epsilon)$ because $\delta\ll\epsilon$, and the
lemma is proved.  $\Box$

\vspace{0.1 in}

As was the case when $\lambda$ was near the origin, only certain
terms are important when $\lambda-iA$ is small, as a direct
consequence of Lemma~\ref{lemma:iApeelfirst} and the fact that
$\exp(-\tilde{I}_k(\lambda))$ is analytic for such $\lambda$
when $k\ge L$.  The important terms when $|\lambda-iA|\le\delta$
are
\begin{equation}
S_1^{(iA)}(\lambda):=\left(\prod_{k=0}^{L-1}\frac{\lambda-\lambda_k^*}
{\lambda-\lambda_k}\right)\exp\,\left(\frac{1}{\hbar_N}
\int_{m_L+\hbar_N/2}^M\left(L_{e(m)}^0(\lambda)-
L_{-e(m)}^0(\lambda)\right)\,dm
\right)\,,
\end{equation}
and
\begin{equation}
T_1^{(iA)}(\lambda):=
\left(\prod_{k=0}^{L-1}\frac{\lambda-\lambda_k^*}
{\lambda-\lambda_k}\right)\exp\,\left(\frac{1}{\hbar_N}
\int_{m_L+\hbar_N/2}^M\left(\overline{L}_{e(m)}^0(\lambda)-
\overline{L}_{-e(m)}^0(\lambda)\right)
dm
\right)
\cdot 2\cos\left(\frac{\theta^0(\lambda)}{\hbar_N}\right)\,.
\end{equation}
Lemma~\ref{lemma:iApeelfirst} states that
$S(\lambda)=S^{(iA)}_1(\lambda)(1+O(\hbar_N/\epsilon))$ and
$T(\lambda)=T^{(iA)}_1(\lambda)(1+O(\hbar_N/\epsilon))$ as
$\hbar_N\rightarrow 0$ for $|\lambda-iA|\le\delta$.

The analogue of Lemma~\ref{lemma:originpeelsecond} says that for
$\lambda-iA$ small the sequence of numbers $\lambda_0,\dots,\lambda_{L-1}$
contributing to $S(\lambda)$ and $T(\lambda)$ can be replaced essentially
by a ``straightened-out'' sequence with uniform density.
\begin{lemma}
\label{lemma:iApeelsecond}
Let $\tilde{\lambda}_k$ for $k=0,1,2,\dots$ be the sequence of
numbers defined by the relation:
\begin{equation}
\tilde{\lambda}_k:=iA + \frac{\hbar_N}{\rho^0(iA)}(k+1/2)\,,
\end{equation}
which results from expanding the Bohr-Sommerfeld relation (\ref{eq:BS})
for $\lambda_k$ near $iA$ and keeping only the dominant terms.  Define
\begin{equation}
\begin{array}{rcl}
\displaystyle
S_2^{(iA)}(\lambda)&:=&\displaystyle
\left(\prod_{k=0}^{L-1}\frac{\lambda-\tilde{\lambda}_k^*}
{\lambda-\tilde{\lambda}_k}\right)\\\\
&&\,\,\,\displaystyle\times\,\,\,
\exp\,\left(\frac{1}{\hbar_N}
\int_{m_L+\hbar_N/2}^M\left(L_{iA+e'(M)(m-M)}^0(\lambda)-
L_{-iA-e'(M)(m-M)}^0(\lambda)\right)\,dm
\right)\,,
\end{array}
\end{equation}
and
\begin{equation}
\begin{array}{rcl}
\displaystyle
T_2^{(iA)}(\lambda)&:=&\displaystyle
\left(\prod_{k=0}^{L-1}\frac{\lambda-\tilde{\lambda}_k^*}
{\lambda-\tilde{\lambda}_k}\right)\\\\
&&\,\,\,\displaystyle\times\,\,\,
\exp\,\left(\frac{1}{\hbar_N}
\int_{m_L+\hbar_N/2}^M\left(\overline{L}_{iA+e'(M)(m-M)}^0(\lambda)-
\overline{L}_{-iA-e'(M)(m-M)}^0(\lambda)\right)\,dm
\right)\\\\
&&\displaystyle\,\,\,\times\,\,\, 2\cos\left(\frac{\pi\rho^0(iA)}{\hbar_N}
(iA-\lambda)\right)\,.
\end{array}
\end{equation}
Then, for $\Im(\lambda)\le A$ and $|\lambda-iA|\le\delta$,
\begin{equation}
T_1^{(iA)}(\lambda)=T_2^{(iA)}(\lambda)\left(1+O\left(\frac{\epsilon^2}
{\hbar_N}\log\left(\frac{\epsilon}{\hbar_N}\right)\right)\right)
\end{equation}
where we suppose that the scale $\epsilon$ is further constrained so
that the relative error is asymptotically small.  If $\lambda$ is additionally
bounded outside of some downward opening sector with vertex at $iA$, then
\begin{equation}
S_1^{(iA)}(\lambda)=S_2^{(iA)}(\lambda)\left(1+O\left(\frac{\epsilon^2}{\hbar_N}\right)\right)\,.
\end{equation}
\end{lemma}

{\em Proof:} The proof of this Lemma follows that of
Lemma~\ref{lemma:originpeelsecond} almost exactly and will not be
repeated here.  The only difference is that the square in that proof should
be replaced here by the rectangle whose top side is $\Im(\lambda)=A$ and
$-\epsilon\le\Re(\lambda)\le\epsilon$ and whose bottom is $\Im(\lambda)=
-i(\tilde{\lambda}_{L-1}+\tilde{\lambda}_L)/2$.  $\Box$

\vspace{0.1 in}

Without any approximation, $S_2^{(iA)}(\lambda)$ and $T_2^{(iA)}(\lambda)$
can be rewritten in a more transparent form by expressing the products in
terms of gamma functions and evaluating the logarithmic integrals exactly.
Introduce a local variable $\zeta$ in terms of a transformation
$\varphi_{iA}$ given by the relation
\begin{equation}
\zeta = \varphi_{iA}(\lambda):=\rho^0(iA)\frac{\lambda-iA}{i\hbar_N}\,,
\label{eq:varphiiA} 
\end{equation}
and let $B$ be the positive constant
\begin{equation}
B:=-\frac{2iA\rho^0(iA)}{\hbar_N}\,.
\end{equation}
In terms of these quantities, one finds that $S_2^{(iA)}(\lambda)$ and
$T_2^{(iA)}(\lambda)$ take a simple form:
\begin{equation}
S_2^{(iA)}(\lambda)=(-i\zeta)^{i\zeta}\Gamma(1/2-i\zeta) \cdot
V(\zeta,B,L) \cdot W(\zeta,B,L)\,,
\label{eq:S2iArewrite}
\end{equation}
and
\begin{equation}
T_2^{(iA)}(\lambda)=\frac{2\pi(i\zeta)^{i\zeta}}{\Gamma(1/2+i\zeta)} 
\cdot V(\zeta,B,L)
\cdot W(\zeta,B,L)\,,
\label{eq:T2iArewrite}
\end{equation}
where
\begin{equation}
V(\zeta,B,L):=\frac{\Gamma(B-i\zeta+1/2)}{\Gamma(B-L-i\zeta+1/2)
\Gamma(L-i\zeta+1/2)}
\end{equation}
and
\begin{equation}
W(\zeta,B,L):=
\frac{(B-L-i\zeta)^{B-L-i\zeta}(L-i\zeta)^{L-i\zeta}}{(B-i\zeta)^{B-i\zeta}}\,.
\end{equation}
Now, $B\gg L\gg |\zeta|$ because $B$ is proportional to $\hbar_N^{-1}$ and
$L$ is of the order of $\epsilon/\hbar$ while $|\zeta|=O(\delta/\hbar_N)$.
So again we can use Stirling's formula to extract the dominant asymptotic
contributions to $S_2^{(iA)}(\lambda)$ and $T_2^{(iA)}(\lambda)$ as $\hbar_N$
tends to zero.  
\begin{lemma}
\label{lemma:iApeelthird}
As $\hbar_N$ tends to zero through positive values, 
\begin{equation}
S_2^{(iA)}(\lambda)=\frac{1}{\sqrt{2\pi}}e^{-i\zeta}(-i\zeta)^{-i\zeta}
\Gamma(1/2-i\zeta)\cdot\left(1+O\left(\frac{\delta^2}{\epsilon\hbar_N}\right)
\right)\,,
\end{equation}
and
\begin{equation}
T_2^{(iA)}(\lambda)=\frac{\sqrt{2\pi}e^{-i\zeta}(i\zeta)^{i\zeta}}
{\Gamma(1/2+i\zeta)}
\cdot\left(1+O\left(\frac{\delta^2}{\epsilon\hbar_N}\right)\right)\,.
\end{equation}
\end{lemma}

{\em Proof:}  Using Stirling's formula, one expands $V(\zeta,B,L)$ to
find
\begin{equation}
V(\zeta,B,L)=\frac{e^{1/2-i\zeta}}{\sqrt{2\pi}}\frac{(B-i\zeta+1/2)^{B-i\zeta}}
{(B-L-i\zeta+1/2)^{B-L-i\zeta}(L-i\zeta+1/2)^{L-i\zeta}}
\left(1+O\left(\frac{\hbar_N}{\epsilon}\right)\right)\,.
\end{equation}
The error here is dominated by the fact $L\sim\epsilon/\hbar_N$ is the
smallest large number involved.  Now we expand the powers that remain
in conjunction with those in $W(\zeta,B,L)$ to find
\begin{equation}
V(\zeta,B,L)W(\zeta,B,L)=\frac{e^{-i\zeta}}{\sqrt{2\pi}}
\left(1+O\left(\frac{\hbar_N}{\epsilon}\right)\right)
\left(1+O\left(\frac{\delta^2}{\epsilon\hbar_N}\right)\right)\,.
\end{equation}
Since $\delta\ll\hbar_N$, the relative error is dominated by
$O(\delta^2/\epsilon\hbar_N)$.  Using this expression for the product $VW$
in (\ref{eq:S2iArewrite}) and (\ref{eq:T2iArewrite}), the lemma is proved.
$\Box$.

\vspace{0.1 in}

In exactly the same way as in our study of the local behavior near the
origin, we may combine Lemmas~\ref{lemma:iApeelfirst},
\ref{lemma:iApeelsecond}, and \ref{lemma:iApeelthird} and choose the
``internal'' scale $\epsilon$ in terms of $\delta$ and $\hbar_N$ to
obtain asymptotics for $S(\lambda)$ and $T(\lambda)$ with optimized
relative error.  

\begin{proposition}
Let $\Im(\lambda)\le A$, with $|\lambda-iA|\le\hbar_N^\alpha$, where
$3/4<\alpha<1$, and let $\lambda$ be bounded outside some fixed symmetrical
sector with vertex at $iA$ and opening downward.  Then
\begin{equation}
S(\lambda)=\frac{1}{\sqrt{2\pi}}e^{-i\zeta}(-i\zeta)^{-i\zeta}\Gamma(1/2-\zeta)
\left(1+O\left(\hbar_N^{4\alpha/3-1}\right)\right)\,,
\end{equation}
where $\zeta=\varphi_{iA}(\lambda)$.
\label{prop:iAasympS}
\end{proposition}

\vspace{0.1 in}

\begin{proposition}
Let $\Im(\lambda)\le A$, with $|\lambda-iA|\le\hbar_N^\alpha$, where
$3/4<\alpha<1$.  Then for all $\nu>0$, however small,
\begin{equation}
T(\lambda)=\frac{\sqrt{2\pi}e^{-i\zeta}(i\zeta)^{i\zeta}}{\Gamma(1/2+i\zeta)}
\left(1+O\left(\hbar_N^{4\alpha/3-1-\nu}\right)\right)\,,
\end{equation}
where $\zeta=\varphi_{iA}(\lambda)$.
\label{prop:iAasympT}
\end{proposition}

\subsubsection{Why a local model near $\lambda=iA$ is not necessary.}  
In particular it follows from these considerations that both
$S(\lambda)$ and $T(\lambda)$ are uniformly bounded functions on their
respective contours in any fixed neighborhood $U$ of $\lambda=iA$.  We
claim that the quotient of the jump matrices for ${\bf N}(\lambda)$
and $\hat{\bf N}_{\rm out}(\lambda)$ is uniformly close to the
identity matrix in $U$ as $\hbar_N\rightarrow 0$.  Since $\hat{\bf
N}_{\rm out}(\lambda)$ is, by definition, analytic throughout $U$, it
suffices to show that the jump matrix ${\bf v}_{\bf N}(\lambda):={\bf
N}_-(\lambda)^{-1}{\bf N}_+(\lambda)$ is uniformly close to the
identity in $U$.  This will be the case if $a_L(\lambda)$,
$a_R(\lambda)$, and $a_M(\lambda)$ are uniformly small on their
respective contours.  Now for $\lambda\in C_M$,
\begin{equation}
a_M(\lambda)=i\exp\,\left(\frac{\tilde{\phi}(\lambda)}{\hbar_N}\right)
T(\lambda)\,,
\end{equation}
so, with $\tilde{\phi}(\lambda)$ being real and strictly negative for
$\lambda\in C_M\cap U$ according to Proposition~\ref{prop:tildephineg}
and $T(\lambda)$ being bounded, we see that $a_M(\lambda)$ is in fact
exponentially small as $\hbar_N$ tends to zero through positive
values.  Similarly, for $\lambda\in C_L$, 
\begin{equation}
a_L(\lambda)=i\exp\,\left(\frac{\tau(\lambda)-i\theta^0(\lambda)}{\hbar_N}
\right)S(\lambda)=i\exp\,\left(\frac{\tilde{\phi}(\lambda)-2i\theta^0(\lambda)}{\hbar_N}\right) S(\lambda)\,,
\end{equation}
where we have used (\ref{eq:tauleft}) and the fact that
$\theta(\lambda)\equiv 0$ on $C_M$ above $\lambda = iA(x)$.  Since
$\theta^0(iA)=0$, it is possible to choose the neighborhood $U$ small
enough (independent of $\hbar_N$) so that
$\Re(\tilde{\phi}(\lambda)-2i\theta^0(\lambda))<0$ throughout $U$.
Since $S(\lambda)$ is bounded, it then follows that for $\lambda\in
C_L\cap U$, $a_L(\lambda)$ is exponentially small as $\hbar_N$ tends
to zero through positive values.  Virtually the same argument using
(\ref{eq:tauright}) in place of (\ref{eq:tauleft}) shows that
$a_R(\lambda)$ is also exponentially small for $\lambda\in C_R\cap U$
(it may be necessary to make $U$ slightly smaller to have
$\Re(\tilde{\phi}(\lambda)+2i\theta^0(\lambda))<0$ throughout $U$).
It follows that ${\bf v}_{\bf N}(\lambda)-{\mathbb I}$ is
exponentially small uniformly for the contours within $U$.

For this reason we expect that the outer model $\hat{\bf N}_{\rm
out}(\lambda)$ will be a good approximation to ${\bf N}(\lambda)$ near
$\lambda=iA$ even though $S(\lambda)-1$ and $T(\lambda)-1$ are not
small.  We do not need to construct a special-purpose local model for
${\bf N}(\lambda)$ in this case.

\subsection{Local analysis near $\lambda=iA(x)$.}  
\label{sec:local_iA(x)}
Let $D$ be a circular disk centered at $\lambda=iA$ of sufficiently
small radius (independent of $\hbar_N$) that $0\not\in D$ and
$iA\not\in D$, and that $L_L$ and $L_R$ each have exactly one
intersection with $\partial D$ (of course $C_M$ will have two
intersection points with $\partial D$).  This situation is possible as
long as $x\neq 0$.  The case of $x=0$ is a degenerate case that we
will not treat in detail here.

Since $x\neq 0$ and therefore $A(x)<A$, it follows from the definition
(\ref{eq:rhodefine}) of $\rho(\eta)$ that as $\lambda$ tends to
$iA(x)$ along $I$, $\theta(\lambda)$ vanishes like
$(\lambda-iA(x))^{3/2}$, and not to higher order.  Since
$\theta(\lambda)$ may be extended from $I$ to be an analytic function
in $D$ except for a branch cut along the part of $C_M$ in $D$ lying
above the center, and since $\rho(\lambda)$ extended to this cut
domain from $I$ is nonzero, it is easy to see that the function
\begin{equation}
\varphi_{iA(x)}(\lambda):= \left(\frac{\theta(\lambda)}{\hbar_N}\right)^{2/3}
\label{eq:varphiiAx}
\end{equation}
defines an invertible conformal mapping of all of $D$ to its image.
Consider the local variable $\zeta$ defined by the relation
$\zeta=\varphi_{iA(x)}(\lambda)$.  The image of $D$ in the
$\zeta$-plane is a neighborhood of $\zeta=0$ that scales with
$\hbar_N$ such that it contains the disk centered at $\zeta=0$ with
radius $C\hbar_N^{-2/3}$ for some constant $C>0$.  The transformation
(\ref{eq:varphiiAx}) maps $I\cap D$ to a ray segment of the positive
real $\zeta$-axis, and takes the portion of $C_M$ in $D$ lying above
$\lambda=iA(x)$ to a ray segment of the negative real $\zeta$-axis.
We suppose that the contours $L_L$ and $L_R$ have been chosen so that
$\varphi_{iA(x)}(L_L\cap D)$ and $\varphi_{iA(x)}(L_R\cap D)$ are
straight ray segments with angles $-\pi/3$ and $\pi/3$ respectively.

For $\zeta\in \varphi_{iA(x)}(D)$, the matrix ${\bf
S}(\zeta):=\sigma_1^{\frac{1-J}{2}} {\bf
O}(\varphi_{iA(x)}^{-1}(\zeta))\sigma_1^{\frac{1-J}{2}}$ satisfies the
following jump relations:
\begin{equation}
{\bf S}_+(\zeta)={\bf S}_-(\zeta)
\left[\begin{array}{cc}0 & i\\\\i&0\end{array}\right]
\end{equation}
for $\zeta\in {\mathbb R}_+\cap \varphi_{iA(x)}(D)$, oriented from
right to left,
\begin{equation}
{\bf S}_+(\zeta)={\bf S}_-(\zeta)
\left[\begin{array}{cc}1 & -ie^{-i\zeta^{3/2}}\\\\0&1\end{array}\right]
\end{equation}
on the part of the ray $\arg(\zeta)=-\pi/3$ in $\varphi_{iA(x)}(D)$,
oriented toward the origin,
\begin{equation}
{\bf S}_+(\zeta)={\bf S}_-(\zeta)
\left[\begin{array}{cc}1 & -ie^{i\zeta^{3/2}}\\\\0&1\end{array}\right]
\end{equation}
on the part of the ray $\arg(\zeta)=\pi/3$ in $\varphi_{iA(x)}(D)$,
oriented toward the origin, and finally
\begin{equation}
{\bf S}_+(\zeta)={\bf S}_-(\zeta)
\left[\begin{array}{cc}1 & 0\\\\ie^{-(-\zeta)^{3/2}} & 1\end{array}\right]
\end{equation}
for $\zeta\in {\mathbb R}_-\cap \varphi_{iA(x)}(D)$, oriented from
right to left.  The jump relation on the negative real $\zeta$-axis
follows from the formula (\ref{eq:ntildemiddle}) which applies because
${\bf O}(\lambda)=\tilde{\bf N}(\lambda)$ here.  In
(\ref{eq:ntildemiddle}) one uses the fact that $\theta(\lambda)\equiv
0$ for $\lambda\in C_M$ above $iA(x)$, and the relation
(\ref{eq:phifromtheta}) giving $\tilde{\phi}(\lambda)$ above $iA(x)$
in terms of the analytic continuation of $\theta(\lambda)$ from $I$,
which one writes in terms of the local coordinate $\zeta$.

As $\zeta\rightarrow\infty$ on all of the rays except for ${\mathbb
R}_+$, the jump matrix for ${\bf S}(\zeta)$ decays exponentially to
the identity matrix.  These jump conditions were precisely the ones
that were neglected in obtaining the outer model.  That is, the
matrix $\tilde{\bf S}(\zeta):=\sigma_1^{\frac{1-J}{2}}\tilde{\bf
O}(\varphi_{iA(x)}^{-1}(\zeta))\sigma_1^{\frac{1-J}{2}}$ defined for
$\zeta\in\varphi_{iA(x)}(D)$ is analytic except on the positive real
$\zeta$-axis, where it satisfies
\begin{equation}
\tilde{\bf S}_+(\zeta)=\tilde{\bf S}_-(\zeta)\left[\begin{array}{cc}
0 & i\\\\i & 0\end{array}\right]\,.
\end{equation}
It follows that the matrix $\tilde{\bf S}(\zeta)$ can be decomposed
into a product of a holomorphic prefactor depending on $\hbar_N$ and a
universal ({\em i.e.} independent of $\hbar_N$) local factor that
takes care of the jump.  We therefore may write
\begin{equation}
\tilde{\bf S}(\zeta)=\tilde{\bf S}^{\rm hol}(\zeta)
\tilde{\bf S}^{\rm loc}(\zeta)
\end{equation}
where $\tilde{\bf S}^{\rm hol}(\zeta)$ is holomorphic in
$\varphi_{iA(x)}(D)$ and where
\begin{equation}
\tilde{\bf S}^{\rm loc}(\zeta):= \frac{1}{\sqrt{2}}
(-\zeta)^{\sigma_3/4}\left[\begin{array}{cc}
1 & 1 \\\\ -1& 1\end{array}\right]=\frac{1}{\sqrt{2}}
\left[\begin{array}{cc} (-\zeta)^{1/4} & (-\zeta)^{1/4}\\\\
-(-\zeta)^{-1/4} & (-\zeta)^{-1/4}\end{array}\right]\,.
\end{equation}
Note that $\tilde{\bf S}^{\rm hol}(\zeta)$ has determinant one.  Its
matrix elements are of size $O(\hbar_N^{-1/6})$ for $\zeta\in
\varphi_{iA(x)}(D)$.  It is easy to write down an explicit formula
for $\tilde{\bf S}^{\rm hol}(\zeta)$ because both $\tilde{\bf S}(\zeta)$
and $\tilde{\bf S}^{\rm loc}(\zeta)$ are known.

\begin{figure}[h]
\begin{center}
\mbox{\psfig{file=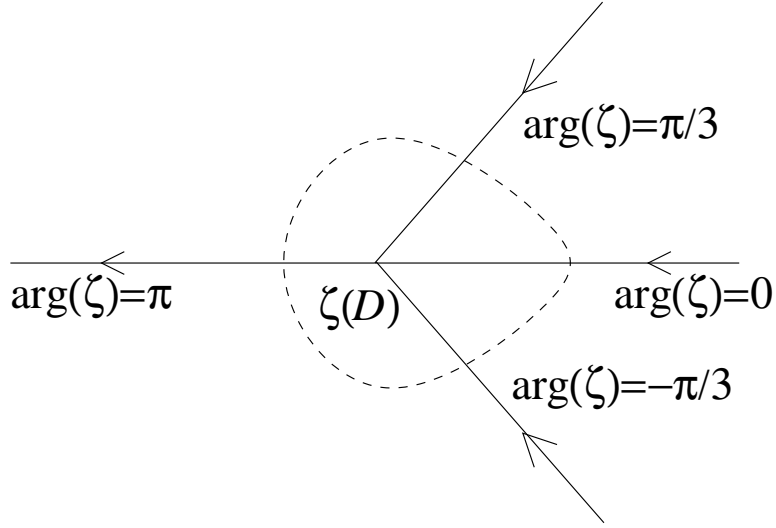,width=4 in}}
\end{center}
\caption{\em The contour $\Sigma_{iA(x)}$ in the $\zeta$-plane for 
Riemann-Hilbert Problem~\ref{rhp:Airy}.  The boundary of the image
$\varphi_{iA(x)}(D)$, expanding as $\hbar_N\rightarrow 0$, is shown as
a dashed curve.}
\label{fig:Airycontour}
\end{figure}

We will now approximate ${\bf S}(\zeta)$ by 
\begin{equation}
\hat{\bf S}(\zeta):=\tilde{\bf S}^{\rm hol}(\zeta){\bf S}^{\rm loc}(\zeta)
\label{eq:Shatdef}
\end{equation}
where ${\bf S}^{\rm loc}(\zeta)$ is the solution of the following
Riemann-Hilbert problem.  Let $\Sigma_{iA(x)}$ be the contour shown in
Figure~\ref{fig:Airycontour}.
\begin{rhp}[Local Model for $iA(x)$]  Find a matrix ${\bf S}^{\rm loc}(\zeta)$
with the following properties:
\begin{enumerate}
\item
{\bf Analyticity:}  ${\bf S}^{\rm loc}(\zeta)$ is analytic for 
$\zeta\in{\mathbb C}\setminus\Sigma_{iA(x)}$.
\item
{\bf Boundary behavior:}  ${\bf S}^{\rm loc}(\zeta)$ assumes continuous
boundary values on $\Sigma_{iA(x)}$.
\item
{\bf Jump conditions:}  The boundary values taken on $\Sigma_{iA(x)}$
satisfy
\begin{equation}
{\bf S}^{\rm loc}_+(\zeta)={\bf S}^{\rm loc}_-(\zeta)
\left[\begin{array}{cc}0 & i\\\\i&0\end{array}\right]\,,\hspace{0.2 in}
\mbox{for}\hspace{0.2 in}\arg(\zeta)=0\,,
\end{equation}
\begin{equation}
{\bf S}^{\rm loc}_+(\zeta)={\bf S}^{\rm loc}_-(\zeta)
\left[\begin{array}{cc}1 & -ie^{-i\zeta^{3/2}}\\\\0&1\end{array}\right]\,,
\hspace{0.2 in}\mbox{for}\hspace{0.2 in}
\arg(\zeta)=-\pi/3\,,
\end{equation}
\begin{equation}
{\bf S}^{\rm loc}_+(\zeta)={\bf S}^{\rm loc}_-(\zeta)
\left[\begin{array}{cc}1 & -ie^{i\zeta^{3/2}}\\\\0&1\end{array}\right]\,,
\hspace{0.2 in}\mbox{for}\hspace{0.2 in}
\arg(\zeta)=\pi/3\,,\hspace{0.2 in}\mbox{and}\,,
\end{equation}
\begin{equation}
{\bf S}^{\rm loc}_+(\zeta)={\bf S}^{\rm loc}_-(\zeta)
\left[\begin{array}{cc}1 & 0\\\\ie^{-(-\zeta)^{3/2}} & 1\end{array}\right]\,,
\hspace{0.2 in}\mbox{for}\hspace{0.2 in}\arg(\zeta)=\pi\,.
\end{equation}
\item
{\bf Normalization:}  ${\bf S}^{\rm loc}(\zeta)$ is normalized at infinity
so that
\begin{equation}
{\bf S}^{\rm loc}(\zeta)\tilde{\bf S}^{\rm loc}(\zeta)^{-1}\rightarrow
{\mathbb I}\hspace{0.2 in}\mbox{as}\hspace{0.2 in}\zeta\rightarrow\infty\,.
\label{eq:Snorm}
\end{equation}
\end{enumerate}
\label{rhp:Airy}
\end{rhp}

We will describe how Riemann-Hilbert Problem~\ref{rhp:Airy} can be
solved explicitly.  First, we make an explicit change of variable 
to a new matrix ${\bf T}(\zeta)$ by setting
\begin{equation}
{\bf S}^{\rm loc}(\zeta)={\bf T}(\zeta)\left[\begin{array}{cc}
e^{-(-\zeta)^{3/2}/2+i\pi/4} & 0 \\\\
0 & e^{(-\zeta)^{3/2}/2-i\pi/4}\end{array}\right]\,.
\label{eq:SlocfromT}
\end{equation}
If ${\bf S}^{\rm loc}(\zeta)$ satisfies Riemann-Hilbert
Problem~\ref{rhp:Airy}, then it follows that ${\bf T}(\zeta)$
obeys the following jump relations:
\begin{equation}
{\bf T}_+(\zeta)={\bf T}_-(\zeta)\left[\begin{array}{cc} 0 & -1\\\\1&0
\end{array}\right]\,,\hspace{0.2 in}\mbox{for}\hspace{0.2 in}
\arg(\zeta)=0\,,
\end{equation}
\begin{equation}
{\bf T}_+(\zeta)={\bf T}_-(\zeta)\left[\begin{array}{cc} 1 & 1\\\\0&1
\end{array}\right]\,,\hspace{0.2 in}\mbox{for}\hspace{0.2 in}
\arg(\zeta)=\pm\pi/3\,,\hspace{0.2 in}\mbox{and}\,,
\end{equation}
\begin{equation}
{\bf T}_+(\zeta)={\bf T}_-(\zeta)\left[\begin{array}{cc} 1 & 0\\\\1&1
\end{array}\right]\,,\hspace{0.2 in}\mbox{for}\hspace{0.2 in}
\arg(\zeta)=\pi\,.
\end{equation}
It is a general fact that $N\times N$ matrix functions that satisfy
piecewise constant jump conditions, like the matrix ${\bf T}(\zeta)$
does, can be expressed in terms of solutions of $N$th order linear
differential equations with meromorphic (and often rational, or even
polynomial) coefficients.  In the $2\times 2$ case, classical special
functions therefore play a key role.  In this case, we see immediately
from the fact that the boundary values taken by ${\bf S}^{\rm
loc}(\zeta)$ on $\Sigma_{iA(x)}$ are continuous that the matrix
\begin{equation}
{\bf Q}(\zeta):=\frac{d{\bf T}}{d\zeta}(\zeta){\bf T}(\zeta)^{-1}=
\frac{d{\bf S}^{\rm loc}}{d\zeta}(\zeta){\bf S}^{\rm loc}(\zeta)^{-1}
-\frac{3}{4}(-\zeta)^{1/2}{\bf S}^{\rm loc}(\zeta)\left[\begin{array}{cc}
1 & 0\\\\0 & -1\end{array}\right]{\bf S}^{\rm loc}(\zeta)^{-1}
\end{equation}
is analytic for $\zeta\in{\mathbb C}^*$.  If we suppose that ${\bf
T}(\zeta)$ has a bounded derivative near the origin in each sector of
${\mathbb C}\setminus\Sigma_{iA(x)}$ --- a hypothesis that must be
verified later --- then we see that in fact ${\bf Q}(\zeta)$ is analytic
at the origin and is consequently an entire function of $\zeta$.

To work out how ${\bf Q}(\zeta)$ behaves for large $\zeta$, we need to
use the normalization condition (\ref{eq:Snorm}) for ${\bf S}^{\rm
loc}(\zeta)$.  We interpret (\ref{eq:Snorm}) to mean both that
\begin{equation}
{\bf S}^{\rm loc}(\zeta)=\left({\mathbb I}+O\left(\frac{1}{\zeta}\right)\right)
\tilde{\bf S}^{\rm loc}(\zeta)
\label{eq:Snorm2}
\end{equation}
and also that
\begin{equation}
\frac{d{\bf S}^{\rm loc}}{d\zeta}(\zeta)=
\left({\mathbb I}+O\left(\frac{1}{\zeta}\right)\right)
\frac{d\tilde{\bf S}^{\rm loc}}{d\zeta}(\zeta) + 
O\left(\frac{1}{\zeta^2}\right)\tilde{\bf S}^{\rm loc}(\zeta)\,.
\label{eq:Snorm2deriv}
\end{equation}
Both (\ref{eq:Snorm2}) and (\ref{eq:Snorm2deriv}) are again hypotheses
that must be verified once we obtain a solution for ${\bf S}^{\rm
loc}(\zeta)$.  They are not true {\em a priori} by virtue of
(\ref{eq:Snorm}) alone; for example the decay rate in (\ref{eq:Snorm})
might not be as fast as $1/\zeta$, and the error term might have rapid
oscillations that would make its derivative larger than $1/\zeta^2$
thus violating (\ref{eq:Snorm2deriv}).  It follows from our 
hypotheses that ${\bf Q}(\zeta)$ must be a polynomial; in fact,
\begin{equation}
{\bf Q}(\zeta)= \frac{3}{4}\left[\begin{array}{cc} 0 & -\zeta\\\\
1 & 0\end{array}\right]\,.
\end{equation}

Under our hypotheses, it then follows that the matrix ${\bf T}(\zeta)$
solves the linear differential equation
\begin{equation}
\frac{d{\bf T}}{d\zeta}(\zeta) = \frac{3}{4}\left[\begin{array}{cc}
0 & -\zeta\\\\1 & 0\end{array}\right]{\bf T}(\zeta)\,.
\end{equation}
Upon introducing the new independent variable
\begin{equation}
\xi := -\left(\frac{3}{4}\right)^{2/3}\zeta
\end{equation}
we see that the elements of the second row of ${\bf T}$ satisfy
Airy's equation:
\begin{equation}
\frac{d^2T_{2k}}{d\xi^2}=\xi T_{2k}\,,
\end{equation}
and that the elements of the first row are given by
\begin{equation}
T_{1k}=-\left(\frac{4}{3}\right)^{1/3}\frac{dT_{2k}}{d\xi}\,.
\end{equation}

So which solutions of Airy's equation are the appropriate ones for our
purposes?  The first observation is that we need to specify different
solutions of Airy's equation in each simply-connected region of the
complex plane where $T_{21}(\xi)$ and $T_{22}(\xi)$ are analytic.  The
assignment of solutions in these regions must be consistent with the
jump conditions and asymptotics for ${\bf T}(\zeta)$.  From the jump
conditions for ${\bf T}(\zeta)$, we can see that in fact $T_{21}(\xi)$
is analytic in ${\mathbb C}\setminus{\mathbb R}$, while $T_{22}(\xi)$
is analytic except when $\arg(\xi)=\pm 2\pi/3$ or $\xi\in{\mathbb
R}_-$.  We now want to use the normalization condition
(\ref{eq:Snorm2}) to find sectors in which $T_{21}$ and $T_{22}$ are
exponentially decaying.  Then we will be able to uniquely identify
these functions with particular solutions of Airy's equation that also
decay.  From the presumed asymptotic relation (\ref{eq:Snorm2}) we
have (in terms of the variable $\xi$),
\begin{equation}
T_{21}(\xi)=-\left(\frac{3}{32}\right)^{1/6}e^{-i\pi/4}\frac{e^{2\xi^{3/2}/3}}
{\xi^{1/4}}\left(1+O\left(\frac{1}{\xi^{1/2}}\right)\right)\,,
\end{equation}
and
\begin{equation}
T_{22}(\xi)=\left(\frac{3}{32}\right)^{1/6}e^{i\pi/4}\frac{e^{-2\xi^{3/2}/3}}
{\xi^{1/4}}\left(1+O\left(\frac{1}{\xi^{1/2}}\right)\right)\,,
\end{equation}
as $\xi\rightarrow\infty$ with $-\pi<\arg(\xi)<\pi$.  These show that
$T_{21}(\xi)$ is exponentially decaying for $-\pi<\arg(\xi)<-\pi/3$
and also for $\pi/3<\arg(\xi)<\pi$, while $T_{22}(\xi)$ is
exponentially decaying for $-\pi/3<\arg(\xi)<\pi/3$.  Significantly,
both matrix elements are analytic throughout the sectors where they
are exponentially decaying for large $\xi$.  As a basis of linearly
independent solutions of Airy's equation we take the functions ${\rm
Ai}(\xi)$ and ${\rm Ai}(\xi e^{2i\pi/3})$.  These decay in different
sectors, and have the asymptotic expansions
\begin{equation}
{\rm Ai}(\xi)=\frac{1}{2\sqrt{\pi}}\frac{e^{-2\xi^{3/2}/3}}{\xi^{1/4}}(1+
O(1/\xi))
\end{equation}
as $\xi\rightarrow\infty$ for $-\pi<\arg(\xi)<\pi$, and
\begin{equation}
{\rm Ai}(\xi e^{2i\pi/3})=\frac{e^{-i\pi/6}}{2\sqrt{\pi}}
\frac{e^{2\xi^{3/2}/3}}{\xi^{1/4}}(1+O(1/\xi))
\end{equation}
as $\xi\rightarrow\infty$ for $-\pi<\arg(\xi)<-\pi/3$.  Comparing the expansion
of ${\rm Ai}(\xi e^{2i\pi/3})$ with that of $T_{21}(\xi)$ in the sector
$-\pi<\arg(\xi)<-\pi/3$, we find that here
\begin{equation}
T_{21}(\xi)=-e^{-i\pi /12}6^{1/6}\sqrt{\pi}{\rm Ai}(\xi e^{2i\pi/3})\,.
\end{equation}
Since $T_{21}$ is analytic in the lower half $\xi$-plane, this relation
holds identically for $\Im(\xi)<0$.  Similarly, comparing the
expansion of ${\rm Ai}(\xi)$ with that of $T_{22}(\xi)$ in the sector
$-\pi/3<\arg(\xi)<\pi/3$, we find that here
\begin{equation}
T_{22}(\xi)=e^{i\pi/4}6^{1/6}\sqrt{\pi}{\rm Ai}(\xi)\,.
\end{equation}
Being as $T_{22}$ is analytic for $-2\pi/3<\arg(\xi)<2\pi/3$, this identity
holds throughout the sector of analyticity.

Restoring the original independent variable $\zeta$, we find that for
$\Im(\zeta)>0$,
\begin{equation}
T_{21}(\zeta)=-e^{-i\pi /12}6^{1/6}\sqrt{\pi}{\rm Ai}
\left(\left(\frac{3}{4}\right)^{2/3}\zeta e^{-i\pi/3}\right)
\end{equation}
and therefore throughout the same domain,
\begin{equation}
T_{11}(\zeta)=e^{i\pi /12}\left(\frac{32}{3}\right)^{1/6}
\sqrt{\pi}{\rm Ai}'\left(\left(\frac{3}{4}\right)^{2/3}\zeta e^{-i\pi/3}\right)\,.
\end{equation}
For $\zeta$ with $\pi/3<\arg(\zeta)\le\pi$ or $-\pi\le\arg(\zeta)<-\pi/3$,
\begin{equation}
T_{22}(\zeta)=e^{i\pi/4}6^{1/6}\sqrt{\pi}{\rm Ai}\left(-\left(\frac{3}{4}
\right)^{2/3}\zeta\right)\,,
\end{equation}
and therefore throughout the same domain,
\begin{equation}
T_{12}(\zeta)=e^{3i\pi/4}\left(\frac{32}{3}\right)^{1/6}
\sqrt{\pi}{\rm Ai}'\left(-\left(\frac{3}{4}\right)^{2/3}\zeta\right)\,.
\end{equation}
The sector of ${\mathbb C}\setminus\Sigma_{iA(x)}$ that is contained
in both of these domains is $\pi/3<\arg(\zeta)<\pi$.  It is sufficient
to have specified the matrix elements of ${\bf T}(\zeta)$ in this
sector, since it may be consistently obtained in the remaining sectors
of ${\mathbb C}\setminus\Sigma_{iA(x)}$ by making use of the jump relations
for ${\bf T}(\zeta)$.  The procedure is consistent because the cyclic
product of these jump matrices is the identity:
\begin{equation}
\left[\begin{array}{cc} 1 & 0\\\\1 & 1\end{array}\right]
\left[\begin{array}{cc} 1 & 1 \\\\ 0 & 1\end{array}\right]^{-1}
\left[\begin{array}{cc} 0  & -1 \\\\ 1 & 0\end{array}\right]^{-1}
\left[\begin{array}{cc} 1 & 1 \\\\ 0 & 1\end{array}\right]^{-1}
={\mathbb I}\,.
\end{equation}
Once ${\bf T}(\zeta)$ is known for $\zeta\in{\mathbb
C}\setminus\Sigma_{iA(x)}$, the original unknown matrix ${\bf S}^{\rm
loc}(\zeta)$ solving Riemann-Hilbert Problem~\ref{rhp:Airy}
is obtained directly by the transformation (\ref{eq:SlocfromT}).  It
suffices to give a formula that holds for $\pi/3<\arg(\zeta)<\pi$.
We find
\begin{equation}
\begin{array}{rcl}
S^{\rm loc}_{11}(\zeta)&=&\displaystyle
e^{i\pi /3}\left(\frac{32}{3}\right)^{1/6}\sqrt{\pi}e^{-(-\zeta)^{3/2}/2}
{\rm Ai}'\left(\left(\frac{3}{4}\right)^{2/3}\zeta e^{-i\pi/3}\right)\,,\\\\
S^{\rm loc}_{12}(\zeta)&=&\displaystyle
i\left(\frac{32}{3}\right)^{1/6}\sqrt{\pi}e^{(-\zeta)^{3/2}/2}
{\rm Ai}'\left(-\left(\frac{3}{4}\right)^{2/3}\zeta\right)\,,\\\\
S^{\rm loc}_{21}(\zeta)&=&\displaystyle
e^{-5i\pi/6}6^{1/6}\sqrt{\pi}e^{-(-\zeta)^{3/2}/2}
{\rm Ai}\left(\left(\frac{3}{4}\right)^{2/3}\zeta e^{-i\pi/3}\right)\,,\\\\
S^{\rm loc}_{22}(\zeta)&=&\displaystyle
6^{1/6}\sqrt{\pi}e^{(-\zeta)^{3/2}/2}
{\rm Ai}\left(-\left(\frac{3}{4}\right)^{2/3}\zeta\right)\,.
\end{array}
\end{equation}

We will have found a solution to Riemann-Hilbert
Problem~\ref{rhp:Airy} if we can verify the two hypotheses ({\em cf.} 
equations (\ref{eq:Snorm2}) and (\ref{eq:Snorm2deriv})) we made
regarding the interpretation of the normalization condition
(\ref{eq:Snorm}) and the differentiability of ${\bf S}^{\rm
loc}(\zeta)$ at $\zeta=0$.  One verifies these directly, using
the explicit formulae given here.  

\subsubsection{The local model for ${\bf N}(\lambda)$ near $\lambda=iA(x)$.}
To build a better model for ${\bf N}(\lambda)$ in the disk $D$ than
$\hat{\bf N}_{\rm out}(\lambda)$, we begin by recalling the exact
relationship between the matrix ${\bf O}(\lambda)$ and the matrix
${\bf S}(\lambda)$:
\begin{equation}
{\bf O}(\lambda)=\sigma_1^{\frac{1-J}{2}}{\bf S}(\varphi_{iA(x)}(\lambda))
\sigma_1^{\frac{1-J}{2}}
\end{equation}
for all $\lambda\in D$.  The matrix $\tilde{\bf N}(\lambda)$ is also
explicitly related to ${\bf O}(\lambda)$.  For $\lambda\in D$ in the part of
the lens between the contours $L_L$ and $C_M$, 
\begin{equation}
\tilde{\bf N}(\lambda)={\bf O}(\lambda)\sigma_1^{\frac{1-J}{2}}
\left[\begin{array}{cc}1 & \displaystyle -i\exp\left(-\frac{i\theta(\lambda)}
{\hbar_N}\right)\\\\0 & 1\end{array}\right]\sigma_1^{\frac{1-J}{2}}\,,
\end{equation}
for $\lambda\in D$ in the part of the lens between the contours $C_M$
and $L_R$,
\begin{equation}
\tilde{\bf N}(\lambda)={\bf O}(\lambda)\sigma_1^{\frac{1-J}{2}}
\left[\begin{array}{cc}1 & \displaystyle i\exp\left(\frac{i\theta(\lambda)}
{\hbar_N}\right)\\\\0 & 1\end{array}\right]\sigma_1^{\frac{1-J}{2}}\,,
\end{equation}
and for all other $\lambda\in D$, we simply have $\tilde{\bf
N}(\lambda)= {\bf O}(\lambda)$.  As we do not expect the difference
between ${\bf N}(\lambda)$ and the formal continuum limit
approximation $\tilde{\bf N}(\lambda)$ to be important in the disk $D$
since it is isolated from the points $\lambda=0$ and $\lambda=iA$, we
can obtain a guess for an approximation for ${\bf N}(\lambda)$ that
should be valid in $D$ simply by substituting $\hat{\bf
S}(\varphi_{iA(x)}(\lambda))$ for ${\bf S}(\varphi_{iA(x)}(\lambda))$
in these formulae.

Putting these steps together, the model for ${\bf N}(\lambda)$ for
$\lambda\in D$ that we will use is defined as follows.  For $\lambda\in D$
in the lens between $L_L$ and $C_M$, set
\begin{equation}
\hat{\bf N}_{\rm endpoint}(\lambda):=
\sigma_1^{\frac{1-J}{2}}
\hat{\bf S}(\varphi_{iA(x)}(\lambda))
\left[\begin{array}{cc}1 & \displaystyle -i\exp\left(-\frac{i\theta(\lambda)}
{\hbar_N}\right)\\\\0 & 1\end{array}\right]\sigma_1^{\frac{1-J}{2}}\,,
\end{equation}
for $\lambda\in D$ in the lens between $C_M$ and $L_R$, set
\begin{equation}
\hat{\bf N}_{\rm endpoint}(\lambda):=
\sigma_1^{\frac{1-J}{2}}\hat{\bf S}(\varphi_{iA(x)}(\lambda))
\left[\begin{array}{cc}1 & \displaystyle i\exp\left(\frac{i\theta(\lambda)}
{\hbar_N}\right)\\\\0 & 1\end{array}\right]\sigma_1^{\frac{1-J}{2}}\,,
\end{equation}
and for all other $\lambda\in D$, set
\begin{equation}
\hat{\bf N}_{\rm endpoint}(\lambda):=
\sigma_1^{\frac{1-J}{2}}\hat{\bf S}(\varphi_{iA(x)}(\lambda))
\sigma_1^{\frac{1-J}{2}}\,,
\end{equation}
where $\hat{\bf S}(\zeta)$ is defined by (\ref{eq:Shatdef}). The most
important properties of the matrix $\hat{\bf N}_{\rm
endpoint}(\lambda)$ in the disk $D$ are the following.
\begin{proposition}
The matrix $\hat{\bf N}_{\rm endpoint}(\lambda)$ is piecewise analytic
in the left and right half-disks of $D$.  On the imaginary axis (which
bisects $D$) oriented in the positive imaginary direction, 
\begin{equation}
\hat{\bf N}_{{\rm endpoint},-}(\lambda)^{-1}
\hat{\bf N}_{{\rm endpoint},+}(\lambda)
 = \tilde{\bf N}_-(\lambda)^{-1}\tilde{\bf N}_+(\lambda)\,,
\end{equation}
that is, the local model has exactly the same jump as $\tilde{\bf
N}(\lambda)$.  For $\lambda\in D$, the matrix function $\hat{\bf
N}_{\rm endpoint}(\lambda)$ is bounded by a constant of order
$\hbar_N^{-1/3}$.  Also, for $\lambda\in \partial D$, 
\begin{equation}
\hat{\bf N}_{\rm endpoint}(\lambda)\hat{\bf N}_{\rm out}(\lambda)^{-1}
={\mathbb I} + O(\hbar_N^{1/3})\,.
\end{equation}
\label{prop:endpointproperties}
\end{proposition}

{\em Proof:} Computing the jump matrix for $\hat{\bf N}_{\rm
endpoint}(\lambda)$ is straightforward.  To show the bound on
$\hat{\bf N}_{\rm endpoint}(\lambda)$, we recall that $\tilde{\bf
S}^{\rm hol}(\varphi_{iA(x)}(\lambda))$ is bounded by a quantity of
order $\hbar_N^{-1/6}$, and note that the other factor of the same
size comes from the factor ${\bf S}^{\rm
loc}(\varphi_{iA(x)}(\lambda))$ via the normalization condition on
this matrix and the fact that $\varphi_{iA(x)}(\lambda)$ grows like
$\hbar_N^{-2/3}$ for $\lambda\in D$.  Similar reasoning using
(\ref{eq:Snorm2}) establishes the error in matching onto $\hat{\bf
N}_{\rm out}(\lambda)$ on the boundary of $D$. $\Box$

\section{The Parametrix and its Error}
We are now in a position to put all of our models together to build a
guess for a uniformly valid approximation of ${\bf N}(\lambda)$.  Such
a guess is called a parametrix.

\subsection{Constructing the parametrix.}
\begin{figure}[h]
\begin{center}
\mbox{\psfig{file=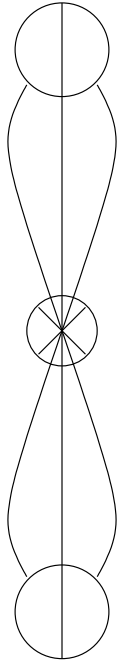,width=0.6 in}}
\end{center}
\caption{\em The contour $\hat{\Sigma}$.  The circles at the top and bottom
of the figure are the boundaries of the disks $D$ and $D^*$
respectively.  The circle at the origin has radius $\hbar_N^\alpha$.
The contours to the left and right of the imaginary axis in the upper
half-plane are portions of the lens boundaries $L_L$ and $L_R$
respectively.  The remaining small segments present in the upper
half-plane for $|\lambda|<\hbar_N^\alpha$ are parts of $C_L$ and
$C_R$.}
\label{fig:sigmahat}
\end{figure}

To build the parametrix $\hat{\bf N}(\lambda)$ as a sectionally
holomorphic matrix function, we simply combine the outer and local
models.  For all $\lambda$ satisfying $|\lambda|\le\hbar_N^\alpha$, where
the parameter $\alpha$ is to be determined later, set
\begin{equation}
\hat{\bf N}(\lambda):=\hat{\bf N}_{\rm origin}(\lambda)\,.
\end{equation}
For $\lambda\in D$, we set
\begin{equation}
\hat{\bf N}(\lambda):=\hat{\bf N}_{\rm endpoint}(\lambda)\,,
\end{equation}
and by symmetry for all $\lambda\in D^*$ we set
\begin{equation}
\hat{\bf N}(\lambda):=\sigma_2\hat{\bf N}_{\rm endpoint}(\lambda^*)^*
\sigma_2\,.
\end{equation}
Finally, for all remaining $\lambda\in{\mathbb C}$, set
\begin{equation}
\hat{\bf N}(\lambda):=\hat{\bf N}_{\rm out}(\lambda)\,.
\end{equation}
The parametrix $\hat{\bf N}(\lambda)$ is holomorphic for $\lambda\in
{\mathbb C}\setminus \hat{\Sigma}$, where $\hat{\Sigma}$ is the
contour illustrated in Figure~\ref{fig:sigmahat}.

\subsection{Estimating the error.}
\begin{figure}[h]
\begin{center}
\mbox{\psfig{file=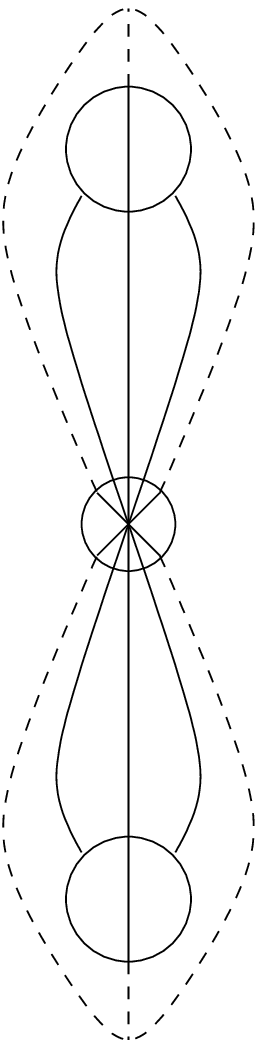,width=1 in}}
\end{center}
\caption{\em The contour $\Sigma_E$.  We have $\hat{\Sigma}\subset\Sigma_E$
and the components of $\Sigma_E\setminus\hat{\Sigma}$ are shown in dashed
lines to make a clear comparison with Figure~\ref{fig:sigmahat}.}
\label{fig:sigmaE}
\end{figure}

To determine the accuracy of the parametrix, we compare it directly with
the original matrix ${\bf N}(\lambda)$.  That is, we consider the error
matrix defined by
\begin{equation}
{\bf E}(\lambda):={\bf N}(\lambda)\hat{\bf N}(\lambda)^{-1}\,.
\label{eq:error}
\end{equation}
This matrix is sectionally analytic in the complex $\lambda$-plane,
with discontinuities on a contour $\Sigma_E$ that is illustrated in
Figure~\ref{fig:sigmaE}.
Note that as a consequence of the symmetry of $\hat{\bf N}(\lambda)$
and ${\bf N}(\lambda)$ under complex conjugation, we have ${\bf E}(\lambda^*)=
\sigma_2{\bf E}(\lambda)^*\sigma_2$.
If the parametrix is indeed a good model for ${\bf N}(\lambda)$, then
we must be able to show that the matrix ${\bf E}(\lambda)$ is
uniformly close to the identity matrix in the whole complex plane.

While we do not know ${\bf E}(\lambda)$ explicitly like we know
$\hat{\bf N}(\lambda)$, we know from the normalization condition of
both factors that
\begin{equation}
{\bf E}(\lambda)\rightarrow{\mathbb I}\hspace{0.2 in}
\mbox{as}\hspace{0.2 in}
\lambda\rightarrow\infty\,.
\end{equation}
It turns out that we can also calculate explicitly the ratio of
boundary values taken by ${\bf E}(\lambda)$ from both sides on each
arc of $\Sigma_E$.  That is, we know the jump matrix for ${\bf
E}(\lambda)$, and can express it explicitly in terms of $\hat{\bf
N}(\lambda)$ and the jump matrix for ${\bf N}(\lambda)$, both of which
are known\footnote{Or at least well-understood.  We have characterized
the parametrix for $|\lambda|\le\hbar_N^\alpha$ in terms of the matrix
function $\hat{F}(\zeta)$ for which we have an existence proof and a
characterization, but not an explicit formula.}.  This means that the
matrix ${\bf E}(\lambda)$ itself is a solution of a particular
Riemann-Hilbert problem for which we know the data.  By solving this
Riemann-Hilbert problem, we will show that indeed ${\bf E}(\lambda)$
is uniformly close to the identity matrix.

There are two kinds of arcs in the contour $\Sigma_E$: ``matching''
arcs of the circles $\partial D$, $\partial D^*$, and
$|\lambda|=\hbar_N^\alpha$ where two different components of the
parametrix have to match well onto each other, and the remaining arcs
within the disks and outside the disks where the jump matrix for
$\hat{\bf N}(\lambda)$ should be a good approximation to that of ${\bf
N}(\lambda)$.

Consider one of the arcs of $\Sigma_E$ oriented in some convenient
way, and as usual let the subscript ``$+$'' (respectively ``$-$'')
denote a boundary value taken on the arc from its left (respectively
right).  We can easily see from the definition (\ref{eq:error}) that
for $\lambda$ on this arc,
\begin{equation}
{\bf E}_+(\lambda)={\bf E}_-(\lambda){\bf v}_{\bf E}(\lambda)
\hspace{0.2 in}\mbox{with}\hspace{0.2 in}
{\bf v}_{\bf E}(\lambda):=\hat{\bf N}_-(\lambda){\bf v}_{\bf N}(\lambda)
\hat{\bf v}_{\hat{\bf N}}(\lambda)^{-1}\hat{\bf N}_-(\lambda)^{-1}
\label{eq:Ejump}
\end{equation}
where ${\bf v}_{\bf N}(\lambda)$ and ${\bf v}_{\hat{\bf N}}(\lambda)$
denote the jump matrices on the arc for ${\bf N}(\lambda)$ and the
parametrix $\hat{\bf N}(\lambda)$ respectively.  If the arc under
consideration is a ``matching'' arc, then the discontinuity in ${\bf
E}(\lambda)$ is wholly due to the mismatch of components of the
parametrix, and the jump matrix ${\bf v}_{\bf N}(\lambda)$ is
therefore replaced with the identity matrix in (\ref{eq:Ejump}).  It then
follows that an equivalent formula for ${\bf v}_{\bf E}(\lambda)$ on
a ``matching'' arc is the following:
\begin{equation}
{\bf v}_{\bf E}(\lambda)=\hat{\bf N}_-(\lambda)\hat{\bf N}_+(\lambda)^{-1}\,,
\hspace{0.2 in}\mbox{for $\lambda$ on a ``matching'' arc of $\Sigma_E$.}
\end{equation}
In this case, the two boundary values represent different components
of the parametrix, for example $\hat{\bf N}_{\rm out}(\lambda)$ would
play the role of $\hat{\bf N}_+(\lambda)$ and $\hat{\bf N}_{\rm
endpoint}(\lambda)$ would play that of $\hat{\bf N}_-(\lambda)$ if the
``matching'' arc under consideration is an arc of $\partial D$,
oriented clockwise.

The key fact that we need now is the following.
\begin{proposition}
The optimal value of the radius parameter $\alpha$ is $\alpha=6/7$.
For this value of $\alpha$, and for all $\nu>0$
arbitrarily small,
\begin{equation}
{\bf v}_E(\lambda)-{\mathbb I}=O(\hbar_N^{1/7-\nu})\,.
\end{equation}
uniformly for all $\lambda\in\Sigma_E$.
\label{prop:errorjumpbound}
\end{proposition}

{\em Proof:}  We begin by considering the ``matching'' arcs.  We take
the circle $\partial D$ to be oriented in the clockwise direction.  Here we
find
\begin{equation}
{\bf v}_{\bf E}(\lambda)=\hat{\bf N}_{\rm endpoint}(\lambda)\hat{\bf N}_{\rm out}(\lambda)^{-1} = {\mathbb I} + O(\hbar_N^{1/3})\,,
\label{eq:vE1}
\end{equation}
with the error estimate coming from Proposition~\ref{prop:endpointproperties}.
An estimate of the same form necessarily holds on the ``matching'' arcs of
$\partial D^*$ according to the conjugation symmetry of ${\bf E}(\lambda)$.
The remaining ``matching'' arcs lie on the circle $|\lambda|=\hbar_N^\alpha$,
which again we take to be oriented in the clockwise direction.  Here we
find
\begin{equation}
{\bf v}_{\bf E}(\lambda)=\hat{\bf N}_{\rm origin}(\lambda)\hat{\bf N}_{\rm out}(\lambda)^{-1} = {\mathbb I} + O(\hbar_N^{1-\alpha})\,,
\label{eq:vE2}
\end{equation}
with the error estimate coming from
Proposition~\ref{prop:originmodelproperties}.

We continue by considering the arcs of $\Sigma_E$ with
$|\lambda|<\hbar_N^\alpha$.  Using the fact recorded in
Proposition~\ref{prop:originmodelproperties} that $\hat{\bf N}_{\rm
origin}(\lambda)$ has determinant one and is uniformly bounded, we see
from (\ref{eq:Ejump}) that the important quantity to estimate is
simply ${\bf v}_{\bf N}(\lambda){\bf v}_{\hat{\bf
N}}(\lambda)^{-1}-{\mathbb I}$, the difference between the jump matrix
ratio and the identity.  First, consider the portion of the contour
$C_L$ with $|\lambda|<1$.  Using
Proposition~\ref{prop:originmodelproperties} and (\ref{eq:CLjumpN}),
we find that here
\begin{equation}
{\bf v}_{\bf N}(\lambda){\bf v}_{\hat{\bf N}}(\lambda)^{-1} = 
\sigma_1^{\frac{1-J}{2}}\left[\begin{array}{cc}
1 & 0 \\\\
\displaystyle
a_L(\lambda)-ie^{i\theta(0)/\hbar_N}e^{(2i+\pi)\zeta}
(-i\zeta)^{-i\zeta}(i\zeta)^{-i\zeta}\frac{\Gamma(1/2+i\zeta)}
{\Gamma(1/2-i\zeta)} & 1\end{array}\right]
\sigma_1^{\frac{1-J}{2}}\,,
\label{eq:CLquotient}
\end{equation}
where $\zeta=\varphi_0(\lambda)$.  Using (\ref{eq:aLmodel}) and the
uniform boundedness of the leading-order term on the right-hand side
of (\ref{eq:aLmodel}) for $|\lambda|<\hbar_N^\alpha$, we see that the
matrix quotient in (\ref{eq:CLquotient}) differs from the identity
matrix by an order $\hbar_N^{4\alpha/3-1}$ amount.  Virtually the same
argument using (\ref{eq:CRjumpN}) and (\ref{eq:aRmodel}) in
conjunction with Proposition~\ref{prop:originmodelproperties}
establishes that on $C_R$ the matrix quotient ${\bf v}_{\bf
N}(\lambda){\bf v}_{\hat{\bf N}}(\lambda)^{-1}$ differs from the
identity by a quantity of order $\hbar_N^{4\alpha/3-1}$.  Next,
consider the contour $L_L$ for $|\lambda|<\hbar_N^\alpha$.  On this
contour, there is no jump for ${\bf N}(\lambda)$, so ${\bf v}_{\bf
N}(\lambda)={\mathbb I}$ in the formula (\ref{eq:Ejump}) for ${\bf
v}_{\bf E}(\lambda)$.  But from the formula for ${\bf v}_{\hat{\bf
N}}(\lambda)$ for $\lambda\in L_L$ given in
Proposition~\ref{prop:originmodelproperties}, we see by ordinary
Taylor expansion that for $|\lambda|<\hbar_N^\alpha$, we have ${\bf
v}_{\bf E}(\lambda)-{\mathbb I}=O(\hbar_N^{2\alpha-1})$ for
$\lambda\in L_L$.  Virtually the same argument yields the same
estimate for ${\bf v}_{\bf E}(\lambda)-{\mathbb I}$ on $L_R$ with
$|\lambda|<\hbar_N^\alpha$.  Finally consider the contour $C_M$ (the
positive imaginary axis) with $|\lambda|<\hbar_N^\alpha$.  Using
(\ref{eq:CMjumpN}) and the jump matrix ${\bf v}_{\hat{\bf N}}(\lambda)$
for $\lambda\in C_M$ recorded in Proposition~\ref{prop:originmodelproperties},
we find that here
\begin{equation}
{\bf v}_{\bf N}(\lambda){\bf v}_{\hat{\bf N}}(\lambda)^{-1}=
\sigma_1^{\frac{1-J}{2}}\left[\begin{array}{cc}
1+W_+Z & ie^{i\theta(\lambda)/\hbar_N}(W_++W_-)Z\\\\
e^{-i\theta(\lambda)/\hbar_N}\left(a_M(\lambda)-i(1+Z)\right) &
1+i(W_++W_-)a_M(\lambda)Z + W_-Z\end{array}
\right]\sigma_1^{\frac{1-J}{2}}\,,
\end{equation}
where $Z$ is given in terms of $\zeta=\varphi_0(\lambda)$
by (\ref{eq:Delta}) and where
\begin{equation}
W_\pm:=1-e^{\pm \pi\zeta + i(\theta(\lambda)-\theta(0))/\hbar_N}\,.
\end{equation}
Note that $W_+$ and $W_-$ are both of order $\hbar_N^{2\alpha-1}$ for
$|\lambda|<\hbar_N^\alpha$ by Taylor expansion arguments.  Also,
$Z$ is uniformly bounded in the disk of radius $\hbar_N^\alpha$,
and $e^{\pm i\theta(\lambda)/\hbar_N}$ both have modulus one for
$\lambda\in C_M$ in this disk.  Also, from (\ref{eq:aMmodel}) we get
$a_M(\lambda)-i(1+Z)=O(\hbar_N^{4\alpha/3-1-\nu})$ for all $\nu>0$
since $\tilde{\phi}(\lambda)\equiv 0$ on this part of $C_M$.  This
error dominates those arising from Taylor approximation, and thus on
$C_M$ with $|\lambda|<\hbar_N^\alpha$ we find that ${\bf v}_{\bf
N}(\lambda){\bf v}_{\hat{\bf N}}(\lambda)^{-1}-{\mathbb
I}=O(\hbar_N^{4\alpha/3-1-\nu})$.  The corresponding estimates hold
on the corresponding contours in the lower half-disk, by conjugation
symmetry.  Putting this information together with the uniform boundedness
of $\hat{\bf N}_{\rm origin}(\lambda)$ and its inverse, we find that for
all $\lambda\in \Sigma_E$ with $|\lambda|<\hbar_N^\alpha$, 
\begin{equation}
{\bf v}_{\bf E}(\lambda)-{\mathbb I}=O(\hbar_N^{4\alpha/3-1-\nu})\,,
\label{eq:vE3}
\end{equation}
for all $\nu>0$.

Now we proceed to study the jump matrix ${\bf v}_{\bf E}(\lambda)$
inside the disk $D$ centered at the endpoint $\lambda=iA(x)$, assuming
$x\neq 0$ is fixed so that $D$ is fixed and bounded away from the
origin and from $\lambda=iA$.  The only contour we need to consider is
the imaginary axis.  The jump matrix ${\bf v}_{\bf E}(\lambda)$ is
given by (\ref{eq:Ejump}).  This time, the conjugating factors of
$\hat{\bf N}_{{\rm endpoint},-}(\lambda)$ and its inverse are not
uniformly bounded in $D$ as $\hbar_N$ tends to zero. According to
Proposition~\ref{prop:endpointproperties} each conjugating matrix
contributes an amplifying factor of $\hbar_N^{-1/3}$.  Also according
to Proposition~\ref{prop:endpointproperties}, we have that ${\bf
v}_{\hat{\bf N}}(\lambda)$ is the same as the jump matrix for
$\tilde{\bf N}(\lambda)$.  Therefore, using (\ref{eq:CMjumpN})
and (\ref{eq:ntildemiddle}), we find that
\begin{equation}
{\bf v}_{\bf N}(\lambda){\bf v}_{\hat{\bf N}}(\lambda)^{-1}=
\sigma_1^{\frac{1-J}{2}}\left[\begin{array}{cc}1 & 0\\\\
e^{-i\theta(\lambda)/\hbar_N}\left(a_M(\lambda)-ie^{i\tilde{\phi}(\lambda)/\hbar_N}\right) & 1\end{array}\right]\sigma_1^{\frac{1-J}{2}}\,.
\end{equation}
Using Proposition~\ref{prop:a_M} and the fact that $\Re(\tilde{\phi})\le 0$
while $e^{-i\theta(\lambda)/\hbar_N}$ has modulus one on $C_M$ within $D$,
we get that ${\bf v}_{\bf N}(\lambda){\bf v}_{\hat{\bf N}}(\lambda)^{-1}=
{\mathbb I}+O(\hbar_N^{1-\mu})$ for all $\mu>0$.  Combining this with 
the bounds on the conjugating factors, we find that within $D$,
\begin{equation}
{\bf v}_{\bf E}(\lambda)-{\mathbb I} = O(\hbar_N^{1/3-\mu})
\label{eq:vE4}
\end{equation}
for all $\mu>0$.  By conjugation symmetry, the same estimate holds for the
jump matrix ${\bf v}_{\bf E}(\lambda)$ when $\lambda\in D^*$.

Finally, we consider the parts of $\Sigma_E$ outside all of the disks,
where we have set $\hat{\bf N}(\lambda):=\hat{\bf N}_{\rm
out}(\lambda)$.  On all of these parts of $\Sigma_E$,
Proposition~\ref{prop:Noutproperties} guarantees that the conjugating
factors $\hat{\bf N}_{{\rm out},-}(\lambda)$ and $\hat{\bf N}_{{\rm
out},-}(\lambda)^{-1}$ are uniformly bounded as $\hbar_N$ tends to
zero.  So it remains to determine the magnitude of the difference
between the quotient ${\bf v}_{\bf N}(\lambda){\bf v}_{\hat{\bf
N}}(\lambda)^{-1}$ and the identity.  First consider the part of $C_M$
between the disk at the origin and the disk $D$.  Using
Proposition~\ref{prop:Noutproperties} to find ${\bf v}_{\hat{\bf
N}}(\lambda)$ and recalling (\ref{eq:CMjumpN}), we find 
\begin{equation}
{\bf v}_{\bf N}(\lambda){\bf v}_{\hat{\bf N}}(\lambda)^{-1}
=\sigma_1^{\frac{1-J}{2}}\left[\begin{array}{cc}
1 & 0\\\\e^{-i\theta(\lambda)/\hbar_N}\left(a_M(\lambda)-i\right) & 1\end{array}
\right]\sigma_1^{\frac{1-J}{2}}\,.
\end{equation}
Using Proposition~\ref{prop:a_M} and the fact that
$\tilde{\phi}(\lambda)\equiv 0$ while $e^{-i\theta(\lambda)/\hbar_N}$
has modulus one here then allows us to conclude that ${\bf v}_{\bf
N}(\lambda){\bf v}_{\hat{\bf N}}(\lambda)^{-1}={\mathbb I}
+O(\hbar_N^{1-\alpha-\mu})$ for all $\mu>0$.  The $\alpha$ appears
because we need the estimate down to the outside boundary of the
shrinking disk at the origin.  Next we look at the contours $L_L$ and
$L_R$.  On these contours there is no jump for ${\bf N}(\lambda)$, and
we see directly from Proposition~\ref{prop:Noutproperties} that ${\bf
v}_{\bf N}(\lambda){\bf v}_{\hat{\bf N}}(\lambda)^{-1}-{\mathbb I}$ is
exponentially small as $\hbar_N$ tends to zero through positive
values.  The next contour we examine is the portion of $C_M$ lying
above the disk $D$.  Here we observe that $\hat{\bf N}_{\rm
out}(\lambda)$ has no jump, and because we have on this contour the
strict inequality $\Re(\tilde{\phi}(\lambda))<0$, the matrix ${\bf
v}_{\bf N}(\lambda)$ is exponentially close to the identity matrix.
To see this, note that here
\begin{equation}
{\bf v}_{\bf N}(\lambda)=\sigma_1^{\frac{1-J}{2}}\left[\begin{array}{cc}
1 & 0 \\\\ a_M(\lambda) & 1\end{array}\right]\sigma_1^{\frac{1-J}{2}}
\end{equation}
because $\theta(\lambda)\equiv 0$.  While the relative error in
replacing $a_M(\lambda)$ by $ie^{\tilde{\phi}(\lambda)/\hbar_N}$ is
not small near $\lambda=iA$, it is bounded.  So the exponential decay
afforded by the strict inequality on the real part of
$\tilde{\phi}(\lambda)$ is maintained.  Virtually the same arguments
show that on the contours $C_L$ and $C_R$ outside of the disk at the
origin, the quotient ${\bf v}_{\bf N}(\lambda){\bf v}_{\hat{\bf
N}}(\lambda)^{-1}$ is again an exponentially small perturbation of the
identity matrix, since on these contours there is again no jump of the
parametrix $\hat{\bf N}_{\rm out}(\lambda)$.  Therefore, for all
$\lambda\in \Sigma_E$ outside all disks, we have
\begin{equation}
{\bf v}_{\bf E}(\lambda)-{\mathbb I}=O(\hbar_N^{1-\alpha-\mu})
\label{eq:vE5}
\end{equation}
for all $\mu>0$.  

We come up with an overall estimate for ${\bf v}_{\bf
E}(\lambda)-{\mathbb I}$ for $\lambda\in\Sigma_E$ by combining the
estimates (\ref{eq:vE1}), (\ref{eq:vE2}), (\ref{eq:vE3}),
(\ref{eq:vE4}), and (\ref{eq:vE5}), and optimizing the error by
choosing the parameter $\alpha$.  The optimal balance among all
$\alpha\in(3/4,1)$ comes from taking $\alpha=6/7$, which gives an overall
error estimate of 
\begin{equation}
{\bf v}_{\bf E}(\lambda)-{\mathbb I}=O(\hbar_N^{1/7-\nu})
\end{equation} 
uniformly for all $\lambda\in \Sigma_E$, for all $\nu>0$.  This proves
the proposition.  $\Box$

The following is then a consequence of the $L^2$ theory of
Riemann-Hilbert problems (see the analogous discussion in
\cite{manifesto}).
\begin{proposition}
For $\hbar_N$ sufficiently small, the Riemann-Hilbert problem for
${\bf E}(\lambda)$ has a unique solution.  Let $R>0$ be sufficiently
large so that $\Sigma_E$ is contained in the circle of radius $R$
centered at the origin.  Then uniformly for all $\lambda$ outside of
this circle, and for all $\mu>0$ however small, the matrix ${\bf
E}(\lambda)$ satisfies
\begin{equation}
{\bf E}(\lambda)-{\mathbb I}=O(\hbar_N^{1/7-\nu})
\end{equation}
with the size of the matrix measured in any matrix norm.
\end{proposition}

{\em Proof:}
Only one thing must be verified in order to deduce existence and
uniqueness from the general theory: the Cauchy-kernel singular
integral operators defined on the contour $\Sigma_E$, which depends on
$\hbar_N$ because of the shrinking boundary of the circle at the
origin, have $L^2(\Sigma_E)$ norms that can be bounded uniformly in
$N$.  But this fact follows in this case from the fact that for $N$
sufficiently large, the circle $|\lambda|=\hbar_N^\alpha$ intersects
only radial straight-line segments (we chose the contours to all be
exactly straight lines in some fixed neighborhood of the origin), so
that the portion of $\Sigma_E$ near the origin simply scales with
$\hbar_N^\alpha$.  The uniform bound we need can then be established
using the fact that the Cauchy operators commute with scaling.  A
similar result was established under more general conditions in
\cite{manifesto}.  Once existence and uniqueness have been
established, the estimate of ${\bf E}(\lambda)$ follows from an
integral representation formula for this matrix:
\begin{equation}
{\bf E}(\lambda)={\mathbb I} + \frac{1}{2\pi i}\int_{\Sigma_E}
(s-\lambda)^{-1}({\bf m}(s)({\bf v}_{\bf E}(s)-{\mathbb I}))\,ds
\label{eq:cauchyformula}
\end{equation}
in which the matrix function ${\bf m}(\lambda)$ for $\lambda\in
\Sigma_E$ is an element of $L^2(\Sigma_E)$ with a norm that is bounded
independently of $\hbar_N$ (and in fact converges to the identity
matrix in $L^2(\Sigma_E)$ as $\hbar_N$ tends to zero). $\Box$

We are now in a position to prove our main result, which we presented
as Theorem~\ref{theorem:mainresult} in the introduction.

\vspace{0.1 in}

\noindent{\em Proof of Theorem~\ref{theorem:mainresult}:}  
We have been setting $t=0$ all along, so the function
$\psi_0^{\hbar_N}(x)$ given by (\ref{eq:psi0define}) can be found from
the matrix ${\bf N}(\lambda)$ by the relation ({\em cf.} equation
(\ref{eq:potentialrecovery}))
\begin{equation}
\psi_0^{\hbar_N}(x)=2i\lim_{\lambda\rightarrow\infty}\lambda N_{12}(\lambda)\,.
\end{equation}
Writing ${\bf N}(\lambda)={\bf E}(\lambda)\hat{\bf
N}(\lambda)=\hat{\bf N}(\lambda) + ({\bf E}(\lambda)-{\mathbb
I})\hat{\bf N}(\lambda)$, we get
\begin{equation}
\psi_0^{\hbar_N}(x)=2i\lim_{\lambda\rightarrow\infty}\lambda
\hat{N}_{12}(\lambda) + 2i\lim_{\lambda\rightarrow\infty}
\lambda (E_11(\lambda)-1)\hat{N}_{12}(\lambda) +
2i\lim_{\lambda\rightarrow\infty}\lambda E_{12}(\lambda)\hat{N}_{22}(\lambda)\,.
\label{eq:controlpsi}
\end{equation}
Now, the first term on the right-hand side of (\ref{eq:controlpsi})
can be evaluated explicitly since near $\lambda=\infty$ we have
$\hat{\bf N}(\lambda)\equiv
\hat{\bf N}_{\rm out}(\lambda)\equiv\tilde{\bf O}(\lambda)$, and we have 
an explicit formula ({\em cf.} equation (\ref{eq:otildesolution})) for
$\tilde{\bf O}(\lambda)$.  We find
\begin{equation}
2i\lim_{\lambda\rightarrow\infty}\lambda\hat{N}_{12}(\lambda)=A(x)\,,
\end{equation}
which is the ``true'' initial data that we started with, before making
any modifications based on the WKB approximation of the spectral data.
When we consider the second term on the right-hand side of
(\ref{eq:controlpsi}), we see that as a consequence of the
normalization of the matrices ${\bf E}(\lambda)$ and $\hat{\bf
N}(\lambda)$,
\begin{equation}
{\bf E}(\lambda)={\mathbb I} + O(1/\lambda)\,,\hspace{0.2 in}
\mbox{and}\hspace{0.2 in}
\tilde{\bf N}(\lambda)={\mathbb I} + O(1/\lambda)\,,
\end{equation}
as $\lambda\rightarrow\infty$.  Therefore the second term on the right-hand
side of (\ref{eq:controlpsi}) vanishes identically.  Finally, for the third
term on the right-hand side of (\ref{eq:controlpsi}) we can again apply the
normalization condition for $\hat{\bf N}(\lambda)$ to obtain
\begin{equation}
2i\lim_{\lambda\rightarrow\infty} \lambda E_{12}(\lambda)\hat{N}_{22}(\lambda) = 2i\lim_{\lambda\rightarrow\infty}\lambda E_{12}(\lambda)\,.
\end{equation}
Putting these steps together, we have
\begin{equation}
\psi_0^{\hbar_N}(x)-A(x) = 2i\lim_{\lambda\rightarrow\infty}
\lambda E_{12}(\lambda)\,.
\end{equation}
The proof of the theorem is finished upon using the integral formula
(\ref{eq:cauchyformula}) for ${\bf E}(\lambda)$ and
Proposition~\ref{prop:errorjumpbound}. $\Box$

\section{Discussion}
Using the new technique of simultaneous interpolation of residues by
two different analytic interpolating functions, combined with
``steepest-descent'' techniques for matrix-valued Riemann-Hilbert
problems, we have established the validity of the formal WKB
approximation of the spectrum in the nonselfadjoint Zakharov-Shabat
eigenvalue problem (\ref{eq:ZS}) in the sense of pointwise convergence
of the potentials.  Strictly speaking, our analysis applies to certain
classes of potential functions whose most important property for our
purposes is their real analyticity, and then we obtain convergence for
all nonzero values of $x$.

In order to extend the result of Theorem~\ref{theorem:mainresult} to
$x=0$, some different steps are required.  Since $A(x)\rightarrow A$
as $x\rightarrow 0$, the local analysis that we carried out
independently for $\lambda\approx iA$ ({\em cf.}
\S~\ref{sec:local_iA}) and 
for $\lambda\approx iA(x)$ ({\em cf.}
\S~\ref{sec:local_iA(x)}) will need to be combined.  Consequently, a
different local model for ${\bf N}(\lambda)$ will need to be
constructed near $\lambda=iA=iA(0)$.  Due to the presence of the gamma
functions in the asymptotics established in \S~\ref{sec:local_iA} and
given in Proposition~\ref{prop:iAasympS} and
Proposition~\ref{prop:iAasympT}, it is likely that the construction of
the local model will require knowledge of the solution of a new
Riemann-Hilbert problem that, like that for the matrix $\hat{\bf
F}(\zeta)$ in \S~\ref{sec:origin}, cannot be solved explicitly.
Nonetheless, one expects that to establish the validity of
Theorem~\ref{theorem:mainresult} for $x=0$ will require only technical
modifications of what we have done here.

Understanding the nature of the WKB approximation at the level of the
potentials is one step in a larger ongoing program to obtain
corresponding information at the level of the (unknown) spectrum
itself.  Indeed, quantifying the difference between the true spectrum
of a given potential $A(x)$ and the WKB approximation of the spectrum,
in terms of motion of eigenvalues, will be necessary before it can be
proven that the rigorous asymptotic analysis of SSEs is relevant to
the problem of semiclassical asymptotics for the initial-value problem
for the focusing nonlinear Schr\"odinger equation (\ref{eq:nls}).  One
imagines that a study of the semiclassical limit for (\ref{eq:nls})
should proceed by first generating from the given initial data
$\psi(x,0)=A(x)$ the corresponding well-defined SSE, and then using
the fact that by combining the results of \cite{manifesto} with
Theorem~\ref{theorem:mainresult} from this paper, one has a complete
picture of the limiting behavior of the SSE for an open interval of
time $t$ that is independent of $\hbar$ and includes $t=0$, and
moreover that according to Theorem~\ref{theorem:mainresult} the SSE is
pointwise close to the given initial data $A(x)$ for $t=0$.  The
problem here is that the focusing nonlinear Schr\"odinger equation is
known to have modulational instabilities whose exponential growth
rates become arbitrarily large in the semiclassical limit.  There is,
therefore, the very real possibility that while the SSE is close to
the initial data $A(x)$ at $t=0$, it is not close to the corresponding
solution of (\ref{eq:nls}) for any positive $t$.  In order to control
the difference for positive time, it is necessary to know in advance
how much the SSE spectral data differs from the true (unknown)
spectral data, as it is the spectral data that is the starting point
for analysis ({\em cf.}  Riemann-Hilbert
Problem~\ref{rhp:meromorphic}).  One can imagine obtaining this sort
of information from the pointwise estimate given in
Theorem~\ref{theorem:mainresult} by Rayleigh-Schr\"odinger
perturbation theory.

\section*{Acknowledgements}
I would like to thank Jinho Baik, Thomas Kriecherbauer, and Ken
McLaughlin for useful feedback.  This work was supported in part by 
the National Science Foundation under award number DMS-0103909.

\end{document}